# Lintasan Bebas Rata-rata Neutrino di Bintang Netron

Tesis
Diajukan sebagai salah satu syarat
untuk memperoleh gelar

Magister Fisika

oleh :

Parada TP Hutauruk
NPM : 6301020285

Program Pascasarjana Fisika
Departemen Fisika
Fakultas Matematika dan Ilmu Pengetahuan Alam
Universitas Indonesia
Depok
2004

# Lembar Persetujuan

| | | |
|---|---|---|
| Judul Tesis | : | Lintasan Bebas Rata-rata Neutrino di Bintang Netron |
| Nama | : | Parada TP Hutauruk |
| NPM | : | 6301020285 |

Tesis ini telah diperiksa dan disetujui oleh dosen pembimbing dan penguji.

Depok, 3 Juni 2004
Mengesahkan,

| | |
|---|---|
| Pembimbing I | Pembimbing II |
| ( Dr. Anto Sulaksono ) | ( Dr. Terry Mart ) |

| | | |
|---|---|---|
| Penguji I | Penguji II | Penguji III |
| ( Dr. L. T. Handoko ) | ( Dr. M. Hikam ) | ( Dr. Rachmat Widodo Adi ) |

| | |
|---|---|
| Ketua Jurusan Fisika | Ketua Program Pascasarjana FMIPA-UI |
| ( Dr. Djarwani Suharso ) | ( Dr. Dedy Suyanto ) |

Tanggal Lulus: 11 Mei 2004

# Kata Pengantar

Puji dan syukur kepada Tuhan Yang Maha Esa atas berkat dan anugerah yang dilimpahkan kepada penulis sehingga tesis yang berjudul: *Lintasan Bebas Rata-rata Neutrino di Bintang Netron* dapat terselesaikan dengan baik setelah menghadapi dan melewati banyak pergumulan. Dan penulis yakin, tesis ini dapat terselesaikan dengan baik karena kemampuan penulis semata tetapi karena kasih Tuhan yang senantiasa menemani penulis dalam setiap pergumulan. Pada kesempatan ini, penulis juga mengucapkan banyak terima kasih kepada :

- Dr. Anto Sulaksono selaku Pembimbing I atas bimbingan, pengalaman yang diberikan, semangat dan perhatiannya serta kesabarannya dan Dr. Terry Mart selaku Pembimbing II atas bimbingan dan kesabarannya serta dukungan yang diberikan sehingga tesis ini dapat terselesaikan.

- Dr. Laksana Tri Handoko, Dr. Muhammad Hikam dan Dr. Rachmat Widodo Adi yang telah bertindak sebagai penguji, atas saran dan kritik yang diberikan. Dr. Dedi Suyanto sebagai Ketua Program Magister Fisika dan ketua sidang magister bagi penulis.

- Ayahanda R. Hutauruk dan Ibunda M. Hutabarat tercinta atas cinta, kesabaran dukungan baik materil maupun spritual serta doa yang diberikan.

- Adik-adik saya ; Johardi, Johanes, Lena, Yanti, Jonni atas cinta, semangat dan doanya yang diberikan.

- Eva Dina Sakti, S. P yang telah memberikan dukungan baik materil dan spritual serta diskusi-diskusi yang luar biasa.

- Prof.Yohanes Surya, PhD dan Joko Prabowo, M. H serta seluruh Pengurus Yayasan Pelita Harapan atas bantuan materil yang diberikan.

- Rekan-rekan junior Grup Fisika Teori; Julio, Novirwan, Nita, Freddy, Nowo, Candy, Mulyadi, Ardi, Anton atas semangat dan kebersamaannya.

- Rekan-rekan di Pascasarjana Fisika; Bu Dian, Bu Evi, Iwan Murthada, Chairul, Caroline, Pak Andreas, Bu Aprillia atas semangat kebersamaan dan dukungannya.

- Pak Parman dan staff pascasarjana di Salemba atas kerelaannya untuk disibukkan oleh penulis.

- Teman-teman KTB ; Maria, S. P, Romaida Sirait, S. P, Linda Siahaan, S. P, Sonia Novalinda Purba, S. P, Ririn, Andi Supriyana, S. Hut atas dukungan dan terutama doa-doa yang diberikan.

- Teman-teman pemuda GKI Pengadilan Bogor atas semangat dan doa yang diberikan.

- Henri Sinaga, S.P, Suroyo , Pandapotan dan Hendra Ganda Siregar, S.P (keluarga besar Riau 10C Belakang) atas kebersamaan dan semangat kekeluargaan yang ada.

- Semua pihak yang telah membantu kelancaran pembuatan tesis ini.

Akhir kata semoga tesis ini bermanfaat dan menjadi berkat bagi anda, walaupun mungkin masih banyak kekurangan dan jauh dari sempurna.

Depok, 3 Juni 2004

Penulis

# Intisari

### Lintasan Bebas Rata-rata Neutrino di Bintang Netron

Telah dihitung tampang lintang diferensial dan lintasan bebas rata-rata neutrino dari interaksi neutrino pada materi padat (*dense matter*) seperti bintang netron. Untuk interaksi neutrino digunakan model standar interaksi lemah, sedangkan untuk model nuklirnya digunakan dua model relativistik yaitu model Kopling Titik (PC) dan Walecka (FR). Hasil prediksi tampang lintang diferensial dan lintasan bebas rata-rata neutrino dari kedua model cukup bervariasi. Bahkan untuk tiap model, tampang lintang dan lintasan bebas rata-rata cukup bervariasi, jika dilihat dari variasi parameter set yang digunakan. Tampang lintang diferensial neutrino meningkat bila kerapatan dari materi netron dan energi awal dari neutrino meningkat. Sebaliknya, lintasan bebas rata-rata neutrino menurun dengan meningkatnya kerapatan dan energi awal neutrino.

# Abstract

### Neutrino Mean Free Path in Neutron Star

Have been calculated neurtino differential cross section and mean free path from neutrino interaction in dense matter as neutron star. The neutrino interaction used weak interaction standard model and for it's nuclear models are used two nuclear relativistic models, these are Point Coupling (RMF-PC) model and Walecka (RMF-FR) model. The prediction result of neutrino differential cross section and mean free pathfrom two models are quite variate. It's also showed by each parameter set that is using. Neutrino differential cross section and mean free path increase, if nucleon density and initial neutrino energy is increased. It's opposite with behavior of mean free path.

# Daftar Isi

# Daftar Gambar

# Daftar Tabel

# Bab 1

# Pendahuluan

## 1.1 Latar Belakang

Dari awal pembentukan bintang netron dalam ledakan supernova sampai fase pendinginan, emisi neutrino memiliki peranan yang sangat penting. *Neutrino Opacity* merupakan besaran yang sangat berperan dalam pengangkutan neutrino. Besaran ini memberikan informasi-informasi penting mengenai interaksi-interaksi yang terjadi dan komposisi dari bintang netron. Untuk menerangkan interaksi yang terjadi dalam bintang netron, maka diperlukan model nuklir yang relatif baik. Model nuklir relativistik merupakan salah satu model yang memenuhi kriteria ini, karena model ini dapat menjelaskan sifat-sifat nuklir pada kerapatan normal dengan baik. Karena kerangkanya relativistik, interpolasi model ini pada kerapatan tinggi cukup masuk akal. Interaksi neutrino di materi juga merupakan fenomena yang banyak terjadi dalam astrofisika seperti supernova dan bintang netron. Referensi [1] telah meneliti pengaruh dari korelasi gas elektron tergenerasi dengan menggunakan RPA (*Random Phase Approximation*).

Model relativistik dengan menggunakan pendekatan medan rata-rata dapat mereproduksi sifat-sifat nuklir dengan sangat baik. Hal ini memberikan dasar yang baik untuk mempelajari eksitasi nuklir dalam model relativistik [2],[3]. Selain model ini, sifat-sifat nuklir pada kerapatan normal juga dapat dijelaskandengan menggunakan model nuklir non-relativistik Skyrme-Force. Namun, jika model ini diinterpolasikan pada kerapatan tinggi, akan terjadi ketidakstabilan terhadap fluktuasi spin sehingga prinsip kausalitas dilanggar (*Causality Violation*). Peyimpangan ini menyebabkan lintasan bebas rata-rata neutrino meningkat ter-hadap kerapatan nuklir [2], [4]. Sebaliknya, hal ini tidak muncul pada model relativistik seperti model Walecka dan Kopling Titik. Model-model nuklir relativistik ini dapat menjelaskan sifat-sifat nuklir dan mereproduksi kembali *ground state* dengan kualitas yang relatif sama pada kerapatan normal. Hal ini menimbulkan keingintahuan, apakah prediksi model-model ini masih relatif sama, jika diinterpolasikan pada daerah dengan kerapatan tinggi. Oleh karena itu, pada penelitian ini akan diselidiki dan dianalisis sifat-sifat model relativistik dengan pendekatan medan rata-rata pada kerapatan tinggi. Observabel-observabel yang digunakan adalah tampang lintang diferensial dan lintasan bebas rata-rata neutrino. Model nuklir relativistik yang digunakan adalah model Walecka linier dan nonlinier serta model Kopling Titik, dimana untuk masing-masing model digunakan parameter set yang berbeda-beda.

Sistematika penulisan adalah sebagai berikut. Bab II membahas tentang materi nuklir simetrik, Bab III akan membahas mengenai hamburan neutrino dalam materi pada bintang netron. Bab IV akan membahas tentang sensitivitas model-model nuklir, tampang lintang diferensial dan lintasan bebas rata-rata neutrino. Bab terakhir berisi kesimpulan.

# Bab 2

# Materi Nuklir Simetrik

Penyusun utama materi nuklir adalah netron dan proton (nukleon). Secara makroskopik, sifat-sifat dari materi nuklir ditentukan oleh jumlah rata-rata dari setiap penyusunnya. Materi nuklir simetrik memiliki jumlah netron dan proton yang sama, sehingga netron sisa, $\delta = 0$ dan $\rho_N = \rho_0$, dimana indeks N menyatakan nukleon. Untuk menjelaskan sifat-sifat dari materi nuklir maka dibuat model. Dalam pembuatan model dilakukan pendekatan-pendekatan agar memudahkan pemodelan.

Dalam bab ini, akan dibahas tentang model nuklir relativistik dengan menggunakan pendekatan medan rata-rata untuk menjelaskan materi nuklir simetrik seperti materi netron dan materi penyusun lainnya pada bintang netron [5].

## 2.1 Model Nuklir Relativistik

### 2.1.1 Model Walecka

Pada model ini, interaksi partikel dapat dibayangkan sebagai pertukaran medan (meson) atau dengan kata lain, medan merupakan derajat kebebasan yang independen. Medan dapat diklasifikasikan berdasarkan momentum angular internal, paritas dan isospin. Adapun medan-medan yang bertanggung jawab terhadap interaksi kuat adalah:

- Meson $\sigma$ : meson skalar dan isoskalar. Meson ini bersifat atraktif.
- Meson $\omega$ : meson vektor-isoskalar. Meson ini bersifat repulsif.

Model relativistik menggunakan medan meson $\sigma$ dan $\omega$ untuk menggambarkan saturasi dari materi nuklir dan sifat-sifat dari inti (*nuclei*). Pada kenyataannya, bahwa massa meson $\sigma$ dan kopling konstannya lebih besar dibanding meson $\omega$ sehingga nukleon dapat terikat. Kerapatan Lagrangian standar untuk model relativistik dapat ditulis sebagai [5]

$$\mathcal{L} = \mathcal{L}_{nukleon} + \mathcal{L}_{mesons} + \mathcal{L}_{interaksi} \tag{2.1}$$

dimana, $\mathcal{L}_{nukleon}$ merupakan Lagrangian dari nukleon, $\mathcal{L}_{mesons}$ merupakan Lagrangian dari meson dan $\mathcal{L}_{interaksi}$ merupakan Lagrangian interaksi [5]. Kerapatan Lagrangian standar untuk model Walecka dapat dinyatakan sebagai [5], [6], [7], [8] dan [9]

$$\begin{aligned} \mathcal{L}_{nukleon} &= \overline{\psi}[i\gamma_\mu\partial^\mu - M]\psi, \\ \mathcal{L}_{meson} &= \frac{1}{2}\partial_\mu\phi\partial^\mu\phi - \frac{1}{2}F_{\mu\nu}F^{\mu\nu} + \frac{1}{2}m_\nu^2 V_\mu V^\mu, \\ \mathcal{L}_{interaksi} &= -\overline{\psi}g_s\phi\psi - \overline{\psi}\gamma_\mu g_\nu V^\mu\psi - U_{NL}(\phi), \end{aligned} \tag{2.2}$$

$$F^{\mu\nu} \equiv \partial^\mu V^\nu - \partial^\nu V^\mu.$$

$\psi$ menyatakan operator medan nukleon, $\phi$ menyatakan harga ekspektasi medan meson skalar ($\sigma$). $V^\mu$ adalah harga ekspektasi medan meson vektor ($\omega$). $M$, $m_\nu$ menyatakan massa nukleon dan massa medan meson vektor ($\sigma$). Sedangkan, $g_S$ dan $g_\nu$ menyatakan konstanta kopling medan meson skalar ($\sigma$) dan medan vektor ($\omega$) dengan medan nukleon ($\psi$). Pada pendekatan ini, kontribusi energi negatif dari nukleon diabaikan (*no sea approximation*). Dilain pihak, untuk menggambarkan interaksi yang lebih realistis, kita perlu menambahkan suku skalar nonlinier. Potensial skalar nonlinier dapat ditulis sebagai berikut

$$U_{NL}(\phi) = \frac{1}{2}m_S^2\phi^2 + \frac{1}{3}b_2\phi^3 + \frac{1}{4}b_3\phi^4. \tag{2.3}$$

Pers.(2.2) akan menjadi persamaan model Walecka linier standar, jika $b_2$ dan $b_3$ pada Pers.(2.3) sama dengan nol. Sedangkan, apabila $b_2 \neq b_3 \neq 0$, maka persamaan modelnya menjadi persamaan model nonlinier. Variabel $m_S$ menyatakan massa medan meson skalar. Sedangkan, $b_2$, $b_3$ menyatakan konstanta kopling nonlinier. Pendekatan diatas menggunakan pendekatan medan rata-rata (*mean fields*).

### 2.1.2 Model Kopling Titik

Lagrangian relativistik Kopling Titik (PC) dibangun atas dasar bentuk bilinier operator kerapatan dan arus oleh medan spinor Dirac dari($\psi$) nukleon [10], [11], [12] dan [13]

$$(\overline{\psi} O_\tau \Gamma \psi); \quad O_\tau \epsilon \{1, \tau_i\}; \quad \Gamma \epsilon \{1, \gamma_\mu, \gamma_5, \gamma_5\gamma_\mu, \sigma_{\mu\nu}\}. \tag{2.4}$$

dimana $\tau_i$ menyatakan matriks isospin Pauli, sedangkan indeks i menyatakan indeks dari matriks Pauli yang ke-i. $\Gamma$ menyatakan matriks Dirac. Suku interaksi pada lagrangian dihasilkan dari suku bilinier diatas. Pada prinsipnya, Lagrangian efektif dapat ditulis dalam bentuk deret arus $\overline{\psi} O_\tau \psi$. Model Kopling Titik (PC) terdiri dari verteks interaksi fermion-empat seperti

- $Isoskalar - skalar$ : $(\overline{\psi}\psi)^2$
- $Isoskalar - vektor$ : $(\overline{\psi}\gamma_\mu\psi)(\overline{\psi}\gamma_\mu\psi)$

dimana vektor dalam ruang isospin dinyatakan oleh tanda panah. Model Kopling Titik dapat didefinisikan oleh kerapatan Lagrangian relativistik secara lengkap dapat ditulis sebagai

$$\mathcal{L} = \mathcal{L}_{nukleon} + \mathcal{L}_{interaksi} + \mathcal{L}_{NL} \tag{2.5}$$

dimana,

$$\begin{aligned}
\mathcal{L}_{nukleon} &= \overline{\psi}(i\gamma_\mu \partial^\mu - M)\psi, \\
\mathcal{L}_{interaksi} &= -\frac{1}{2}\alpha_S(\overline{\psi}\psi)(\overline{\psi}\psi) - \frac{1}{2}\alpha_V(\overline{\psi}\gamma_\mu\psi)(\overline{\psi}\gamma^\mu\psi), \\
\mathcal{L}_{NL} &= -\frac{1}{3}\beta_S(\overline{\psi}\psi)^3 - \frac{1}{4}\gamma_S(\overline{\psi}\psi)^4, \\
&- \frac{1}{4}\gamma_V[(\overline{\psi}\gamma_\mu\psi)(\overline{\psi}\gamma^\mu\psi)]^2.
\end{aligned} \tag{2.6}$$

Dengan pendekatan medan rata-rata,

$$\begin{aligned}
&< \overline{\psi}\psi >= \rho_S, \\
&< \overline{\psi}\gamma_0\psi >= \rho_0
\end{aligned} \tag{2.7}$$

dimana $\rho_S$ dan $\rho_0$ menyatakan nilai ekspektasi dari kerapatan dan arus. Dengan pendekatan medan rata-rata, maka kontribusi dari energi negatif tidak disertakan.

Dalam Pers.(2.6) diatas, $\psi$ menyatakan medan nukleon, M menyatakan massa nukleon, indeks "S" and "V" menyatakan interaksi skalar dan vektor. Variabel $\alpha_S$ dan $\alpha_V$ menyatakan konstanta kopling medan skalar dan vektor. Sedangkan, $\beta_S$, $\gamma_S$ dan $\gamma_V$ menyatakan konstanta kopling nonlinier. Lagrangian Kopling Titik $\mathcal{L}$ terdiri dari $\mathcal{L}_{nukleon}$ dan $\mathcal{L}_{interaksi}$ merupakan Lagrangian interaksi, sedangkan $\pounds_{NL}$ merupakan Lagrangian suku orde yang lebih tinggi. Relasi model ini dengan model Walecka dapat dilihat dalam Ref. [11]. Model ini juga menggunakan pendekatan medan rata-rata.

### 2.1.3 Parameter Set Dari Model Nuklir

Telah diketahui dari Bab. 2.1 terdahulu bahwa model-model relativistik baik model Walecka maupun Kopling Titik memiliki beberapa konstanta kopling. Konstanta kopling tersebut ditentukan dengan memparameterisasi model terhadap data eksperimen. Observabel-observabel yang digunakan untuk memfit adalah energi ikat ($E_B$), jari-jari($R$) dan ketebalan permukaaan ($\sigma_1$) dari inti (*finite nuclei*). Hal ini dapat dilihat pada Ref. [9]. Variasi dari model ditentukan dari parameter-parameter yang digunakan. Untuk model Walecka, parameter liniernya adalah ($m_\sigma$, $g_\sigma$, $g_\omega$), nonlinier adalah ($m_\sigma$, $g_\sigma$, $g_\omega$, $b_2$, $b_3$). Sedangkan untuk model Kopling Titik, parameter liniernya adalah ($m_S$, $m_V$, $\alpha_S$, $\alpha_V$) dan nonlinier adalah ($\beta_S$, $\delta_S$, $\delta_V$). Parameterisasi dilakukan dengan cara meminimalkan:

$$\chi^2 = \sum_n [\frac{O_n^{exp} - O_n^{teori}}{\Delta O_n}]^2 \tag{2.8}$$

Nilai $x^2$ yang mendekati satu menunjukkan parameter model sudah baik. Indeks n merujuk pada penjumlahan seluruh observabel ($O_n$) yang dipilih. Dalam analisis data eksperimen, $\Delta O_n$ merupakan kesalahan (*error*) statistik pada data. Secara detail, prosedur ini dapat dilihat pada Ref. [9],[14] dan [15]. Parameter-parameter yang diperoleh dari hasil pencocokkan (*fitting*) untuk model Walecka maupun Kopling Titik dapat dilihat pada tabel. 2.1 dan tabel. 2.2.

Tabel 2.1: Parameter-parameter set dari model Walecka linier dan nonlinier dengan MFT [18].

| Set | M (MeV) | $m_\sigma$ (MeV) | $m_\omega$ (MeV) | $m_\rho$ (MeV) | $g_\sigma$ | $g_\omega$ | $g_\rho$ | $b_2$ ($\mathrm{fm}^{-1}$) | $b_3$ | $c_3$ |
|---|---|---|---|---|---|---|---|---|---|---|
| LW | 939.0 | 550.0 | 783.0 | 763.0 | 9.57 | 11.67 | 0.00 | 0.00 | 0.00 | 0.00 |
| L1 | 938.0 | 550.0 | 783.0 | 763,0 | 10.30 | 12.60 | 0.00 | 0.00 | 0.00 | 0.00 |
| LZ | 938.9 | 551.31 | 780.0 | 763.0 | 11.19 | 13.83 | 5.44 | 0.00 | 0.00 | 0.00 |
| NL1 | 938.0 | 492.25 | 795.539 | 763.0 | 10.14 | 13.28 | 4.98 | -12.17 | -36.26 | 0.00 |
| NL2 | 938.0 | 504.890 | 780.0 | 763.0 | 9.11 | 11.49 | 5.39 | -2.30 | 13.78 | 0.00 |
| NLSH | 939.0 | 526.059 | 783.0 | 763.0 | 10.44 | 12.94 | 4.38 | -6.91 | -15.83 | 0.00 |
| NLZ | 938.9 | 488.67 | 780.0 | 763.0 | 10.05 | 12.91 | 4.85 | -13.51 | -40.22 | 0.00 |
| TM1 | 938.0 | 511.198 | 783.0 | 770.0 | 10.03 | 12.61 | 4.63 | -7.23 | 0.62 | 71.31 |

Tabel 2.2: Parameter-parameter set dari model Kopling Titik dengan MFT [14].

| Set | $\alpha_S$ $(MeV^{-2})$ | $\alpha_V$ $(MeV^{-2})$ | $\beta_S$ $(MeV^{-5})$ | $\gamma_S$ $(MeV^{-8})$ | $\gamma_V$ $(MeV^{-8})$ |
|---|---|---|---|---|---|
| PCF1 | $-3.83602\times10^{-4}$ | $2.59352\times10^{-4}$ | $0.7\times10^{-11}$ | $2.90443\times10^{-17}$ | $0.38795\times10^{-17}$ |
| PCLA | $-4.50763\times10^{-4}$ | $3.42665\times10^{-4}$ | $1.10931\times10^{-11}$ | $5.73359\times10^{-17}$ | $-4.38733\times10^{-17}$ |

## 2.2 Pendekatan Medan Rata-rata

### 2.2.1 Materi Nuklir Berdasarkan Model Walecka

Karena konstanta kopling dari persamaan medan dalam model Walecka besar, maka solusi perturbasi tidak dapat lagi digunakan. Dengan keadaan yang seperti ini, maka dibutuhkan solusi yang nonperturbatif untuk mempelajari implikasinya terhadap kerapatan Lagrangian dari model tersebut [16]. RMFT (*Relativistic Mean Field Theory*) merupakan salah satu solusi nonperturbatif untuk menyelesaikan permasalahan ini [4], [17].

Pendekatan ini dapat dibayangkan sebagai berikut; pertimbangkan sebuah kotak yang besar diisi dengan nukleon yang uniform pada temperatur nol. Asumsikan, kita berada pada kerangka diam terhadap benda tersebut, maka fluks nukleon sama dengan nol (B=0). Dengan kata lain, jumlah nukleon kekal, hal ini dapat dinyatakan ke dalam persamaan:

$$\rho_B = \frac{B}{V},$$

dan,

$$B = \int d^3x B^0,$$

Sebelumnya, dengan menggunakan persamaan Euler-Langrange, maka Pers.(2.2) diatas dapat diturunkan menjadi persamaan medan sebagai

$$\partial_\nu F^{\nu\mu} + m_\nu^2 V^\mu = g_\nu \overline{\Psi} \gamma^\mu \Psi, \quad (2.9)$$

$$(\partial_\mu \partial^\mu + m_s^2)\phi + b_2\phi^2 + b_3\phi^3 = g_s \overline{\Psi}\Psi, \quad (2.10)$$

$$[\gamma^\mu(\imath\partial_\mu - g_V V_\mu) - (M - g_s\phi)]\Psi = 0. \quad (2.11)$$

Jika kotak yang besar tersebut menyusut, maka kerapatan nukleon semakin meningkat. Hal ini menyebabkan suku sumber pada bagian kanan Pers.(2.9) dan (2.10) menjadi sangat besar [16], [17]. Ini mengakibatkan operator medan meson dapat didekati sebagai medan meson klasik dan sumber dapat didekati dengan nilai rata-ratanya (harga ekspektasinya) [16]. Dengan demikian, maka dapat ditulis sebagai

$$\begin{aligned} \phi \rightarrow \langle \phi \rangle &= \phi_0, \\ V^{\mu} \rightarrow \langle V^{\mu} \rangle &= (V_0, 0). \end{aligned}$$

Ketika medan meson diganti dengan medan klasik, maka Lagrangian untuk model Walecka pada Pers.(2.2) menjadi [5], [6], [7], [8], [16], [17], [19] dan [20]

$$\begin{aligned} \mathcal{L}_{Walecka} &= \frac{1}{2} m_{\nu}^2 V_0^2 - \frac{1}{2} m_s^2 \phi_0^2 - \frac{1}{3} b_2 \phi_0^3 - \frac{1}{4} b_3 \phi_0^4, \\ &+ \overline{\Psi}[\imath \partial^{\mu} \gamma_{\mu} - g_{\nu} V_0 \gamma_0 - M^{\star}]\Psi. \end{aligned} \tag{2.12}$$

dimana $g_s, g_\nu$ merupakan konstanta kopling yang merupakan nilai yang diperoleh dari hasil fitting pada observabel-observabel nuklir dan,

$$M^{\star} \equiv M - g_s \phi_0$$

merupakan massa efektif nukleon. Dengan demikian, persamaan medan pada Pers.(2.9), (2.10) dan (2.11) menjadi

$$V_0 = \frac{g_{\nu}}{m_{\nu}^2} \langle \Psi^{\dagger} \Psi \rangle \equiv \frac{g_{\nu}}{m_{\nu}^2} \rho_0 \tag{2.13}$$

$$m_s^2 \phi_0 + b_2 \phi_0^{+} b_3 \phi_0^3 = g_s \langle \overline{\Psi} \Psi \rangle \tag{2.14}$$

$$[\imath \gamma_{\mu} \partial^{\mu} - \gamma_0 g_{\nu} V_0 + M^{\star}]\Psi(x,t) = 0 \tag{2.15}$$

Jika suku nonlinier pada $\phi_0$ sama dengan nol, maka Pers.(2.14) akan menghasilkan ekspresi eksak kerapatan skalar sebagai [16], [17], [19] dan [20]

$$\phi_0 = \frac{g_s}{m_s^2} \langle \overline{\Psi} \Psi \rangle \equiv \frac{g_s}{m_s^2} \rho_S$$

Pers.(2.15) serupa dengan persamaan Dirac bebas, hanya energinya tergeser sebesar $g_\nu V_0$ dan massa nukleon tergeser sebesar $g_s \phi_0$ akibat dari kehadiran medan vektor dan skalar, $V_0$ dan $\phi_0$. Massa nukleon menjadi massa nukleon efektif menjadi

$$M^{\star} = M - g_s \phi_0 \tag{2.16}$$

dan energi per nukleonnya dapat dinyatakan sebagai

$$E = g_\nu V_0 \pm (\mathbf{P}^2 + M^{\star 2})^{\frac{1}{2}} \tag{2.17}$$

Untuk materi nuklir yang seragam, keadaan dasar diperoleh dengan mengisi level-level energi sampai ke momentum Fermi ($k_F$) dengan degenerasi spin-isospin ($\gamma = 4$ untuk materi nuklir dan $\gamma = 2$ untuk materi netron) [20]. Secara singkat, persamaan umum kerapatan energi untuk materi nuklir model Walecka dapat dinyatakan sebagai [6], [16], [17], [20] dan [21]

$$\rho_0 = \rho_B = \frac{\gamma}{(2\pi)^3}\int_0^{k_F} d^3k = \frac{\gamma}{6\pi^2}k_F^3, \tag{2.18}$$

$$\begin{aligned}
\epsilon &= \frac{3}{8k_F^3}\{k_F\sqrt{k_F^2+M^{\star 2}}(2k_F^2+M^{\star 2}) \\
&- M^{\star 4}ln(\frac{k_F+\sqrt{k_F^2+M^{\star 2}}}{M^\star})\} \\
&+ \frac{1}{2}\frac{g_\nu^2}{m_\nu^2}\rho_0 + \{\frac{1}{2}m_s^2\phi^2 + \frac{1}{3}b_2\phi^3 + \frac{1}{4}b_3\phi^4\}\frac{1}{\rho_0}.
\end{aligned} \tag{2.19}$$

dimana $\rho_B$ pada Pers.(2.18) merupakan persamaan kerapatan nukleon [16], [17]. Disamping itu, persamaan kerapatan skalar juga dapat didefenisikan sebagai

$$\rho_S = \frac{\gamma}{(2\pi)^3}\int_0^{k_F} d^3k\frac{M^\star}{(\mathbf{K}^2+M^{\star 2})^{\frac{1}{2}}}. \tag{2.20}$$

Sifat-sifat dari materi nuklir dapat dilihat dari hasil plot massa efektif ($M^\star$) terhadap momentum Fermi ($k_F$) dan energi ($E$) terhadap momentum Fermi ($k_F$). Hasilnya dapat dilihat pada Gambar. 2.1.

### 2.2.2 Materi Nuklir Berdasarkan Model Kopling Titik

Dari Lagrangian efektif model Kopling Titik pada Pers.(2.6) dapat diketahui persamaan medan. Persaman medan dapat diperoleh dengan menggunakan cara yang sama seperti dilakukan pada Pers.(2.9), (2.10) dan (2.11). Disamping itu, kerapatan energi model Kopling Titik dapat dinyatakan sebagai berikut [10], [11], [12], [13], [22], [23] dan [49]

$$\epsilon = \frac{1}{\rho_0}[\frac{\gamma}{16\pi^2}\{k_F\sqrt{k_F^2+M^{\star 2}}(2k_F^2+M^{\star 2}) - M^{\star 4}ln(\frac{k_F+\sqrt{k_F^2+M^{\star 2}}}{M^\star})\},$$

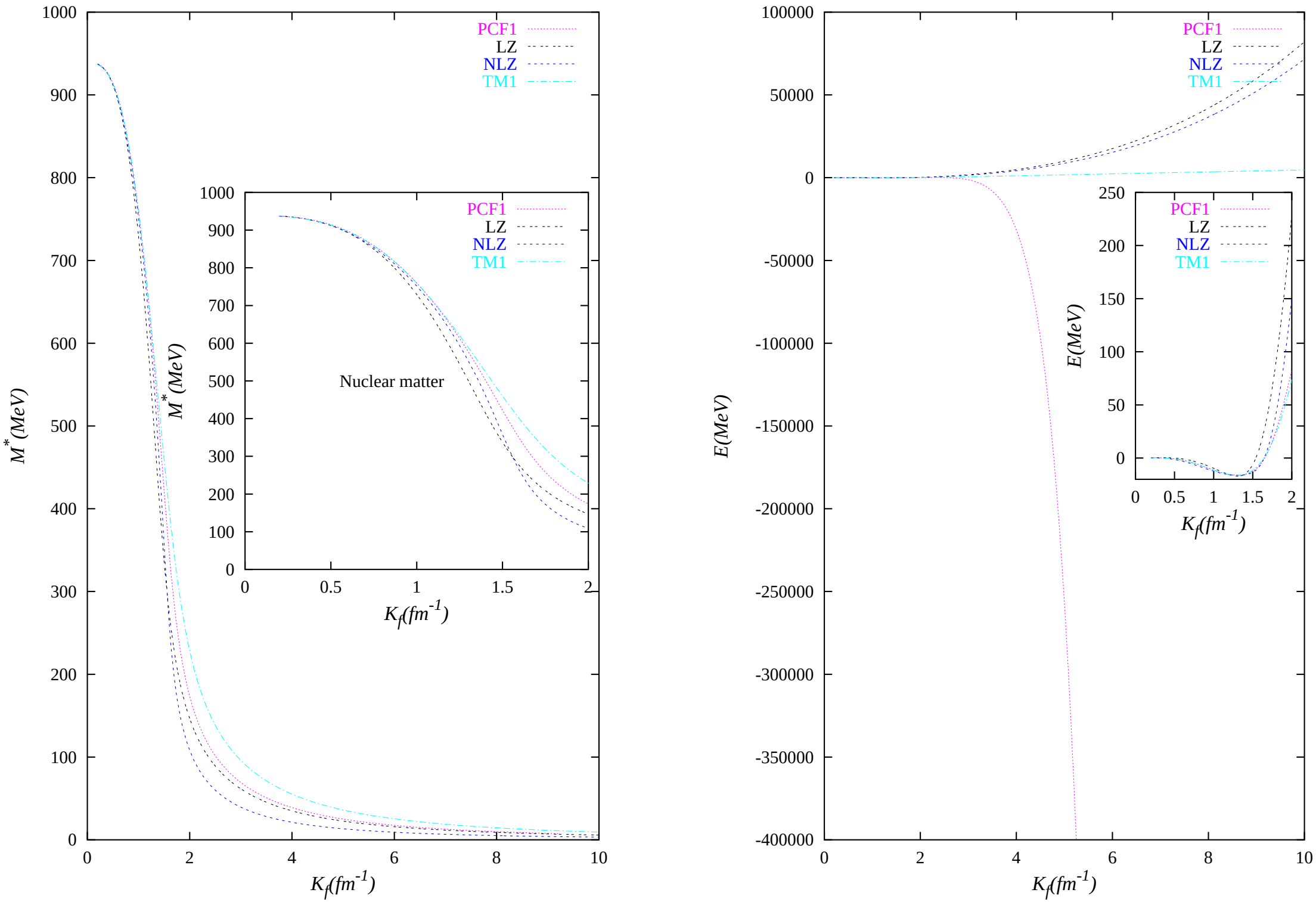

Gambar 2.1: Persamaan keadaan materi nuklir dengan menggunakan beberapa parameter set.

$$
\begin{aligned}
& - \frac{1}{2}\alpha_S \rho_S^2 + \frac{1}{2}\alpha_V \rho_0 2, \\
& - \frac{2}{3}\beta_S \rho_S^3 - \frac{3}{4}\gamma_S \rho_S^4 + \frac{1}{4}\gamma_V \rho_0^4]. \qquad (2.21)
\end{aligned}
$$

dimana $\alpha_S$ menyatakan konstanta kopling skalar dan $\alpha_V$ menyatakan konstanta kopling vektor. Sedangkan, $\gamma_S$ dan $\gamma_V$ merupakan konstanta kopling nonlinier skalar dan vektor. Dalam Pers.(2.21), variabel-variabel seperti; $\rho_S$, $E$, $M^\star$ dan $\rho_B$ didefinisikan sama seperti variabel yang digunakan pada model Walecka, hanya konstanta koplingnya yang berbeda.

Hasil plot $M^\star$ terhadap $k_F$ diperlihatkan oleh Gambar. 2.1, panel sebelah kiri memberikan sifat-sifat materi nuklir model Walecka linier dan nonlinier yang direpresentasikan masing-masing oleh parameter set LZ dan NLZ, TM1. Sedangkan, model Kopling Titik diwakili oleh parameter set PCF1. Terlihat bahwa kedua model memiliki kecendrungan yang sama, baik pada kerapatan normal (gambar pada panel bagian dalam) maupun kerapatan tinggi. Hasil plot $E$ terhadap $k_F$ pada panel sebelah kanan menggunakan model nuklir dan parameter set serupa dengan panel sebelah kiri. Terlihat pada gambar panel bagian dalam, kedua model juga memiliki kecendrungan yang sama pada kerapatan normal. Dan ketika kerapatan diekstrapolasi pada kerapatan tinggi, model Walecka linier dan nonlinier dengan parameter set masing-masing LZ dan NLZ, TM1 memiliki sifat-sifat materi nuklir yang serupa. Perbedaan sifat-sifat materi nuklir yang signifikan diberikan oleh model Kopling Titik dengan parameter set PCF1. Pada model ini diperoleh bahwa energi ikat menurun secara drastik pada $k_F$ antara 4 hingga 6 fm$^{-1}$.

### 2.2.3 Materi Netron

Persamaan keadaan materi netron memiliki bentuk yang sama dengan materi nuklir. Perbedaannya terletak pada spin dan isospin yang dilambangkan dengan $\gamma$ dimana, $\gamma = 4$ untuk materi nuklir, $\gamma = 2$ untuk materi netron. Dengan mensubstitusikan $\gamma = 2$ ke dalam Pers.(2.18) dan (2.19) menjadi persamaan keadaan materi netron untuk model Walecka [6] , [7], [8] dan [24]. Demikian juga, persamaan keadaan materi nuklir model Kopling Titik pada Pers.(2.21) dapat diubah menjadi persamaan keadaan materi netron dengan cara mensubsitusi $\gamma = 4$ menjadi $\gamma = 2$. Sifat-sifat dari materi netron dapat dilihat dari variasi $M^\star$ terhadap $k_F$ dan $E$ terhadap $k_F$ yang dapat dilihat pada Gambar. 2.2

Hasil plot $M^\star$ terhadap $k_F$ diberikan pada Gambar. 2.2, panel sebelah kiri menggunakan model Walecka linier dan nonlinier dengan parameter set LZ dan NLZ, TM1. Model lain adalah model Kopling Titik dengan parameter set PCF1. Terlihat pada panel sebelah kiri, kedua model memiliki kecendrungan yang serupa, baik pada kerapatan normal (gambar panel bagian dalam) maupun pada kerapatan tinggi (gambar bagian luar). Pada panel sebelah kanan juga terlihat model linier dan nonlinier dengan parameter set yang serupa dengan panel sebelah kiri memberikan kecendrungan yang relatif serupa, baik pada kerapatan normal maupun kerapatan tinggi.

Sedangkan, model Kopling Titik dengan parameter set PCF1 memberikan kecendrungan yang sama dengan model Walecka linier dan nonlinier pada kerapatan normal. Dan, ketika kerapatan diekstrapolasi pada kerapatan tinggi, tampak model Kopling Titik memberikan perbedaan sifat-sifat materi nuklir yang signifikan,

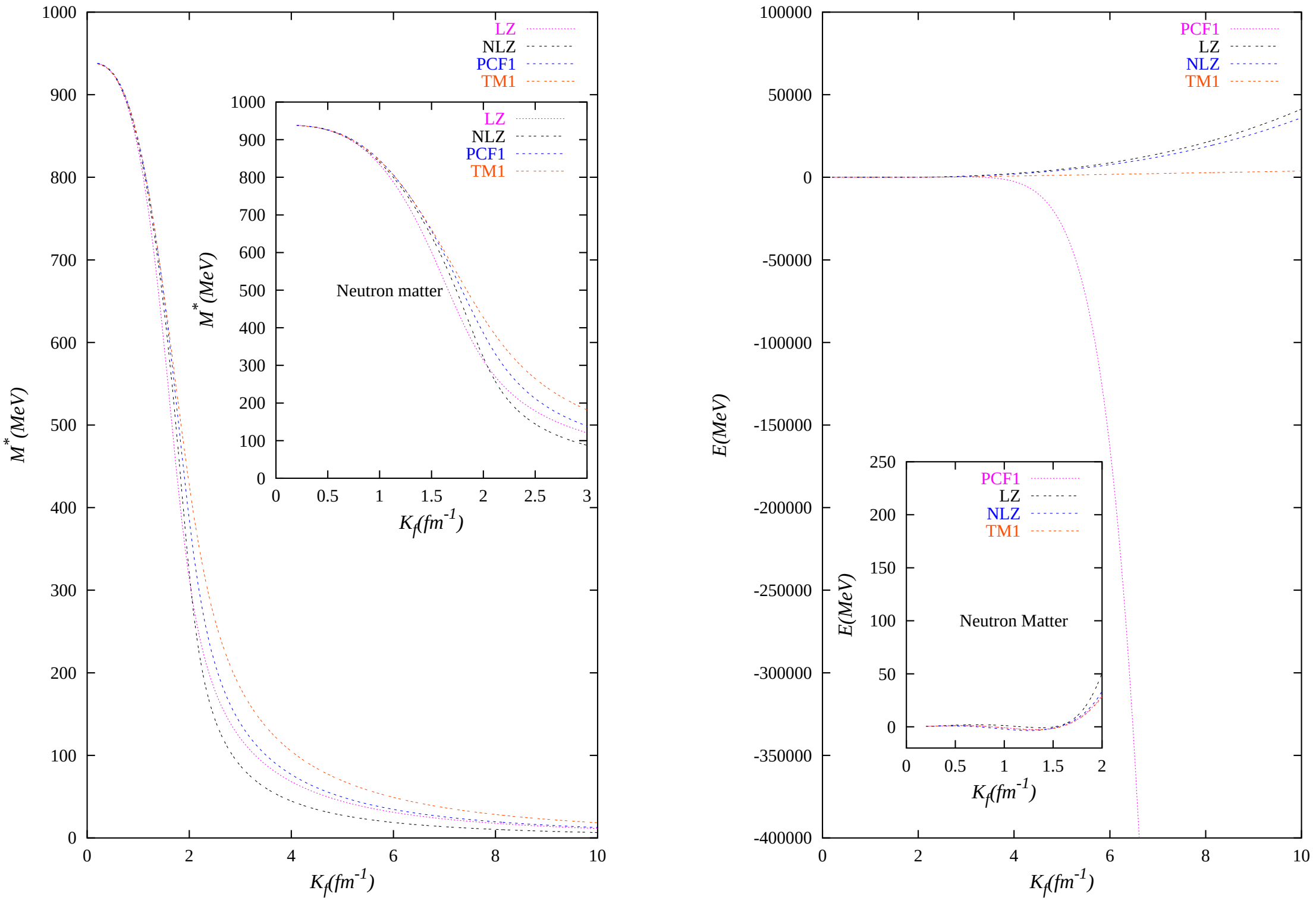

Gambar 2.2: Persamaan keadaan materi neutron dengan menggunakan beberapa parameter set.

dimana energi ikat menurun secara drastis pada $k_F$ antara 6 hingga 8 fm$^{-1}$. Bila Gambar. 2.2 dibandingkan dengan Gambar. 2.1 diatas, terlihat perbedaan yang signifikan antara sifat-sifat materi nuklir dan materi netron pada kedua model, ketika kerapatan diekstrapolasi hingga kerapatan tinggi.

### 2.2.4 Bintang Netron (*Neutron Star*)

Bintang netron merupakan salah satu contoh bintang yang waktu hidupnya telah hampir habis. Bintang-bintang menjadi tua disebabkan pembakaran gas Hidrogen. Pada kenyataannya, bintang-bintang melakukan reaksi fusi (menggabungkan) inti Hidrogen menjadi inti Helium. Reaksi nuklir ini melepaskan energi radiasi dan panas. Disamping menghasilkan energi radiasi dan panas, reaksi ini juga mengakibatkan tekanan yang sangat besar. Adapun rantai reaksi pembentukkan inti Hidrogen menjadi inti Helium adalah sebagai

$$^1\mathrm{H} + ^1\mathrm{H} \rightarrow ^2\mathrm{H} + \mathrm{e}^+ + \nu + 1.44\ \mathrm{MeV}$$

$$^2\mathrm{H} +^1 \mathrm{H} {\rightarrow} ^3\mathrm{He} + \gamma + 5.49 \text{ MeV}$$
$$^3\mathrm{He} +^3 \rightarrow ^4\mathrm{He} +^1 \mathrm{H} +^1 \mathrm{H} + 12.85 \text{ MeV} \quad (2.22)$$

Pembakaran berhenti ketika seluruh Hidrogen diubah menjadi Helium. Selanjutnya, reaksi yang terjadi adalah sebagai

$$^4\mathrm{He} +^8 \rightarrow ^8\mathrm{Be} + \gamma - 95 \text{ KeV}$$
$$^8\mathrm{Be} +^4 \mathrm{He} {\rightarrow} ^{12}\mathrm{C} + \gamma + 7.4 \text{ MeV} \quad (2.23)$$

yang terus berlanjut hingga bintang-bintang hanya terdiri dari Besi, Silikon dan elemen-elemen berat lain. Jika keadaan ini sudah tercapai, maka pembakaran akan berhenti [25]. Jika massa bintang lebih besar dibandingkan massa matahari maka energi rata-rata dari elektron meningkat. Ketika elektron mendapatkan energi relativistik, maka ekspresi kita terhadap tekanan degenerasi berubah secara drastis dimana energi elektron bukan lagi $p^2/2m_e$ tetapi $pc$. Konsekuensi dari tekanan yang sangat besar ini mengakibatkan terjadinya reaksi sebagai berikut

$$\mathrm{e}^- + \mathrm{p} \rightarrow \mathrm{n} + \nu \quad (2.24)$$

yang mengakibatkan terciptanya neutrino. Disamping itu, atom sudah tidak ada lagi karena strukturnya berubah. Hal ini disebabkan tidak lagi proton dan elektron tetapi semuanya menjadi lautan netron (*Neutron Sea*) atau orang lebih mengenalnya sebagai bintang netron (*Neutron Star*).

Bintang netron terdiri dari materi netron murni dengan kerapatan inti yang lebih besar dibandingkan inti atom. Kerapatannya kira-kira tujuh kali lebih besar dibandingkan dengan materi nuklir. Struktur dari bintang netron dapat ditentukan dengan menyelesaikan persamaan relativitas umum Tolman-Oppenheimer-Volkoff (TOV) [24]:

$$\frac{dP}{dr} = \frac{-\frac{G\mu\epsilon}{r^2}(1+\frac{P}{\epsilon})(1+\frac{4\pi r^3 P}{\mu})}{(1-\frac{2G\mu}{r})}. \quad (2.25)$$

dimana $m(r) = \int_0^r 4\pi r'^2 \epsilon(r') dr'$ merupakan massa dari bintang netron yang merupakan fungsi dari jari-jari. P menyatakan tekanan dan $\epsilon$ menyatakan kerapatan energi sebagai fungsi jari-jari bintang netron. Untuk dapat menyelesaikan persamaan TOV, maka perlu diketahui terlebih dahulu persamaan tekanan yang merupakan fungsi dari kerapatan energi ($P = P(\epsilon)$). Penjelasan detail mengenai TOV dapat dilihat pada Ref.[24].

Materi dari bintang netron memiliki penyusun yang sebagian besar terdiri dari netron. Pada kenyataannya di dalam bintang netron masih terdapat proton, elektron dan muon meskipun dalam jumlah yang relatif kecil. Konsentrasi dari setiap partikel di bintang netron ditentukan berdasarkan teori kesetimbangan $\beta$ pada temperatur nol, hal ini dapat dilihat pada Ref. [2], [26], [27]. Persamaan potensial kimia dan fraksi dari setiap partikel pada kesetimbangan $\beta$ dinyatakan sebagai

$$\begin{aligned} \hat{\mu} &= \mu_n - \mu_p = \mu_e - \mu_\nu, \\ Y_i &= \frac{\rho_i}{\rho_B}. \end{aligned} \tag{2.26}$$

dimana $Y_i$ menyatakan fraksi dari partikel ke-i, indeks i melambangkan netron, proton, elektron dan muon. Disini, fraksi netron, proton, elektron dan muon ditentukan dengan pendekatan kualitatif berdasarkan perhitungan pada Ref. [2]. Nilai dari setiap fraksi dapat didekati sebagai

$$\begin{aligned} Y_n &= \frac{\rho_n}{\rho_B} \approx 1, \\ Y_p &= \frac{\rho_p}{\rho_B} \approx 0.1, \\ Y_e &= \frac{\rho_e}{\rho_B} \approx 0.1, \\ Y_\mu &= \frac{\rho_\mu}{\rho_B} \approx 0.1. \end{aligned} \tag{2.27}$$

# Bab 3

# Hamburan Neutrino dengan Materi di Bintang Netron

Emisi neutrino dari inti menyebabkan terjadinya hamburan antara neutrino dengan penyusun materi nuklir pada bintang netron. Proses interaksi neutrino dan materi via reaksi arus netral dapat dilihat pada Ref. [2], [28] dan [52]. Lagrangian interaksi untuk reaksi hamburan neutrino didasarkan pada teori Weinberg-Salam-Glashow. Untuk alih momentum yang lebih kecil dari massa boson tera lemah maka lagrangian interaksinya dapat dinyatakan sebagai

$$\mathcal{L}^j_{int} = \frac{G_F}{\sqrt{2}}[\overline{\nu}\gamma^\mu(1-\gamma_5)\nu](\overline{\psi_j}J^j_\mu\psi_{j)}, \tag{3.1}$$

dimana $\nu$ melambangkan spinor neutrino. Indeks $j$ menunjukkan jenis dari partikel target (netron, proton, elektron, muon). Konstanta $G_F$ menyatakan konstanta kopling lemah. Sedangkan, $\psi_j$ menyatakan spinor dari setiap penyusun nukleon (p, n, $e^-$, $\mu^-$).

Dalam hamburan neutrino dengan materi, fraksi dari setiap partikel dari penyusun materi diperhitungkan, hal ini disebabkan fraksi dari setiap partikel sangat mempengaruhi tampang lintang diferensial neutrino dan juga lintasan bebas rata-rata neutrino. Fraksi atau konsentrasi dari setiap partikel dapat ditentukan melalui netralitas muatan pada kesetimbangan $\beta$ pada temperatur nol. Hal ini dapat dilihat pada Pers.(2.26).

## 3.1 Tampang Lintang Diferensial Neutrino

Melalui hamburan neutrino dengan materi pada bintang netron, maka dapat ditentukan tampang lintang diferensial neutrino per volume yang diekspresikan sebagai [1], [2], [4]

$$\frac{1}{V}\frac{d^3\sigma}{d^2\Omega' dE'_\nu} = -\frac{G_F}{32\pi^2}\frac{E'_\nu}{E_\nu} Im(L_{\mu\nu}\Pi^{\mu\nu}). \tag{3.2}$$

dimana $E_\nu$ dan $E'_\nu$ menyatakan energi neutrino awal dan akhir. Persamaan yang lengkap dapat dilihat pada Pers.(D.1), (D.2) dan (D.3) pada lampiran. Tensor neutrino $L_{\mu\nu}$ dapat ditulis sebagai:

$$\begin{aligned} L_{\mu\nu} &= 8[2k_\mu k_\nu + (k.q)g_{\mu\nu} - (k_\mu q_\nu + q_\mu k_\nu), \\ &\mp i\epsilon_{\mu\nu\alpha\beta}k^\alpha q^\beta]. \end{aligned} \tag{3.3}$$

dengan k$(k_0, \vec{k})$ momentum empat awal neutrino dan $q(q_0, \vec{q})$ alih momentum empat. $G_F = 1.023\times 10^{-5}/M^2$ adalah konstanta kopling lemah dan $\Pi^{\mu\nu}$ adalah tensor polarisasi dari medium. Tensor polarisasi berhubungan dengan struktur materi nuklir. $\Pi_{\mu\nu}$ merupakan gambaran materi nuklir dan sistem lepton serta dapat didefinisikan untuk setiap partikel target. Polarisasi untuk setiap partikel target dinyatakan [1], [31], [32], [33], [34], [35], [36], [37], [38], [39], [40], [41], [45] dan [48]

$$\Pi^j_{\mu\nu}(q) = -i\int \frac{d^4p}{(2\pi)^4} tr[G^j(p)J^j_\mu G^j(p+q)J^j_\nu]. \tag{3.4}$$

dimana $j = n, p, e^-, \mu^-$. $G^j(p)$ merupakan propagator target dan $p(p_0, \vec{p})$ merupakan momentum-empat dari nukleon. $p_F^{n,p}$ adalah momentum Fermi netron dan proton, propagator dinyatakan sebagai [39], [54]

$$\begin{aligned} G^{n,p}(p) &= (P^*_p + M^*_p)\{\frac{1}{P^{*2}_p - M^{*2}_p + i\epsilon} + \frac{i\pi}{E^*_p} \\ &\times \delta(p^*_0 - E^*_p)\theta(p_F^{n,p} - |\vec{p}|)\}. \end{aligned} \tag{3.5}$$

dimana fungsi $\delta$ dihubungkan dengan konservasi energi sedangkan fungsi $\theta$ menunjukkan batas atas momentum pada level Fermi ($p_F^{n,p}$) untuk tiap nukleon. Indeks n dan p melambangkan partikel target yaitu proton dan netron.

## 3.2 Lintasan Bebas Rata-rata Neutrino

Setelah mengetahui formula untuk tampang lintang diferensial neutrino, maka persamaan lintasan bebas rata-rata neutrino dapt ditentukan. Lintasan bebas rata-rata neutrino berbanding terbalik dengan tampang lintang $(\frac{1}{\lambda}\equiv\sigma)$ , dimana $\lambda$ menunjukkan lintasan bebas rata-rata dan $\sigma$ menunjukkan tampang lintang, secara lengkap dapat dilihat pada Ref. [2], [29], [30] dan [53]. Dari hubungan tersebut, maka lintasan bebas rata-rata neutrino dapat dihitung dengan mudah pada keadaan nondegenerasi, oleh sebab itu faktor *Pauli Blocking* dapat diabaikan. Dengan demikian, lintas bebas rata-rata neutrino yang merupakan fungsi dari energi awal neutrino dapat diintegralkan terhadap alih momentum $(\vec{q}, q_0)$ yang secara eksplisit sebagai

$$\frac{1}{\lambda(E_\nu)} = \int_{q_0}^{2E_\nu - q_0} d|\vec{q}| \int_0^{2E_\nu} dq_0 \frac{|\vec{q}|}{E'_\nu E_\nu} 2\pi \frac{1}{V} \frac{d^3\sigma}{d^2\Omega' dE'_\nu}. \tag{3.6}$$

dimana $E_\nu$ menunjukkan energi awal neutrino, $\lambda$ menunjukkan lintasan bebas rata-rata neutrino, $E'_\nu$ menunjukkan energi akhir neutrino dan $\sigma$ menunjukkan tampang lintang neutrino. Vektor $|\overrightarrow{q}|$ merupakan alih momentum, sedangkan $q_0$ merupakan alih energi.

# Bab 4

# Hasil dan Pembahasan

## 4.1 Tampang Lintang Differensial Neutrino

### 4.1.1 Sensitivitas Model

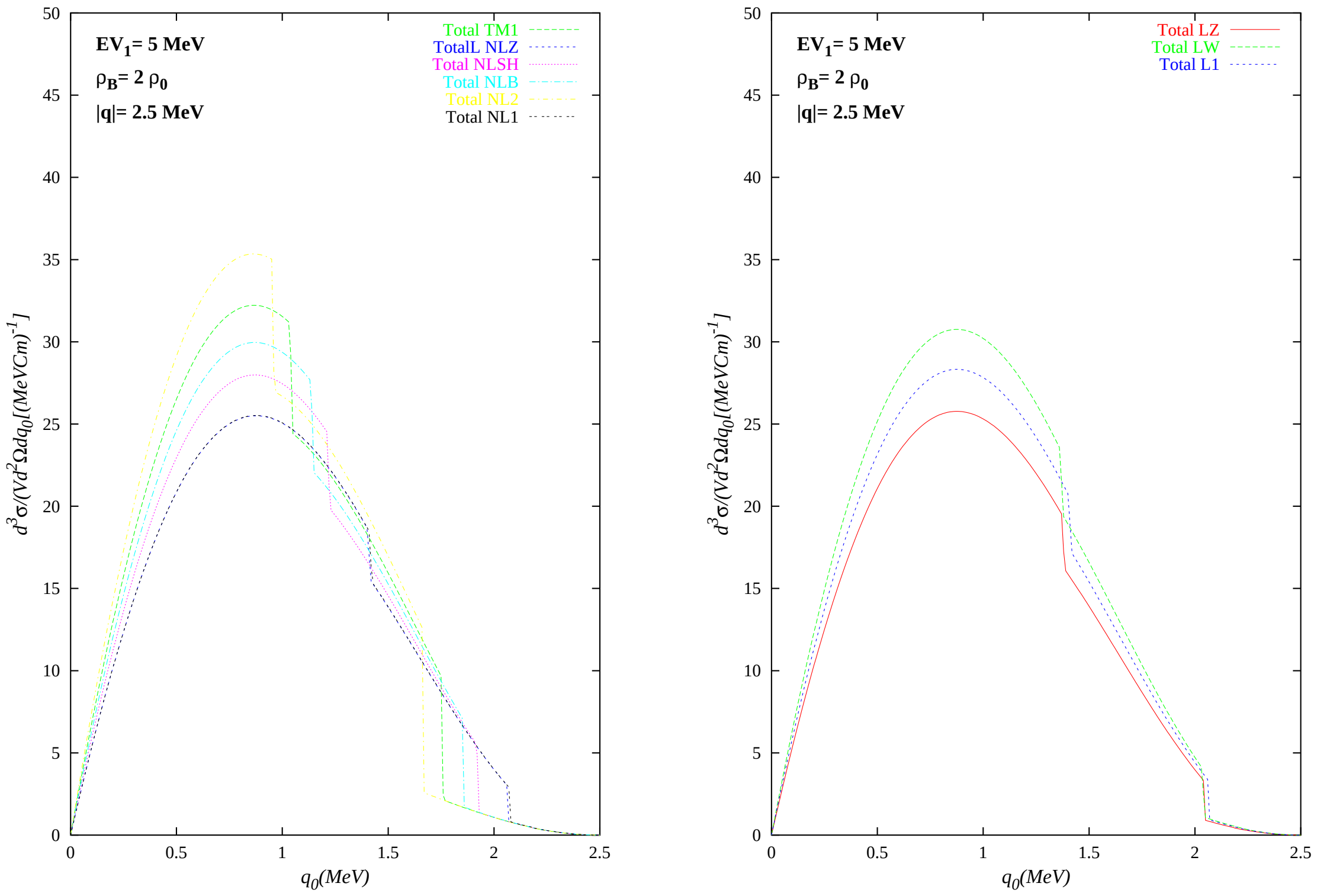

Gambar 4.1: Sensitivitas model Walecka nonlinier dan linier (dari kiri ke kanan) terhadap beberapa parameter set.

Gambar. 4.1 dibuat dengan menggunakan energi awal neutrino $E_\nu = 5$ MeV, alih momentum $|q| = 2.5$ MeV dan untuk kerapatan sebesar dua kali kerapatan saturasi nuklir normal. Panel sebelah kiri untuk model Walecka nonlinier dan panel sebelah kanan untuk yang linier. Untuk model nonlinier digunakan parameter set TM1, NLZ, NLSH, NLB, NL2, NL1, sedangkan model linear digunakan parameter set L1, LW dan LZ [14] dan [18]. Terlihat pada Gambar. 4.1, baik untuk model Walecka nonlinier maupun linier, tampang lintang sangat sensitif terhadap variasi parameter set yang digunakan. Tetapi, perbedaan tampang lintang pada model nonlinier tampak lebih bervariasi dibandingkan model linier. Perbedaan kedua model diatas dapat dilihat pada daerah tampang lintang maksimum tetapi juga pada daerah $q_0$ (alih energi) antara 1.5 hingga 2.5 MeV. Diprediksi bahwa perbedaan tersebut disebabkan penambahan suku nonlinier pada Lagrangian model Walecka linier.

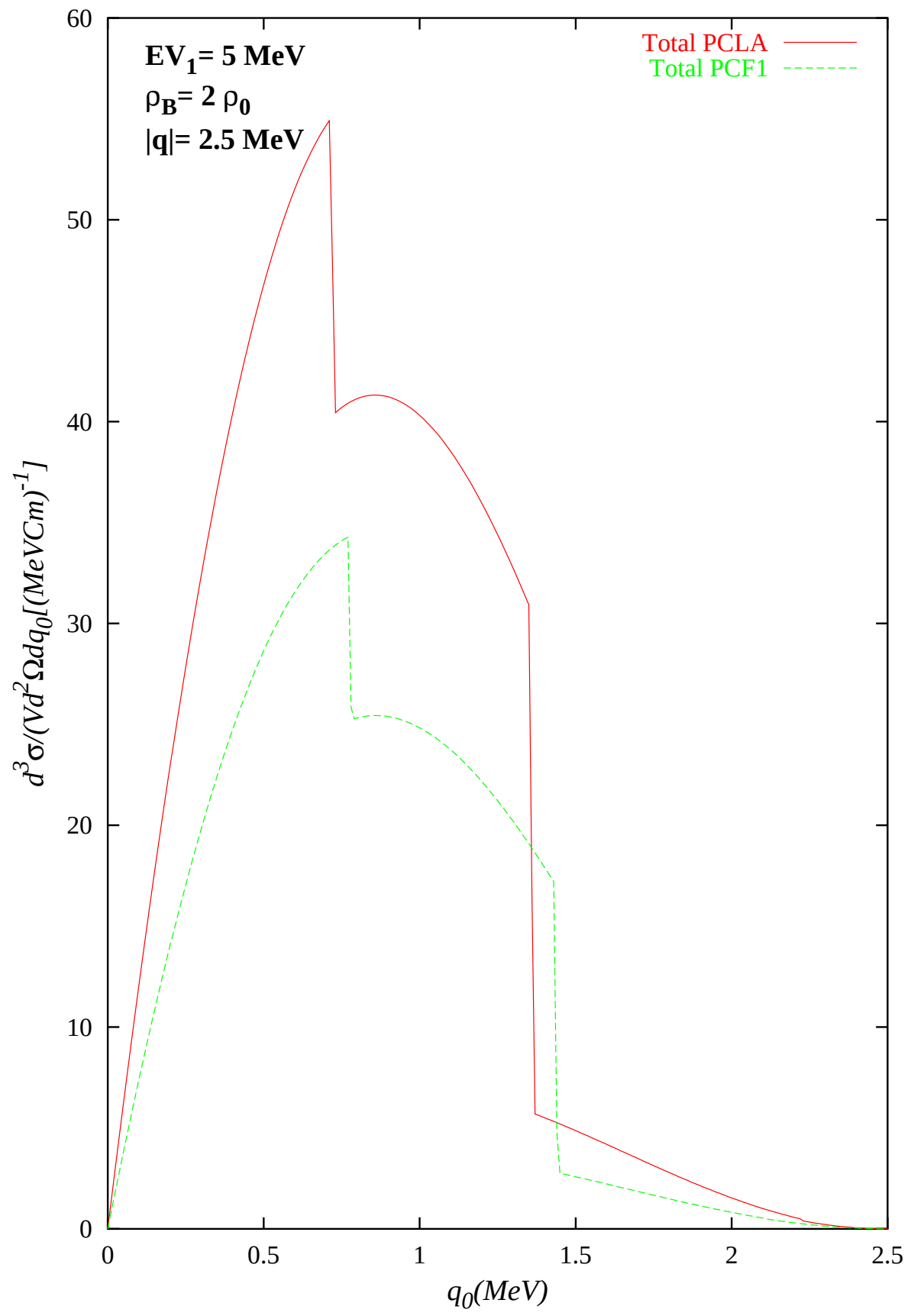

Gambar 4.2: Sensitivitas model Kopling Titik dengan parameter set yang berbeda.

Gambar. 4.2 serupa dengan Gambar. 4.1 hanya model yang digunakan adalah Kopling Titik dengan menggunakan parameter set PCF1 dan PCLA. Terlihat pada gambar, perbedaan tampang lintang dari kedua model cukup drastik. Ini menunjukkan bahwa tampang lintang diferensial neutrinopada model Kopling Titik sangat sensitif terhadap parameter set yang digunakan.

### 4.1.2 Variasi energi awal neutrino, alih momentum dan kerapatan

Pada bagian ini akan dipelajari tampang lintang neutrino dengan menggunakan model nuklir yang berbeda dan parameter set yang berbeda-beda pula untuk energi awal neutrino, kerapatan nukleon dan alih momentum bervariasi. Adapun variasi yang digunakan adalah sebagai berikut:

1. Untuk alih momentumnya sebesar 2.5 MeV akan dipasangkan dengan masing-masing energi awal neutrino, $E_\nu$, yang divariasi dari 5 sampai 10 MeV dan masing-masing kerapatan nukleonnya dua sampai lima kali kerapatan nuklir normal.

2. Untuk alih momentumnya sebesar 5 MeV akan dipasangkan dengan masing-masing energi awal neutrino, $E_\nu$, yang divariasi dari 5 sampai 10 MeV dan masing-masing kerapatan nukleonnya dua sampai lima kali kerapatan nuklir normal. Disamping itu, parameter set yang digunakan dapat dilihat pada Tabel. 2.2.

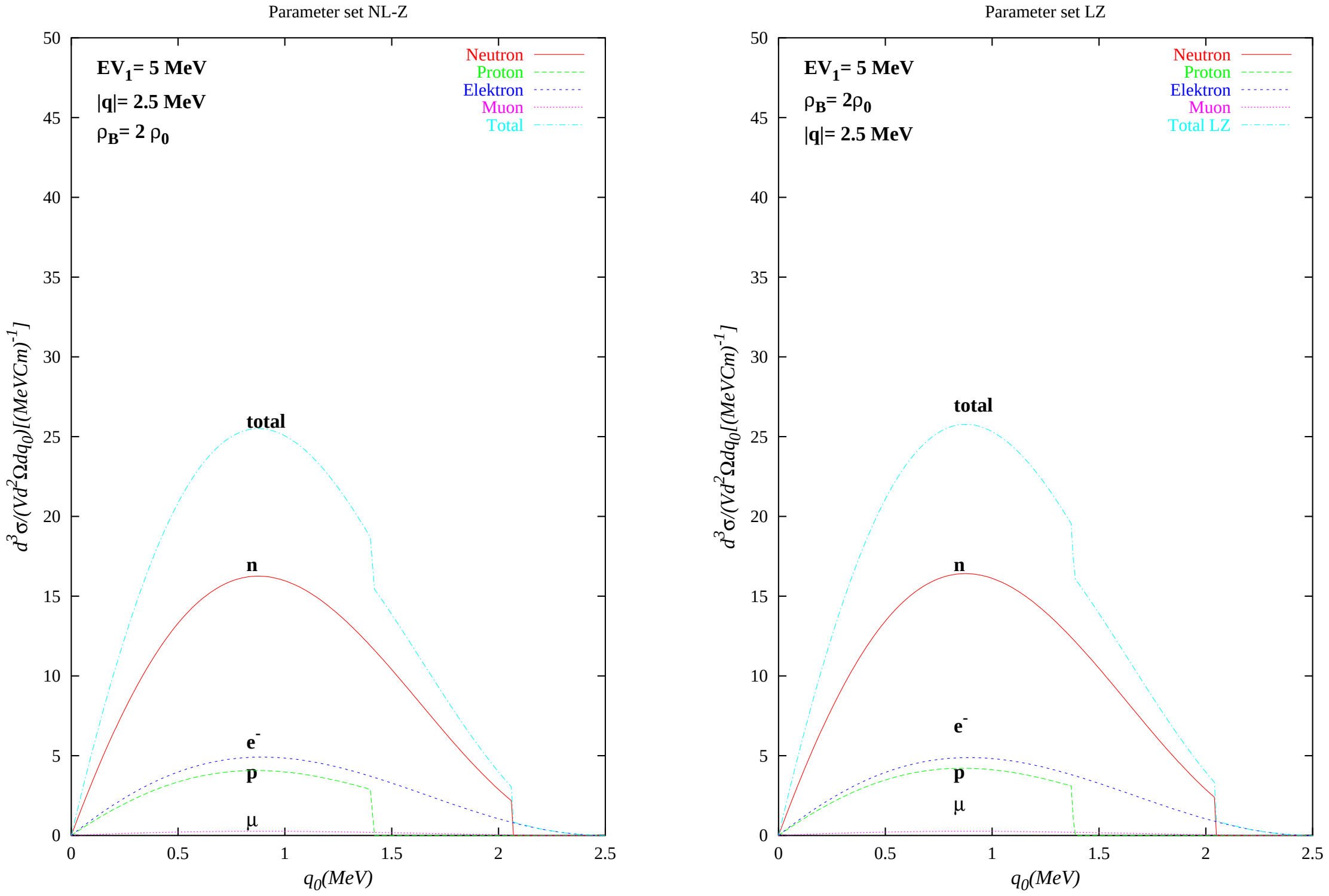

Gambar 4.3: Kiri ke kanan, kontribusi dari netron, proton, elektron dan muon terhadap tampang lintang diferensial neutrino pada energi awal neutrino sebesar 5 MeV, alih momentum sebesar 2.5 MeV dan kerapatannya dua kali kerapatan nuklir normal, masing-masing.

Gambar 4.3 memperlihatkan kontribusi dari setiap partikel target terhadap tampang lintang diferensial neutrino. Pengamatan dilakukan pada energi awal neutrino,$E_\nu = 5$ MeV, alih momentum,$|q| = 2.5$ MeV dan kerapatan nukleonnya dua kali kerapatan nuklir normal. Model yang digunakan adalah model Walecka (FR) nonlinier dan linier yang direpresentasikan oleh parameter set NLZ dan LZ. Parameter yang digunakan dapat dilihat pada Tabel I pada Ref. [14], [18].

Melalui tampang lintang diferensial neutrino dapat dilihat bahwa kontribusi dari netron lebih dominan dibandingkan penyusun materi lainnya, baik untuk model Walecka nonlinier maupun linier. Hal ini disebabkan fraksi netron dalam bintang netron lebih besar dibandingkan dengan fraksi partikel penyusun yang lain. Dengan fraksi netron yang besar maka kemungkinan neutrino menumbuk neutron semakin banyak. Fraksi dari setiap partikel target dapat dilihat pada Pers.(2.27).

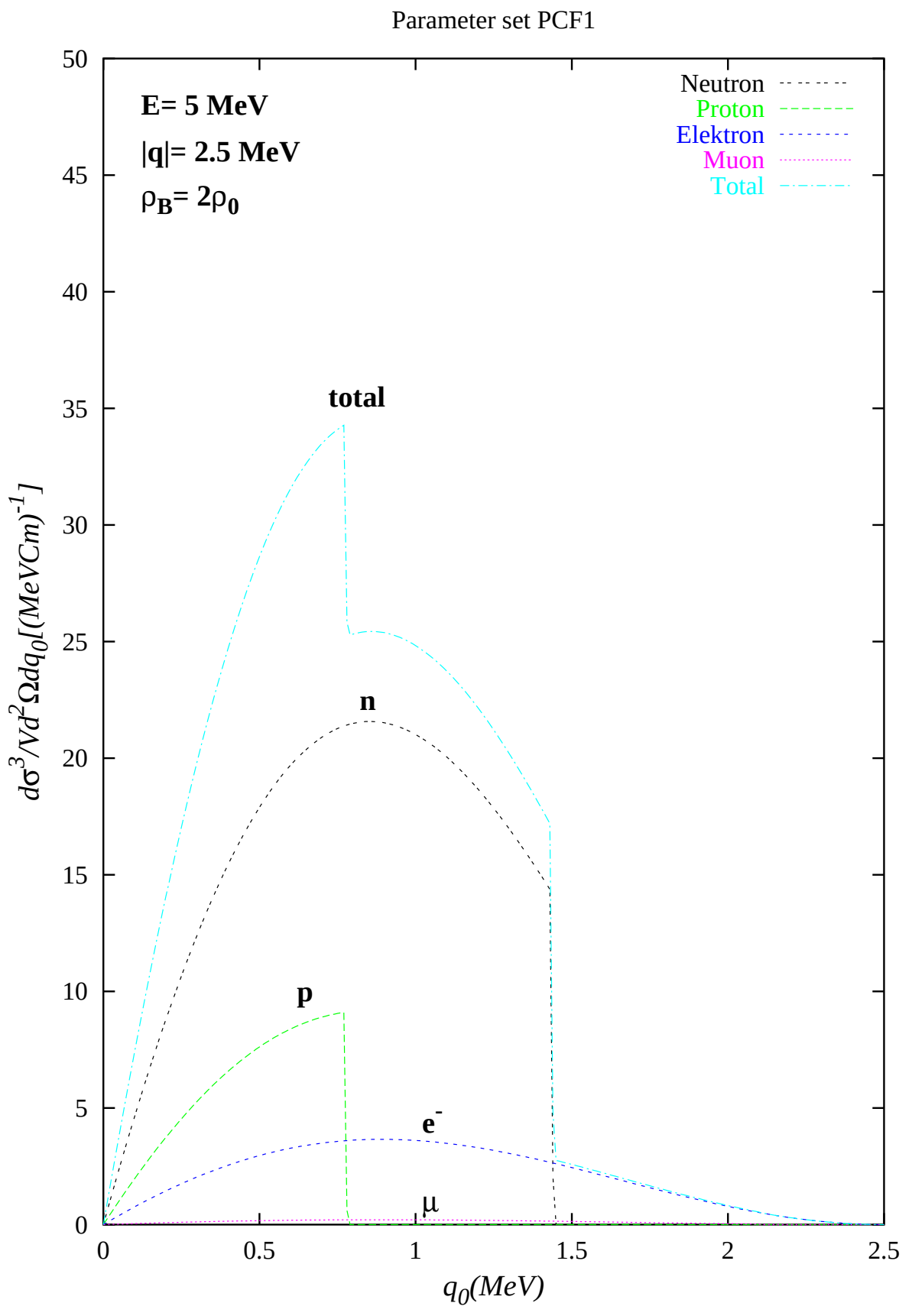

Gambar 4.4: Kontribusi dari netron, proton, elektron dan muon terhadap tampang lintang diferensial neutrino pada $E_\nu = 5$ MeV, $|q| = 2.5$ MeV dan kerapatannya dua kali kerapatan nuklir normal.

Pada Gambar. 4.4 diberikan kontribusi dari setiap partikel target terhadap tampang lintang diferensial neutrino dengan menggunakan model Kopling Titik (PC). Parameter set PCF1 digunakan sebagai representer parameter set yang lain untuk model ini. Terlihat pada tampang lintang diferensial neutrino kontribusi proton meningkat secara drastis, namun jangkauan alih momentum $q_0$ semakin kecil dibandingkan dengan jangkauan proton pada model Walecka nonlinier dan linier pada Gambar. 4.3. Peningkatan jumlah proton yang drastis tersebut menyebabkan tampang lintang diferensial total neutrino semakin tajam. Jadi, perbedaan kecendrungan yang sangat signifikan antara tampang lintang diferensial model Walecka dan Kopling Titik disebabkan perbedaan kontribusi proton pada tiap model. Selain itu, perbedaan kecendrungan dari kedua model juga dapat dilihat pada daerah $q_0$ dari 0.5 hingga 1.5 MeV.

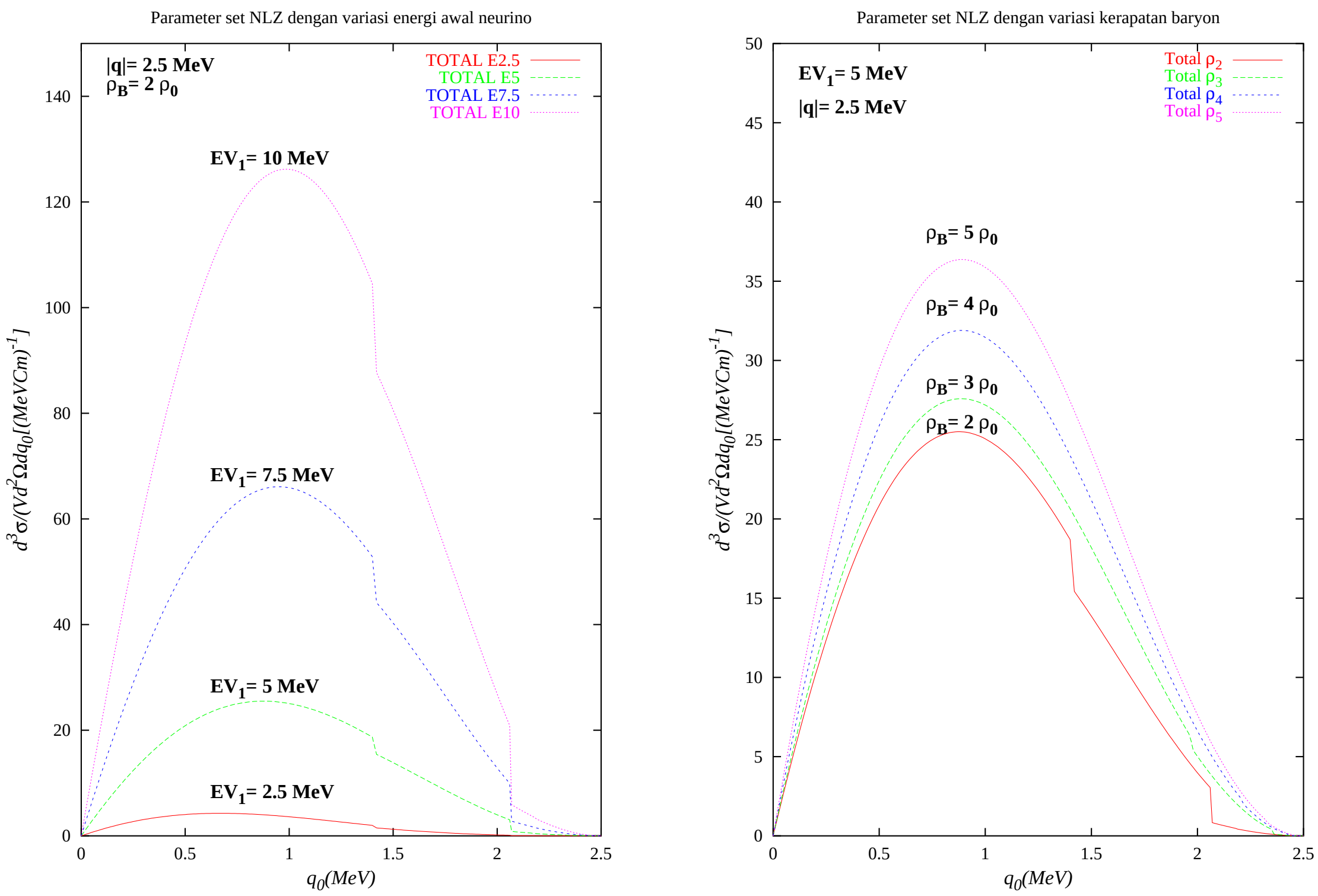

Gambar 4.5: Dari kiri ke kanan, tampang lintang diferensial neutrino dengan energi awal neutrino yang bervariasi pada $|q| = 2.5$ MeV dan kerapatannya dua kali kerapatan nuklir normal (kiri) serta (kanan) kerapatan yang bervariasi pada $E_\nu = 5$ MeV dan $|q| = 2.5$ MeV.

Gambar. 4.5, panel sebelah kiri dilakukan energi awal neutrino, $E_\nu$, yang bervariasi dari 2.5 sampai 10 MeV, kerapatan nukleonnya dua kali kerapatan nuklir normal dan alih momentum, $|q| = 2.5$ MeV. Pada panel sebelah kanan dilakukan variasi kerapatan nukleon dari dua sampai lima kali kerapatan nuklir normal, energi awal neutrino sebesar 5 MeV dan alih momentum sebesar 2.5 MeV. Dari panel sebelah kiri diprediksi bahwa tampang lintang diferensial neutrino semakin besar jika energi awal neutrino diperbesar.

Sedangkan, dari panel sebelah kanan diprediksi bahwa tampang lintang diferensial juga semakin besar bila kerapatan nukleon diperbesar. Dengan kata lain, tampang lintang diferensial menjadi lebih besar, jika materi nuklir semakin padat. Tampang lintang diferensial pada Gambar. 4.5 merupakan tampang lintang diferensial total neutrino. Tampang lintang diferensial neutrino pada gambar ini menggunakan model Walecka nonlinier dengan parameter set NLZ [14], [18].

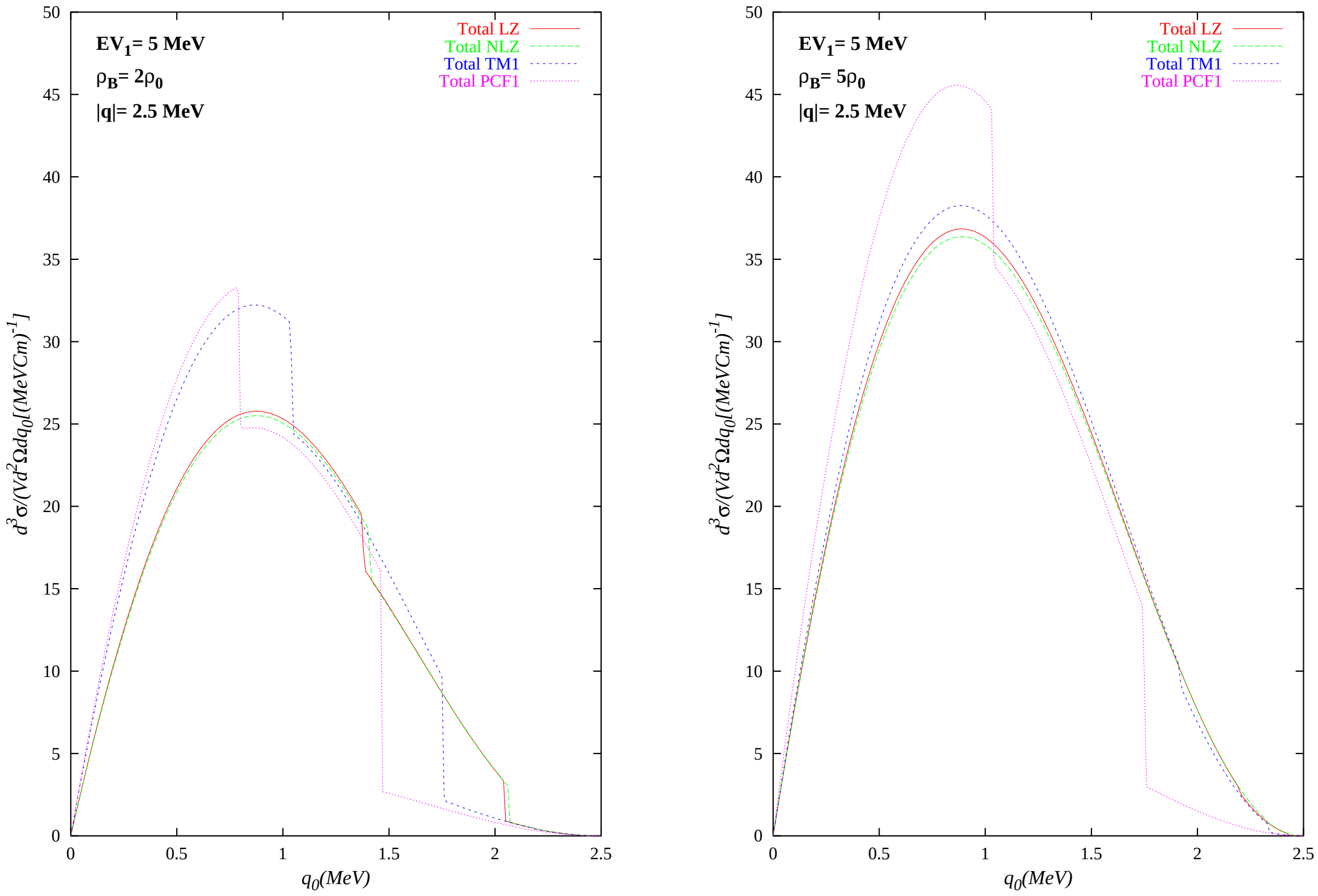

Gambar 4.6: Dari kiri ke kanan, tampang lintang diferensial neutrino dimana energi awal neutrino sebesar 5 MeV, alih momentum, $|q| = 2.5$ MeV dan kerapatan nukleonnya dua kali kerapatan nuklir normal. Panel sebelah kanan menggunakan energi awal neutrino dan alih momentum yang serupa dengan panel sebelah kiri, tetapi kerapatan nukleonnya diubah menjadi lima kali kerapatan normal.

Panel sebelah kiri pada Gambar. 4.6 menggunakan energi awal neutrino, $E_\nu = 5$ MeV, kerapatan nukleon dua kali kerapatan nuklir normal dan alih momentum, $|q| = 2.5$ MeV. Sedangkan, panel sebelah kanan menggunakan energi awal neutrino dan alih momentum yang sama dengan panel sebelah kiri, namun kerapatan nukleonnya berbeda. Kerapatan nukleon pada panel sebelah kanan adalah lima kali kerapatan nuklir normal. Dari panel sebelah kiri diprediksi bahwa tampang lintang diferensial neutrino untuk model Walecka nonlinier dan linier tidak berbeda secara signifikan, kecuali parameter set TM1. Selain model Walecka nonlinier dengan parameter set TM1, perbedaan tampang lintang diferensial neutrino yang signifikan juga diperlihatkan oleh model Kopling Titik dengan parameter set PCF1.

Namun, yang menarik disini adalah parameter set TM1, dimana tampang lintang diferensial dari model Walecka nonlinier dengan parameter set ini menjadi lebih besar dibandingkan parameter set NLZ pada kerapatan $2\rho_0$. Namun, ketika kerapatan nukleon dibuat lima kali kerapatan nuklir normal dengan energi awal neutrino dan alih momentum yang sama, tampang diferensial neutrino untuk parameter TM1 semakin besar, namun kecendrungannya menjadi hampir sama dengan parameter set NLZ.

Disamping itu, Gambar. 4.6 juga memperlihatkan bahwa tampang lintang diferensial neutrino untuk model Kopling Titik dengan parameter set PCF1 semakin meningkat secara drastis, bila kerapatan nukleon diperbesar. Tampak bahwa nilai dari tampang lintang diferensial neutrino baik model Walecka linier dan nonlinier maupun Kopling Titik berbeda sangat signifikan pada kerapatan tinggi. Namun, tampang lintang diferensial neutrino dari kedua model memberikan sifat-sifat yang sama pada kerapatan tinggi, dimana tampang lintang diferensial neutrino dari kedua model meningkat pada kerapatan tinggi. Hal ini menunjukkan bahwa perbedaan kedua model menjadi semakin signifikan pada kerapatan tinggi.

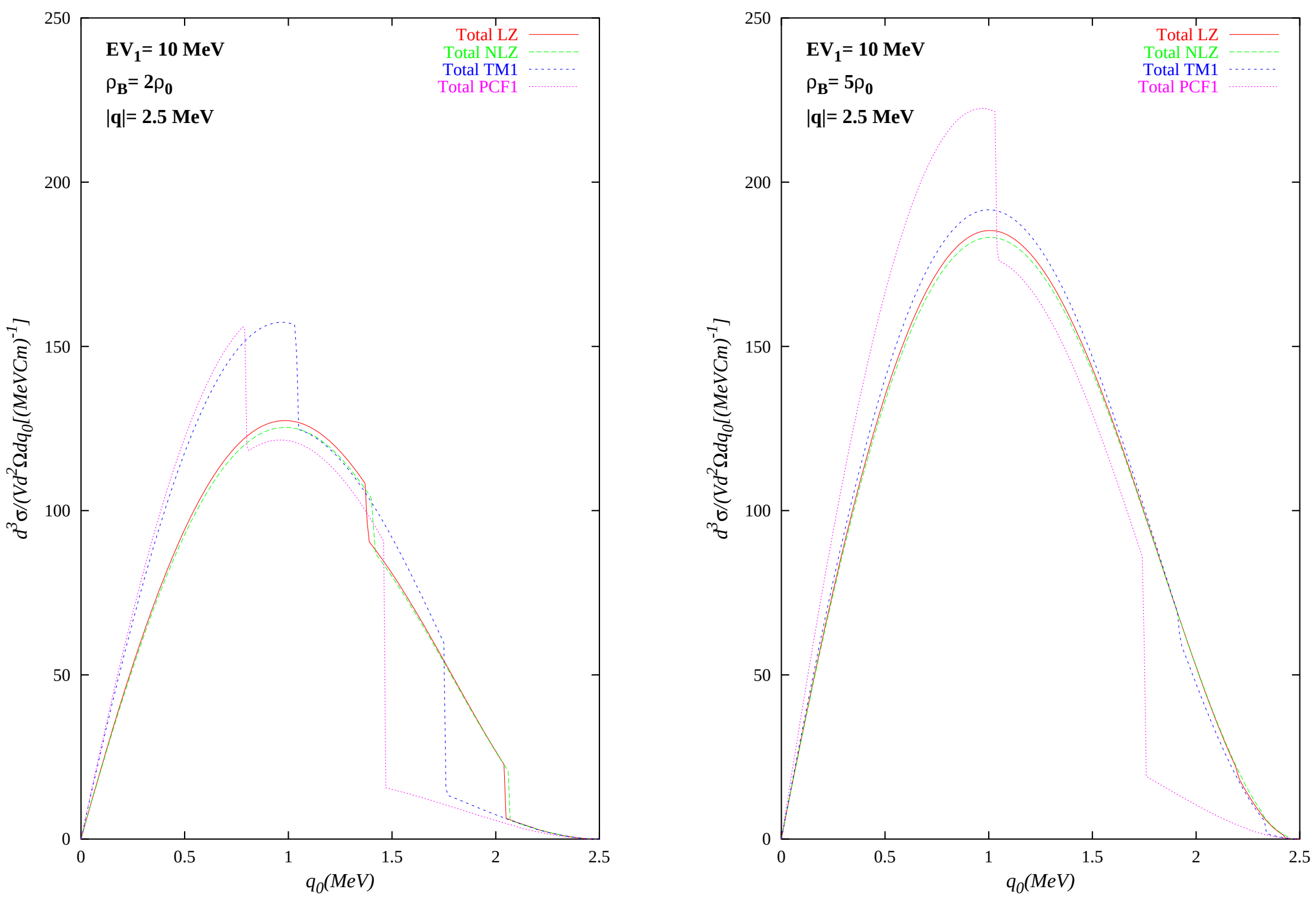

Gambar 4.7: Dari kiri ke kanan, tampang lintang diferensial neutrino dimana alih momentum, $|q| = 2.5$ MeV, energi awal neutrino, $E_\nu = 10$ MeV dan kerapatannya dua kali kerapatan nuklir normal (kiri). Pada panel sebelah kanan, energi awal neutrino sebesar 10 MeV, alih momentum sebesar 2.5 MeV dan kerapatannya lima kali kerapatan nuklir normal.

Pada Gambar. 4.7, panel sebelah kiri dibuat pada $E_\nu = 10$ MeV, $|q| = 2.5$ MeV dan $\rho_B = 2\rho_0$. Panel sebelah kanan, dibuat $E_\nu = 10$ MeV, $|q| = 2.5$ MeV dan kerapatannya diperbesar menjadi lima kali kerapatan nuklir normal. Dari panel sebelah kiri diberikan bahwa nilai tampang lintang diferensial neutrino meningkat secara drastis baik untuk model Walecka maupun Kopling Titik dibandingkan dengan Gambar. 4.6. Panel sebelah kanan memperlihatkan bahwa tampang lintang diferensial neutrino semakin besar, bila kerapatan nukleonnya diperbesar. Dibandingkan dengan Gambar. 4.6, kecendrungan yang dimiliki hampir sama untuk kedua model. Namun, nilai dari tampang lintang diferensial neutrino meningkat lima kali lebih besar, jika energi awal neutrino dinaikkan sebesar 10 MeV. Hal ini menunjukkan bahwa perbedaan dari kedua model menjadi sangat signifikan, bila energi awal neutrino dan kerapatan nukleon diperbesar.

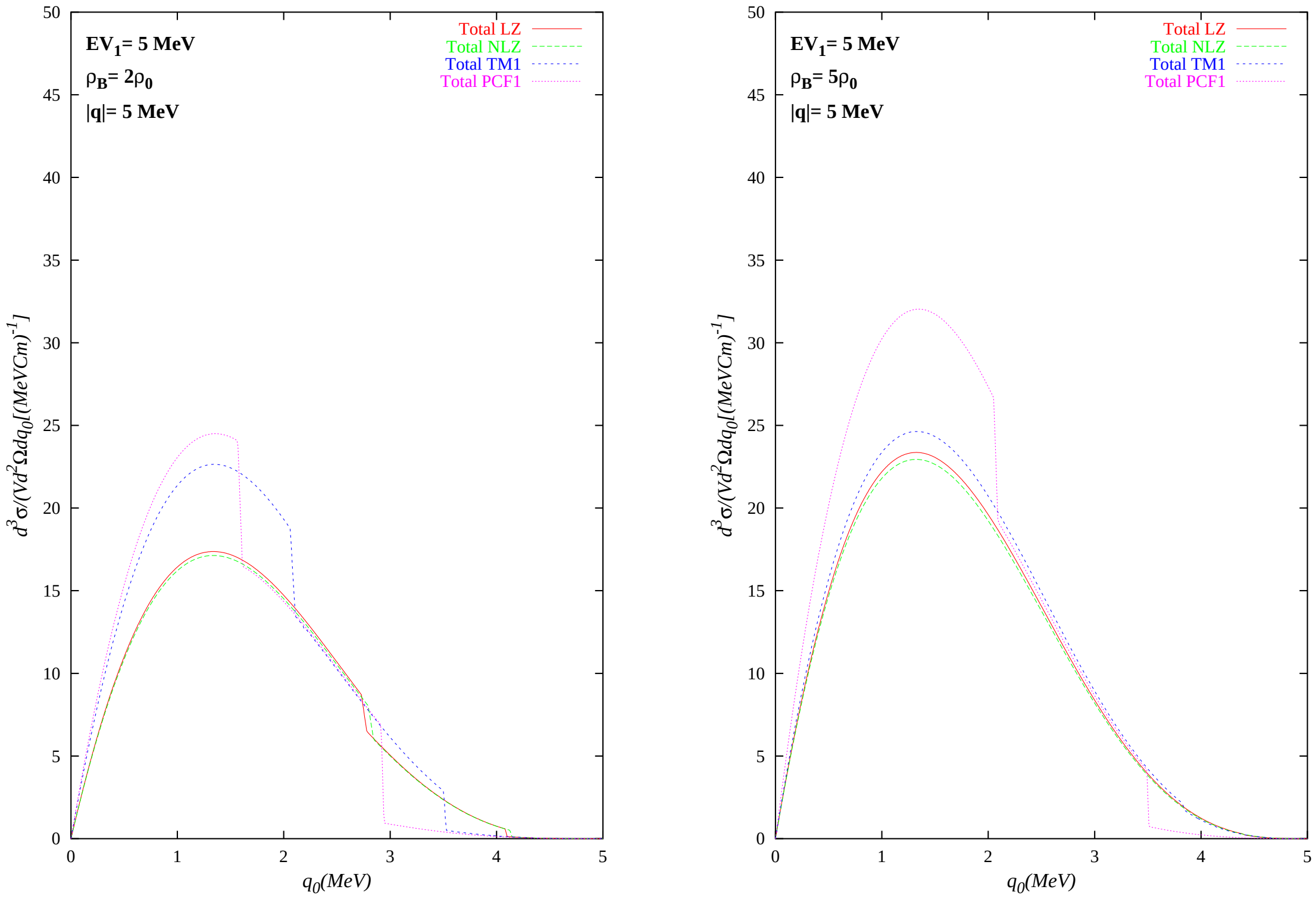

Gambar 4.8: Kiri ke kanan, tampang lintang diferensial neutrino dimana, energi awal neutrino sebesar 5 MeV, alih momentum, $|q| = 5$ MeV dan kerapatan nukleonnya dua kali kerapatan nuklir normal. Kanan, energi awal neutrino, $E_\nu = 5$ MeV, alih momentum sebesar 5 MeV dan kerapatan nukleonnya lima kali kerapatan nuklir normal.

Gambar. 4.8, panel sebelah kiri dibuat pada energi awal neutrino, $E_\nu = 5$ MeV, kerapatannya dua kali kerapatan nuklir normal dan alih momentumnya $|q| = 5$ MeV. Sedangkan, pada panel sebelah kanan, dibuat energi awal neutrinonya, $E_\nu = 5$ MeV, kerapatannya lima kali kerapatan nuklir normal dan alih momentum, $|q| = 5$ MeV. Tampang lintang diferensial neutrino menurun, ketika momentum transfernya dinaikkan sebesar 5 MeV, bila dibandingkan dengan Gambar. 4.6. Namun, tampang lintang diferensial neutrino panel sebelah kanan pada Gambar. 4.8 semakin besar, jika kerapatan dari panel sebelah kiri dinaikkan menjadi lima kali kerapatan nuklir normal. Hal ini memperlihatkan bahwa perbedaan dari kedua model juga menjadi signifikan, bila momentum transfer dan kerapatannya diperbesar.

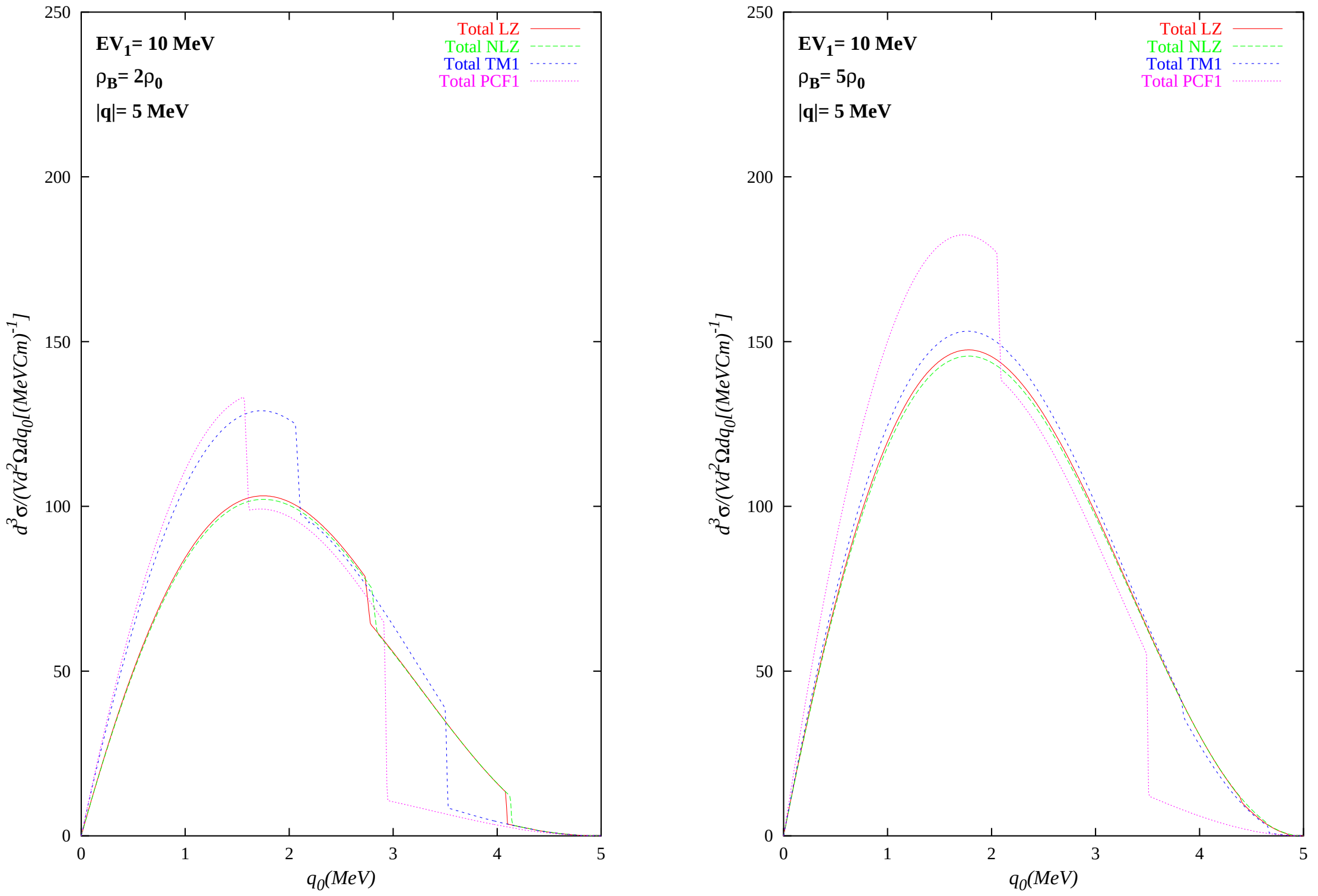

Gambar 4.9: Kiri ke kanan, tampang lintang diferensial neutrino dengan energi awal neutrino sebesar 10 MeV, alih momentum, $|q| = 5$ MeV dan kerapatannya dua kali kerapatan nuklir normal (kiri). Panel sebelah kanan, energi awal neutrino $E_\nu = 10$ MeV, alih momentum sebesar 5 MeV dan kerapatannya lima kali kerapatan nuklir normal.

Gambar. 4.9 dari kiri ke kanan, memiliki energi awal dan kerapatan yang sama dengan Gambar. 4.7. Namun, alih momentum diperbesar menjadi sebesar 5 MeV. Bila dibandingkan dengan Gambar. 4.7, Gambar. 4.9 memperlihatkan bahwa tampang lintang diferensial neutrino menurun, ketika alih momentumnya dinaikkan menjadi sebesar 5 MeV . Namun, ketika kerapatan nukleonnya diperbesar menjadi lima kali kerapatan nuklir normal maka tampang lintang diferensial neutrino pada Gambar. 4.9 semakin besar. Secara umum, tampang lintang diferensial neutrino menjadi lebih besar, jika kerapatan nukleon dan energi awal neutrino diperbesar. Sebaliknya, akan menurun jika alih momentumnya diperbesar.

## 4.2 Lintasan Bebas Rata-rata Neutrino

### 4.2.1 Model Walecka linier

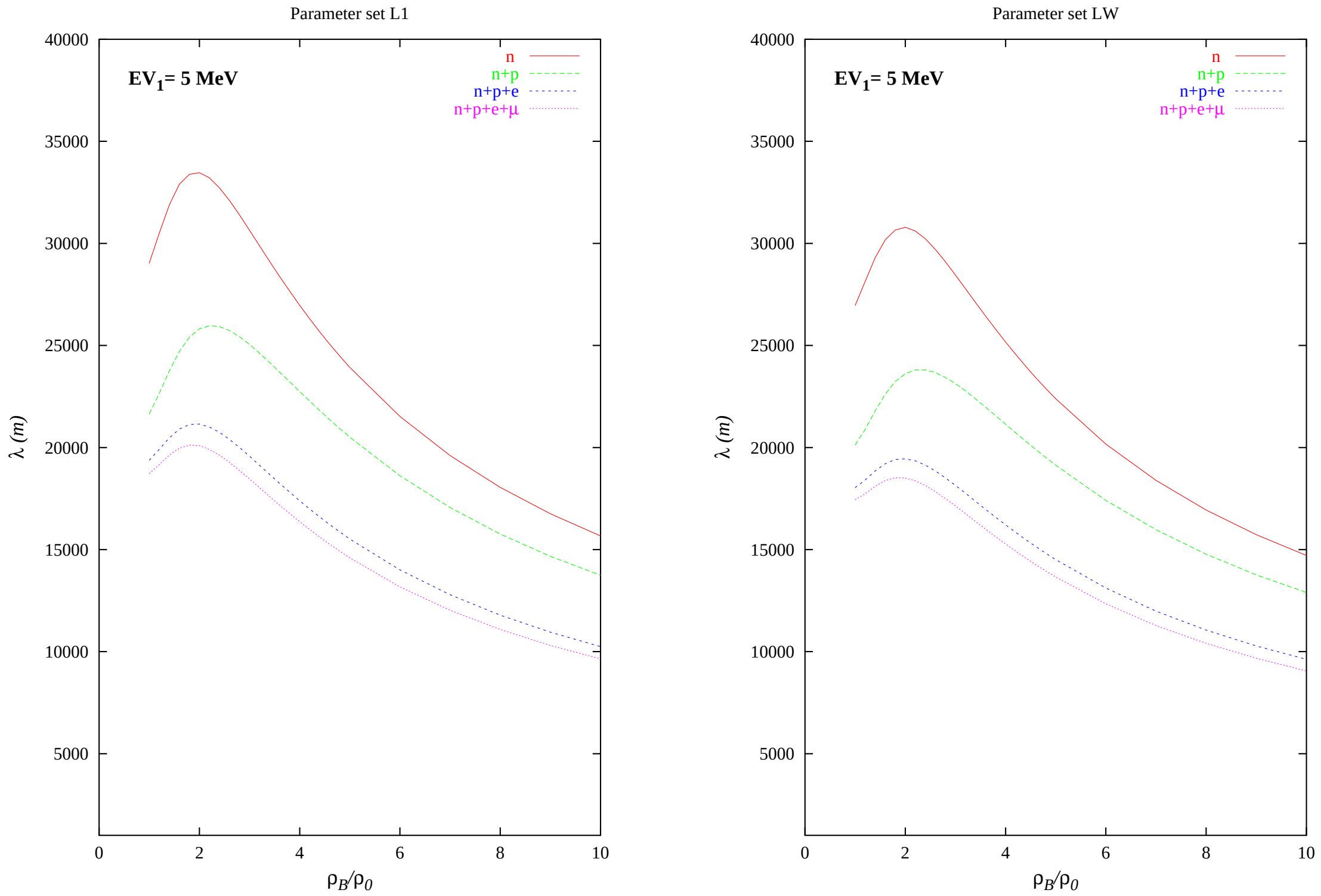

Gambar 4.10: Kiri ke kanan, lintasan bebas rata-rata neutrino dengan menggunakan model Walecka linier untuk parameter set L1 dan LW . Hal ini dilakukan pada energi awal neutrino,$E_\nu = 5$ MeV.

Lintasan bebas rata-rata neutrino yang diperlihatkan oleh Gambar. 4.10 merupakan kontribusi dari setiap partikel target dengan menggunakan parameter set L1 dan LW. Dari sini, dapat dlihat bahwa lintasan bebas rata-rata neutrino kontribusi netron lebih dominan. Bila lintasan bebas rata-rata neutrino kontribusi proton ditambahkan terhadap lintasan bebas rata-rata kontribusi netron, maka lintasan bebas rata-ratanya menurun. Demikian pula, ketika lintasan bebas rata-rata neutrino kontribusi elektron dan muon ditambahkan pada lintasan bebas rata-rata kontribusi neutron dan proton, maka lintasan bebas rata-rata total neutrino semakin menurun.

Lintasan bebas rata-rata yang ditunjukkan pada Gambar. 4.10- 4.11 merupakan lintasan bebas rata-rata neutrino kontribusi dari setiap partikel target.

Dari gambar ini secara keseluruhan, dapat dilihat bahwa lintasan bebas rata-rata neutrino kontribusi dari neutron lebih besar. Gambar ini dibuat dengan energi awal neutrino sebesar 5 MeV. Dari lintasan bebas rata-rata total neutrino terlihat cara masing-masing komponen berkontribusi mempunyai kecendrungan yang serupa untuk ketiga parameter set. Informasi mengenai parameter set dapat dilihat pada Ref. [18]. Lintasan bebas rata-rata neutrino pada Gambar. 4.11 panel sebelah kanan memberikan bahwa lintasan bebas rata-rata, $\lambda$, pada kisaran antara 1500-2500 m. Meskipun tiap-tiap parameter set memberikan harga yang berbeda tetapi kecendrungan dari lintasan bebas rata-ratanya relatif serupa.

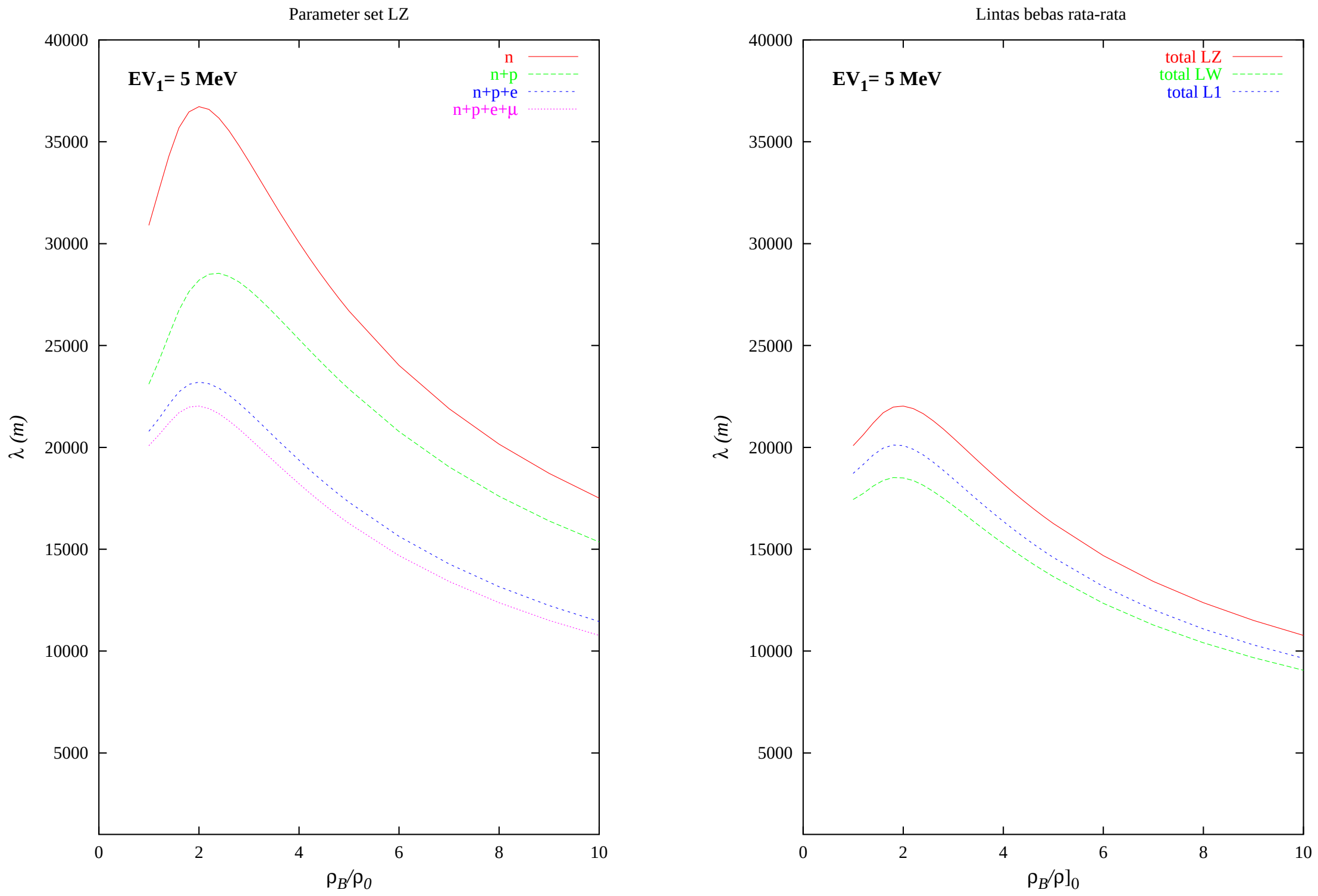

Gambar 4.11: Kiri ke kanan, lintasan bebas rata-rata neutrino menggunakan model Walecka linier untuk parameter set LZ (kiri). Sebelah kanan memperlihatkan lintasan bebas rata-rata total neutrino untuk setiap parameter set secara keseluruhan. Hal ini dilakukan pada energi awal neutrino, $E_\nu = 5$ MeV.

### 4.2.2 Model Walecka nonlinier

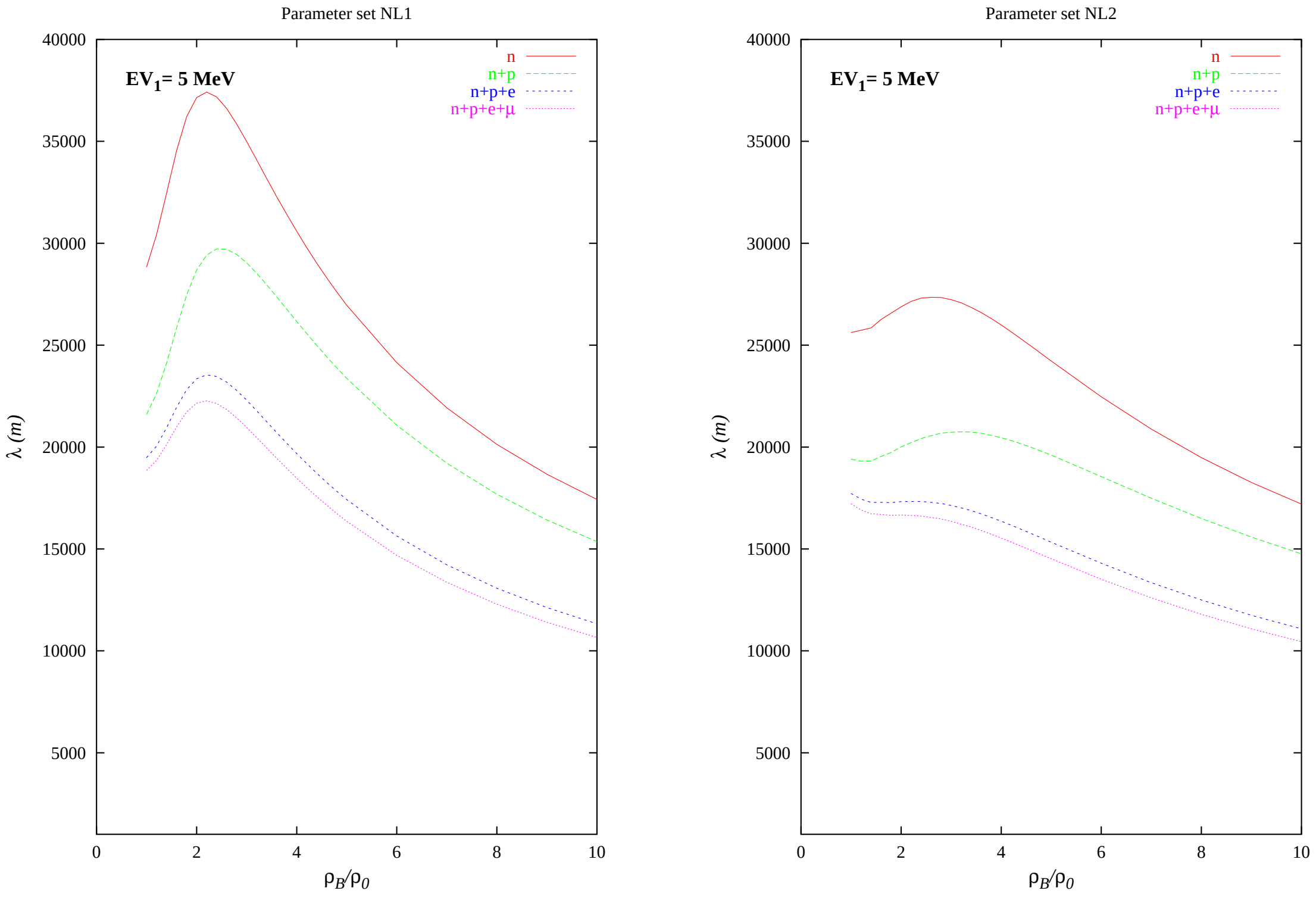

Gambar 4.12: Dari kiri ke kanan, lintasan bebas rata-rata neutrino menggunakan model Walecka nonlinier untuk parameter set NL1 dan NL2. Ini dilakukan pada energi awal neutrino, $E_\nu = 5$ MeV.

Gambar. 4.12- 4.13 dibuat pada energi awal neutrino, $E_\nu = 5$ MeV. Model nuklir yang digunakan adalah model Walecka nonlinier dengan parameter set yang berbeda-beda yakni NL1, NL2, NLZ dan NLSH. Informasi mengenai parameter set ini juga dapat dilihat secara lengkap pada Ref. [14], [18]. Bila dibandingkan Gambar. 4.10- 4.11 yang menggunakan model Walecka linier, lintasan bebas rata-rata neutrino berbeda cukup signifikan baik untuk lintasan bebas rata-rata total neutrino maupun kontribusi dari masing-masing komponen penyusun bintang netron. Perbedaan yang signifikan ini diprediksi disebabkan oleh penambahan suku orde yang lebih tinggi pada Lagrangian model Walecka linier.

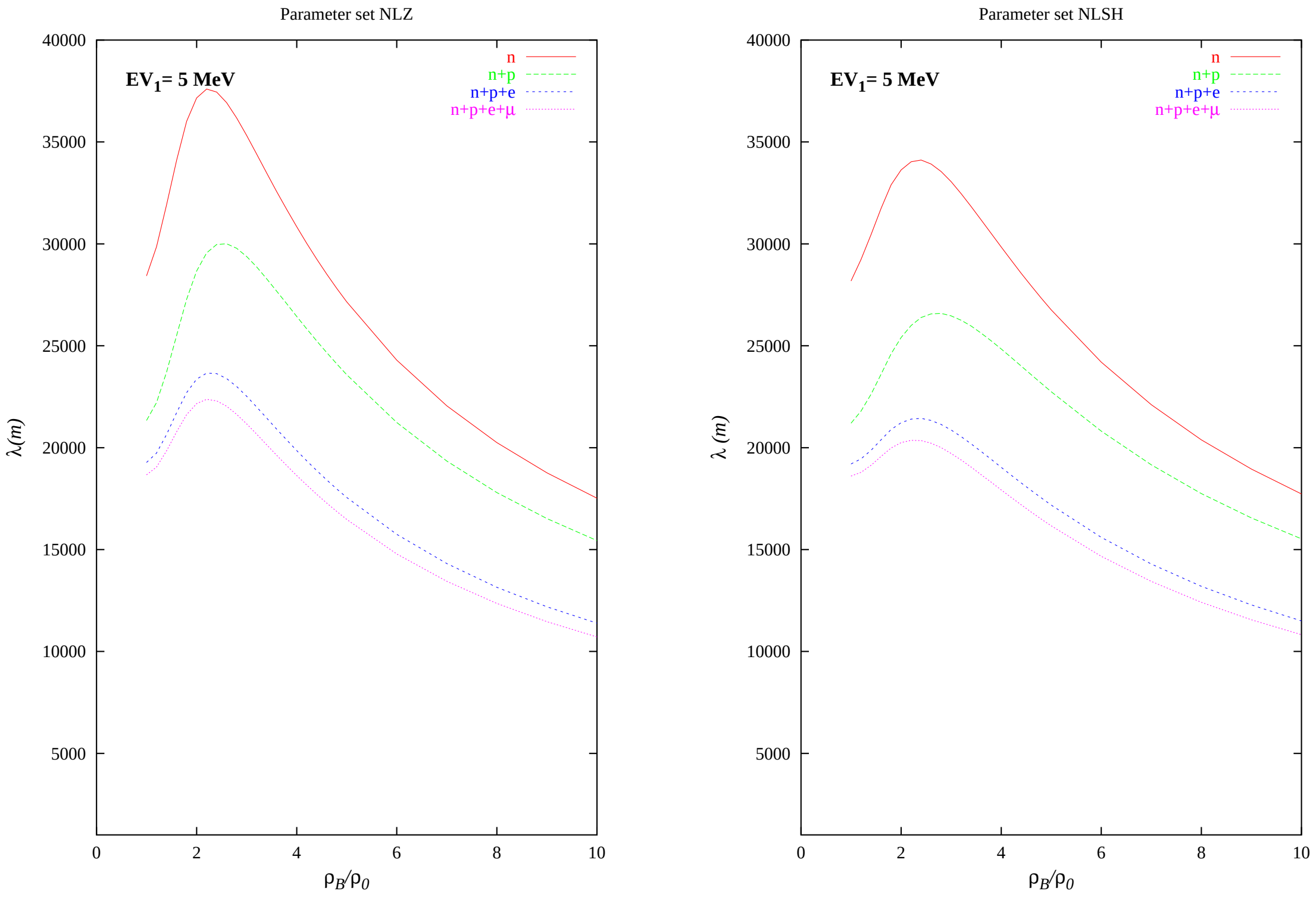

Gambar 4.13: Kiri ke kanan, lintasan bebas rata-rata neutrino dengan menggunakan model Walecka nonlinier untuk parameter set NLZ dan NLSH. Ini dilakukan pada energi awal neutrino, $E_\nu = 5$ MeV.

Dari Gambar. 4.12- 4.13 dapat dilihat bahwa lintasan bebas rata-rata neutrino juga cukup bervariasi. Dari lintasan bebas rata-rata ini dapat diprediksi bahwa model nuklir sangat sensitif terhadap parameter set yang digunakan.

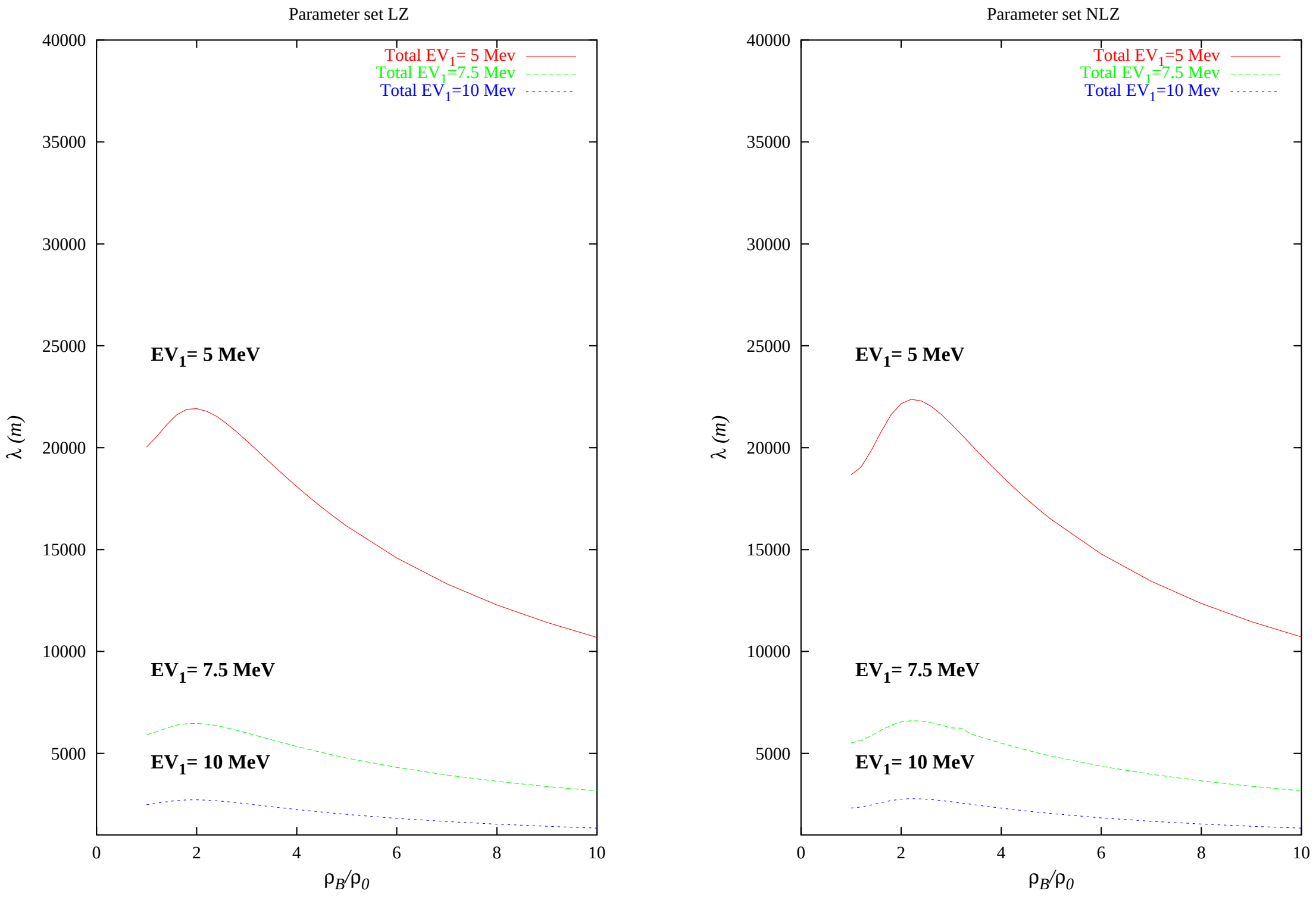

Gambar 4.14: Kiri ke kanan, lintasan bebas rata-rata neutrino menggunakan model Walecka linier dan nonlinier untuk parameter set LZ (kiri) dan NLZ (kanan). Energi awal neutrino divariasikan dari 5 hingga 10 MeV.

Gambar. 4.14 dibuat pada energi awal neutrino yang bervariasi, yakni 5, 7.5 dan 10 MeV. Model nuklir yang digunakan adalah model Walecka linier dan nonlinier yang direpresentasikan oleh parameter set LZ dan NLZ, berturut-turut. Kedua gambar ini (kiri dan kanan) memperlihatkan bahwa semakin besar energi awal neutrino maka lintasan bebas rata-rata neutrino semakin kecil. Dengan kata lain, energi awal neutrino yang semakin besar mengakibatkan neutrino semakin banyak menumbuk materi sehingga energi neutrino semakin berkurang mengakibatkan lintasan bebas rata-rata neutrino semakin kecil. Terlihat baik model nonlinier maupun linier mempunyai kecendrungan yang sama terhadap variasi energi awal neutrino. Namun hal ini, tidak demikian halnya pada model Kopling Titik.

### 4.2.3 Model Kopling Titik

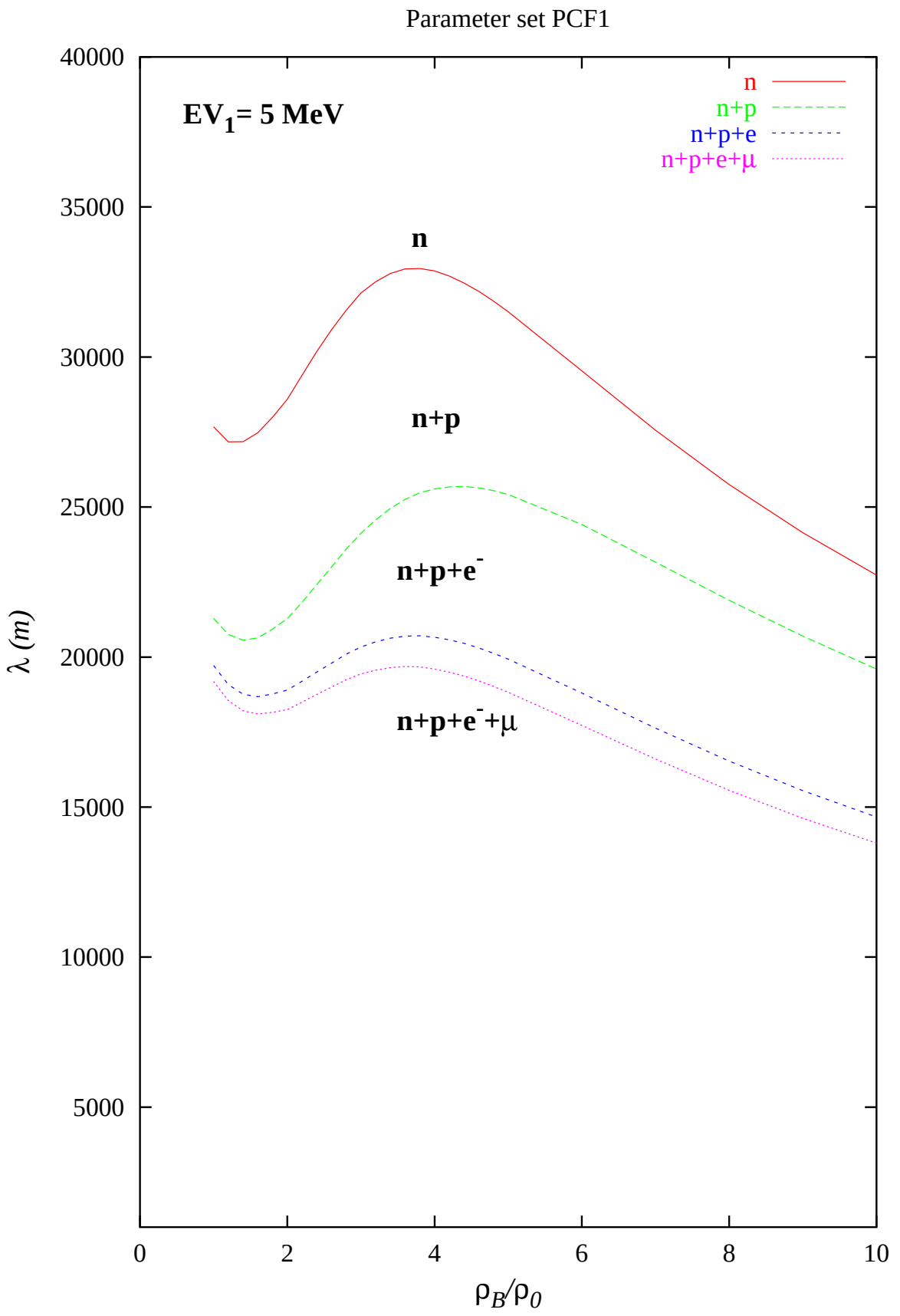

Gambar 4.15: Lintasan bebas rata-rata neutrino menggunakan model Kopling Titik. Hal ini dilakukan pada energi awal neutrino, $E_\nu = 5$ MeV.

Model nuklir yang digunakan pada Gambar. 4.15 adalah model Kopling Titik dengan parameter set PCF1. Gambar ini dibuat pada energi awal neutrino, $E_\nu = 5$ MeV. Gambar ini memperlihatkan bahwa lintasan bebas rata-rata neutrino memiliki kecendrungan yang berbeda dibandingkan model Walecka linier maupun nonlinier. Namun, sama seperti model Walecka linier dan nonlinier, Gambar. 4.15 memperlihatkan bahwa lintasan bebas rata-rata neutrino kontribusi netron masih lebih dominan, dibandingkan kontribusi partikel penyusun materi yang lain.

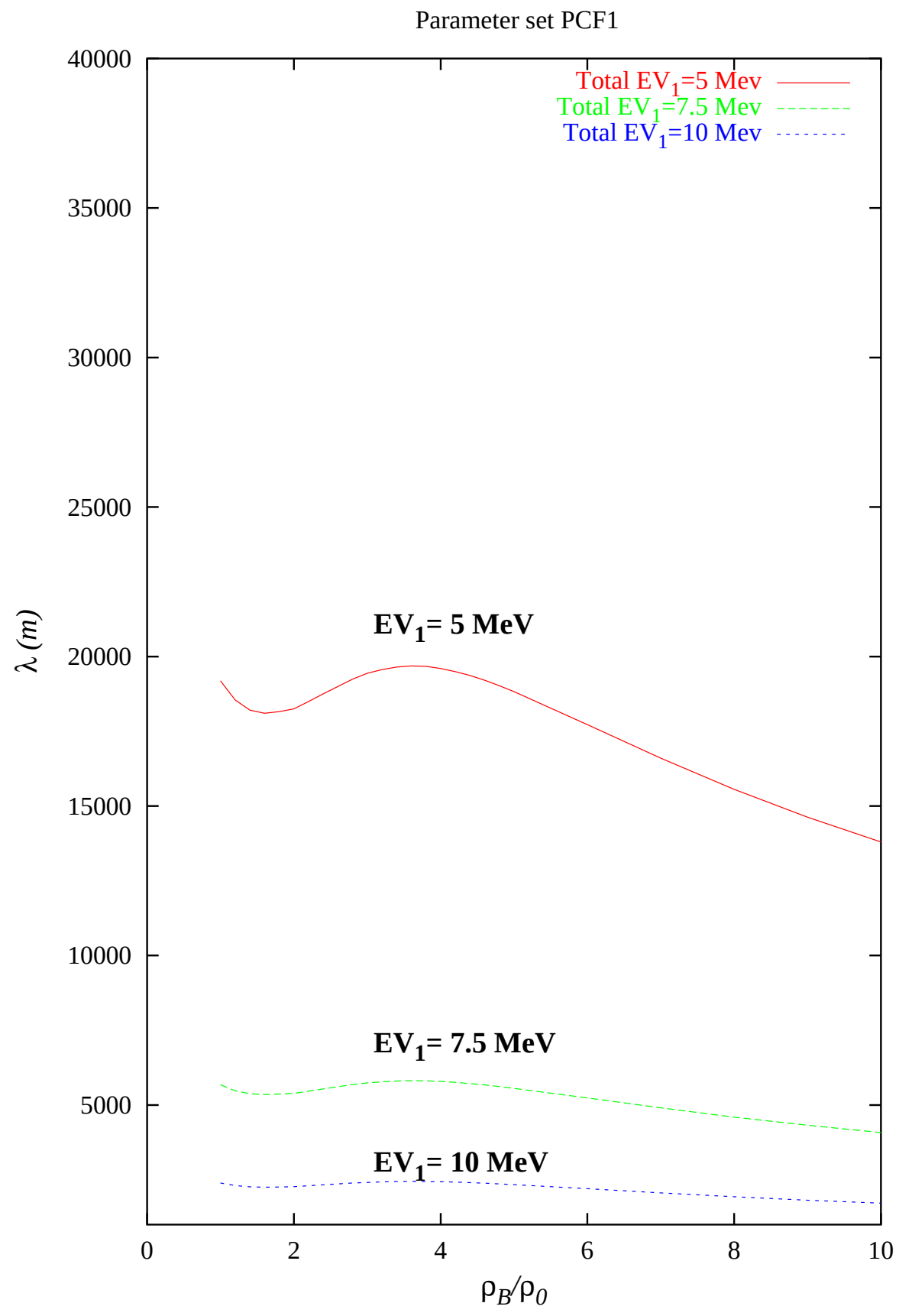

Gambar 4.16: Lintasan bebas rata-rata neutrino mengunakan model Kopling Titik dengan parameter set PCF1. Energi awal neutrino divariasikan dengan interval 5 hingga 10 MeV.

Lintasan bebas rata-rata neutrino pada Gambar. 4.16 memiliki kecendrungan yang berbeda dengan Gambar. 4.14. Gambar. 4.16 memperlihatkan bahwa lintasan bebas rata-rata neutrino menurun bila energi awal neutrino divariasi $E_\nu$ dari 5 hingga 10 MeV. Semakin besar energi awal neutrino, maka semakin kecil lintasan bebas rata-rata neutrino yang diberikan.

Dari Gambar. 4.14 dan 4.16 dapat diprediksi bahwa lintasan bebas rata-rata neutrino menurun, bila energi awal neutrino dan kerapatan nukleon diperbesar.

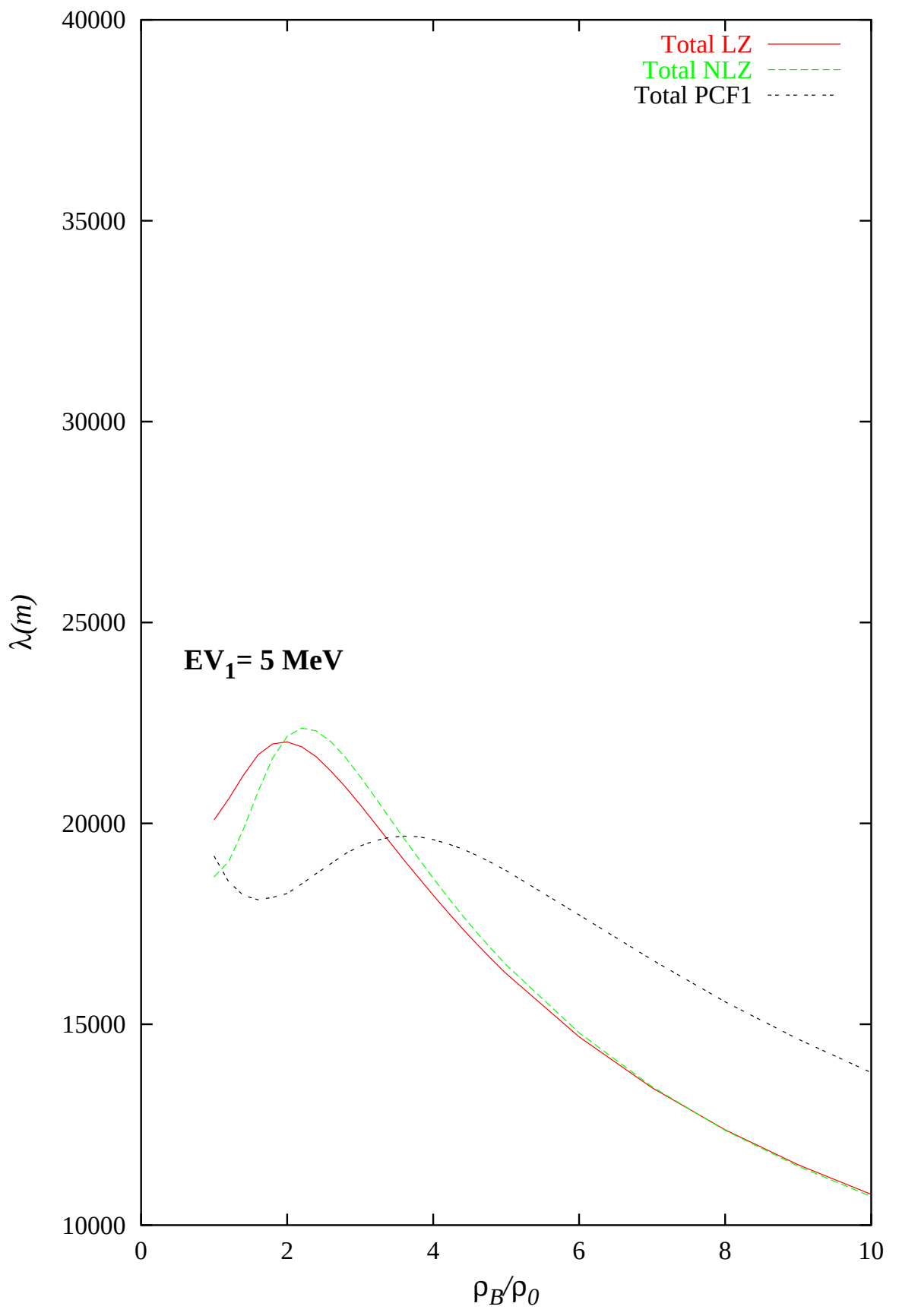

Gambar 4.17: Lintasan bebas rata-rata neutrino untuk model nuklir yang berbeda. Dilakukan pada energi awal neutrino, $E_\nu = 5$ MeV.

Gambar. 4.17 dibuat pada energi awal neutrino sebesar 5 MeV. Gambar ini memperlihatkan bahwa lintas bebas rata-rata dari model Kopling Titik dengan parameter set PCF1 memiliki kecendrungan yang berbeda dibandingkan model Walecka linier dan nonlinier, dalam hal ini direpresentasikan oleh parameter set LZ dan NLZ. Dari kedua model ini dapat diprediksi bahwa lintasan bebas rata-rata neutrino memiliki perbedaan yang cukup signifikan untuk setiap model nuklir yang digunakan (PC vs Walecka).

# Bab 5

# Kesimpulan

Dari hasil pembahasan diatas dapat disimpulkan sebagai berikut:

- Pada bintang netron, tampang lintang diferensial neutrino semakin besar, sebaliknya lintasan bebas rata-rata semakin menurun ketika energi awal neutrino dan kerapatan nukleon diperbesar.
- Untuk energi awal dan kerapatan nukleon tetap dengan alih momentum bervariasi diperoleh tampang lintang menurun bila alih momentumnya diperbesar.
- Perbedaan antara model Walecka dan Kopling Titik juga sangat signifikan baik pada tampang lintang diferensial maupun lintasan bebas rata-rata neutrino.
- Model nuklir sangat sensitif terhadap parameter set yang digunakan. Hal ini diperlihatkan oleh perbedaan tampang lintang diferensial dan lintasan bebas rata-rata neutrino untuk masing-masing parameter set yang digunakan. Perbedaan ini sangat tampak pada model Kopling Titik.

# Lampiran A

# Polarisasi

## A.1 Polarisasi

$$\Pi^j_{\mu\nu}(q) \;=\; -i\int \frac{d^4p}{(2\pi)^4} Tr[G^j(p)J^j_\mu G^j(p+q)J^j_\nu], \qquad (A.1)$$

dimana, indeks j menunjukkan indeks dari partikel target, namun untuk perhitungan selanjutnya indeksnya tidak akan ditulis.

$$\begin{aligned} G(p) &= G_F(p) + G_D(p), \\ G_F(p) &= \frac{\not p + M}{p^2 - M^2 + i\epsilon} \equiv g_F(p)(\not p + M), \\ G_D(p) &= \frac{i\pi}{E_p}\delta(p_0 - E_p)\Theta(p_F - |p|)(\not p + M) \equiv g_D(p)(\not p + M). \end{aligned}$$

Variabel $G(p)$ merupakan propagator partikel target, kemudian, bila $G(p)$ disubsitusikan ke Pers.(A.1) diatas, maka akan diperoleh

$$\begin{aligned} \Pi^D_{\mu\nu}(q) &= -i\int \frac{d^4p}{(2\pi)^4} Tr[(G_F(p) + G_D(p))J_\mu(G_F(p+q) + G_D(p+q))J_\nu] \\ &= -i\int \frac{d^4p}{(2\pi)^4} Tr[G_F(p)J_\mu G_F(p+q)J\nu + G_F(p)J_\mu G_D(p+q)J\nu, \\ &+ G_D(p)J_\mu G_F(p+q)J_\nu + G_D(p)J_\mu G_D(p+q)J_\nu]. \qquad (A.2) \end{aligned}$$

Dengan pendekatan medan rata-rata maka persamaan yang mengandung suku $G_F(p)J_\mu G_F(p+q)J_\nu$ yang divergen dapat diabaikan sehingga persamaannya menjadi

$$\Pi^D_{\mu\nu} \;=\; -i\int \frac{d^4p}{(2\pi)^4} Tr[G_F(p)J_\mu G_D(p+q)J_\nu + G_D(p)J_\mu G_F(p+q)J_\nu$$

$$
\begin{aligned}
&+ \quad G_D(p)J_\mu G_D(p+q)J_\nu],\\
&= \quad -i\int \frac{d^4p}{(2\pi)^4}Tr[G_F(p)J_\mu G_D(p+q)J_\nu + G_D(p)J_\mu G_F(p+q)J_\nu\\
&+ \quad \frac{1}{2}G_D(p)J_\mu G_D(p+q)J_\nu + \frac{1}{2}G_D(p)J_\mu G_D(p+q)J_\nu], \qquad \text{(A.3)}
\end{aligned}
$$

$$
\begin{aligned}
\Pi^D_{\mu\nu} &= -i\int \frac{d^4p}{(2\pi)^4}Tr[\frac{1}{2}G_D(p)J_\mu G_D(p+q)J_\nu + G_F(p)J_\mu G_D(p+q)J_\nu]\\
&- i\int \frac{d^4p}{(2\pi)^4}Tr[\Big[\frac{1}{2}G_D(p)J_\mu G_D(p+q)J_\nu + G_D(p)J_\mu G_F(p+q)J_\nu].\text{(A.4)}
\end{aligned}
$$

Perhatikan suku pertama pada Pers.(A.4) diatas. Dengan menggunakan sedikit manipulasi matematis dan menggunakan teorema Trace maka akan diperoleh

$$
-i\int \frac{d^4p}{(2\pi)^4}Tr[\frac{1}{2}G_D(p)J_\mu G_D(p+q)J_\nu + G_F(p)J_\mu G_D(p+q)J_\nu].
$$

Bila dilakukan pergeseran dari $p{\rightarrow}p-q$, maka didapatkan

$$
-i\int \frac{d^4p}{(2\pi)^4}Tr[\frac{1}{2}G_D(p-q)J_\mu G_D(p)J_\nu + G_F(p-q)J_\mu G_D(p)J_\nu].
$$

Dengan mengunakan teorema Trace maka persamaan diatas dapat ditulis sebagai

$$
Tr[\gamma_\mu\gamma_\nu\gamma_\rho\gamma_\sigma] \quad = \quad Tr[\gamma_\sigma\gamma_\rho\gamma_\mu\gamma_\nu].
$$

maka, Pers. A.1 menjadi

$$
-i\int \frac{d^4p}{(2\pi)^4}Tr[\frac{1}{2}G_D(p)J_\mu G_D(p-q)J_\nu + G_D(p)J_\mu G_F(p-q)J_\nu].
$$

Dengan demikian, Pers.(A.4) diatas dapat ditulis sebagai

$$
\begin{aligned}
\Pi^D_{\mu\nu} &= -i\int \frac{d^4p}{(2\pi)^4}Tr[\frac{1}{2}G_D(p)J_\mu G_D(p+q)J_\nu + G_D(p)J_\mu G_F(p+q)J_\nu]\\
&- i\int \frac{d^4p}{(2\pi)^4}Tr[\frac{1}{2}G_D(p)J_\mu G_D(p-q)J_\nu + G_D(p)J_\mu G_F(p-q)J_\nu]. \text{ (A.5)}
\end{aligned}
$$

persamaan ini dapat ditulis dalam bentuk yang lebih sederhana sebagai berikut

$$
\begin{aligned}
\Pi^D_{\mu\nu} &= -i\int \frac{d^4p}{(2\pi)^4}Tr[\frac{1}{2}G_D(p)J_\mu G_D(p+q)J_\nu + G_D(p)J_\mu G_F(p+q)J_\nu]\\
&+ [q{\rightarrow}-q]. \qquad \text{(A.6)}
\end{aligned}
$$

dimana verteks J dapat ditulis secara eksplisit sebagai

$$
\begin{aligned}
J_\mu &= \gamma_\mu(C_V - C_A)\gamma_5,\\
J_\nu &= \gamma_\nu(C_V - C_A)\gamma_5,
\end{aligned}
$$

Tabel A.1: Kopling vektor dan aksial arus netral yang merupakan kontribusi dari sudut Weinberg $\Theta_W$ dan konstanta kopling aksial nukleon $g_A$ [2].

| Reaksi | $C_V$ | $C_A$ |
|---|---|---|
| $\nu_i$ + n $\rightarrow \nu_i$ + n | -$\frac{1}{2}$ | -$\frac{g_A}{2}$ |
| $\nu_i$ + p $\rightarrow \nu_i$ + p | $\frac{1}{2}$ - $2sin^2\Theta_W$ | $\frac{g_A}{2}$ |
| $\nu_e + e^- \rightarrow \nu_e + e^-$ | $\frac{1}{2} + 2sin^2\Theta_W$ | $\frac{1}{2}$ |
| $\nu_e + \mu^- \rightarrow \nu_e + \mu^-$ | -$\frac{1}{2} + 2sin^2\Theta_W$ | -$\frac{1}{2}$ |

dimana $C_V$, $C_A$ menyatakan konstanta kopling vektor dan skalar yang berhubungan dengan sudut Weinberg $\Theta_W$, dimana $sin^2\Theta_W = 0.223$. Kopling vektor dan aksial dari arus netral dapat dilihat secara lengkap pada tabel. A.1 diatas. Dengan mensubsitusikan verteks J diatas ke Pers.(A.6), maka

$$\begin{aligned}
\Pi^D_{\mu\nu} &= -i\int\frac{d^4p}{(2\pi)^4}Tr[g_D(p)(\not p + M)(\gamma_\mu C_V - C_A\gamma_\mu\gamma_5)g_F(p+q) \\
&\times (\not p + \not q + M)(\gamma_\nu C_V - C_A\gamma_\nu\gamma_5) \\
&+ \frac{1}{2}g_D(p)(\not p + M)(\gamma_\mu C_V - C_A\gamma_\nu\gamma_5)g_D(p+q) \\
&\times (\not p + \not q + M)(\gamma_\nu C_V - C_A\gamma_\nu\gamma_5)] + [q \rightarrow -q], \\
&= -i\int\frac{d^4p}{(2\pi)^4}Tr[g_D(p)g_F(p+q) + \frac{1}{2}g_D(p)g_D(p+q)] \\
&\times (\not p + M)(C_V\gamma_\mu - C_A\gamma_\mu\gamma_5)(\not p + \not q + M)(C_V\gamma_\nu - C_A\gamma_\nu\gamma_5) \\
&+ [q \rightarrow -q], \\
&= -i\int\frac{d^4p}{(2\pi)^4}[g_D(p)g_F(p+q) + \frac{1}{2}g_D(p)g_D(p+q)]\mathbf{F}_{\mu\nu}(\mathbf{p},\mathbf{p}+\mathbf{q}) \\
&+ [q \rightarrow -q]. \qquad (A.7)
\end{aligned}$$

## A.2 Perhitungan Trace $F_{\mu\nu}(p, p+q)$

$$\begin{aligned}
F_{\mu\nu}(p,p+q) &= Tr[(\not p + M)(C_V\gamma_\mu - C_A\gamma_\mu\gamma_5)(\not p + \not q + M)(C_V\gamma_\nu - C_A\gamma_\nu\gamma_5)], \\
&= Tr[(\not p + M)C_V\gamma_\mu(\not p + \not q + M)C_V\gamma_\nu \\
&- (\not p + M)C_A\gamma_\mu\gamma_5(\not p + \not q + M)C_V\gamma_\nu
\end{aligned}$$

$$
\begin{aligned}
&- \quad (\not{p} + M)C_V\gamma_\mu(\not{p} + \not{q} + M)C_A\gamma_\nu\gamma_5 \\
&+ \quad (\not{p} + M)C_A\gamma_\mu\gamma_5(\not{p} + \not{q} + M)C_A\gamma_\nu\gamma_5], \\
&= \quad C_V^2 Tr[(\not{p} + M)\gamma_\mu(\not{p} + \not{q} + M)\gamma_\nu] \\
&- \quad C_V C_A(\not{p} + M)\gamma_\mu\gamma_5(\not{p} + \not{q} + M)\gamma_\nu \\
&- \quad C_V C_A(\not{p} + M)\gamma_\mu(\not{p} + \not{q} + M)\gamma_\nu\gamma_5 \\
&+ \quad C_A^2(\not{p} + M)\gamma_\mu\gamma_5(\not{p} + \not{q} + M)\gamma_\nu\gamma_5, \\
&= \quad C_V^2 \mathbf{F}^{\mathbf{V}}_{\mu\nu}(\mathbf{p}, \mathbf{p} + \mathbf{q}) - 2C_A C_V \mathbf{F}^{\mathbf{V-A}}_{\mu\nu}(\mathbf{p}, \mathbf{p} + \mathbf{q}) \\
&+ \quad C_A^2 \mathbf{F}^{\mathbf{A}}_{\mu\nu}(\mathbf{p}, \mathbf{p} + \mathbf{q}).
\end{aligned} \tag{A.8}
$$

Dari Pers.(A.8) diatas, dapat dilihat bahwa $F_{\mu\nu}$ dapat disederhanakan dibagi menjadi tiga komponen yakni vektor, aksial dan vektor-aksial yang merupakan komponen dari $F_{\mu\nu}(p, p + q)$.

### A.2.1 Bagian Vektor

$$
\begin{aligned}
F^V_{\mu\nu}(p, p + q) &= Tr[(\not{p} + M)\gamma_\mu(\not{p} + \not{q} + M)\gamma_\nu], \\
&= Tr[\not{p}\gamma_\mu\not{p}\gamma_\nu] + Tr[\not{p}\gamma_\mu\not{q}\gamma_\nu] \\
&+ Tr[\not{p}\gamma_\mu M\gamma_\nu] + Tr[M\gamma_\mu\not{p}\gamma_\nu], \\
&+ Tr[M\gamma_\mu\not{q}\gamma_\nu] + M^2 Tr[\gamma_\mu\gamma_\nu] \\
&= Tr[\gamma_\mu p^\mu\gamma_\rho\gamma_\nu p^\nu\gamma_\sigma] + Tr[\gamma_\mu p^\mu\gamma_\rho\gamma\nu q^\nu\gamma_\sigma] \\
&+ Tr[\gamma_\mu p^\mu\gamma_\rho M\gamma_\nu] + Tr[M\gamma_\mu\gamma_\nu P^\nu\gamma_\sigma] \\
&+ Tr[M\gamma_\mu\gamma_\nu q^\nu\gamma_\nu] + M^2 Tr[\gamma_\mu\gamma_\nu],
\end{aligned} \tag{A.9}
$$

$$
\begin{aligned}
F^V_{\mu\nu}(p, p + q) &= p^\mu p^\nu Tr[\gamma_\mu\gamma_\rho\gamma_\nu\gamma_\sigma] + p^\mu q^\nu Tr[\gamma_\mu\gamma_\rho\gamma_\nu\gamma_\sigma] \\
&+ Mp^\mu Tr[\gamma_\mu\gamma_\rho\gamma_\nu] + Mp^\nu Tr[\gamma_\mu\gamma_\nu\gamma_\sigma \\
&+ Mq^\nu Tr[\gamma_\mu\gamma_\nu\gamma_\sigma] + M^2 Tr[\gamma_\mu\gamma_\nu].
\end{aligned} \tag{A.10}
$$

Dengan menggunakan teorema Trace:

$$
Tr[\gamma_\mu\gamma_\rho\gamma_\nu] = 0,
$$

$$
Tr[\gamma_\mu\gamma_\rho\gamma_\nu\gamma_\sigma] = 4[g_{\mu\rho}g_{\nu\sigma} - g_{\mu\nu}g_{\rho\sigma} + g_{\mu\sigma}g_{\rho\nu}].
$$

Pers.(A.10) diatas dapat dinyatakan sebagai

$$\begin{aligned} F^V_{\mu\nu}(p,p+q) &= 4p^\mu p^\nu[g_{\mu\rho}g_{\nu\sigma} - g_{\mu\nu}g_{\rho\sigma} + g_{\mu\sigma}g_{\rho\nu}] \\ &+ 4p^\mu q^\nu[g_{\mu\rho}g_{\nu\sigma} - g_{\mu\nu}g_{\rho\sigma} + g_{\mu\sigma}g_{\rho\nu}] + 4M^2 g_{\mu\nu}, \\ &= 4p_\rho p_\sigma - 4p_\nu p^\nu g_{\rho\sigma} + 4p_\sigma p_\rho \\ &+ 4p_\rho q_\sigma - 4p_\nu q^\nu g_{\rho\sigma} + 4p_\sigma p_\rho + 4M^2 g_{\mu\nu}, \end{aligned} \tag{A.11}$$

$$F^V_{\mu\nu}(p,p+q) = 4[2p_\mu p_\nu + p_\mu q_\nu + p_\nu q_\mu - p.q g_{\mu\nu}]. \tag{A.12}$$

### A.2.2 Bagian Vektor-Aksial

$$\begin{aligned} F^{V-A}_{\mu\nu}(p,p+q) &= Tr[\frac{1}{2}(\not p + M)\gamma_\mu\gamma_5(\not p + \not q + M)\gamma_\nu] + Tr[\frac{1}{2}(\not p + M)\gamma_\mu(\not p + \not q + M)\gamma_\nu\gamma_5], \\ &= \frac{1}{2}Tr[(\not p + M)\gamma_\mu\gamma_5(\not p + \not q + M)\gamma_\nu] + \frac{1}{2}Tr[(\not p + M)\gamma_\mu(\not p + \not q + M)\gamma_\nu\gamma_5], \\ &= \frac{1}{2}Tr[(\not p + M)\gamma_\mu\gamma_5(\not p + \not q + M)\gamma_\nu] + \frac{1}{2}Tr[(\not p + M)\gamma_\mu\gamma_5(\not p + \not q - M)\gamma_\nu], \\ &= \frac{1}{2}Tr[\not p\gamma_\mu(\not p + \not q)\gamma_5\gamma_\nu] + \frac{1}{2}M^2 Tr[\gamma_\mu\gamma_5\gamma_\nu] \\ &+ \frac{1}{2}Tr[\not p\gamma_\mu(\not p + \not q)\gamma_5\gamma_\nu] - \frac{1}{2}M^2 Tr[\gamma_\mu\gamma_5\gamma_\nu], \\ &= Tr[\not p\gamma_\mu(\not p + \not q)\gamma_5\gamma_\nu], \\ &= p^\rho p^\sigma Tr[\gamma_\rho\gamma_\mu\gamma_\sigma\gamma_5\gamma_\nu] + p^\rho q^\sigma Tr[\gamma_\rho\gamma_\mu\gamma_\sigma\gamma_5\gamma_\nu], \\ &= Tr[\gamma_\rho\gamma_\mu\gamma_\sigma\gamma_5\gamma_\nu][p^\rho p^\sigma + p^\rho q^\sigma], \\ &= -Tr[\gamma_\rho\gamma_\mu\gamma_\sigma\gamma_\nu\gamma_5][p^\rho p^\sigma + p^\rho q^\sigma], \\ &= -4i\epsilon_{\rho\mu\sigma\nu}[p^\rho p^\sigma + p^\rho q^\sigma], \end{aligned} \tag{A.13}$$

maka persamaan akhir dari vektor-aksial dapat ditulis sebagai

$$F^{V_A}_{\mu\nu}(p,p+q) = -4i\epsilon_{\mu\nu\rho\sigma}p^\rho q^\sigma. \tag{A.14}$$

### A.2.3 Bagian Aksial

$$\begin{aligned} F^A_{\mu\nu}(p,p+q) &= Tr[(\not p + M)\gamma_\mu\gamma_5(\not p + \not q + M)\gamma_\nu\gamma_5] \\ &= Tr[(\not p + M)\gamma_\mu\gamma_5(\not p + \not q - M)\gamma_5\gamma_\nu], \\ &= Tr[\not p\gamma_\mu\not p\gamma_\nu] + Tr[\not p\gamma_\mu\not q\gamma_\nu] \end{aligned}$$

$$
\begin{aligned}
&- \; Tr[\not{p}\gamma_\mu M\gamma_\nu] + Tr[M\gamma_\mu\not{p}\gamma_\nu] \\
&+ \; Tr[M\gamma_\mu\not{q}\gamma_\nu] - M^2 Tr[\gamma_\mu\gamma_\nu], \\
&= \; p^\mu p^\nu Tr[\gamma_\mu\gamma_\rho\gamma_\nu\gamma_\sigma] \\
&+ \; p^\mu q^\nu Tr[\gamma_\mu\gamma_\rho\gamma_\nu\gamma_\sigma] - 4M^2 g_{\mu\nu}. \qquad \text{(A.15)}
\end{aligned}
$$

Hasil akhir persamaan aksial diperoleh sebagai

$$
F^A_{\mu\nu}(p, p+q) \;=\; 4[2p_\mu p_\nu + p_\mu q_\nu + p_\nu q_\mu - p.q g_{\mu\nu} - 2M^2 g_{\mu\nu}]. \qquad \text{(A.16)}
$$

Setelah menghitung suku pertama dalam polarisasi maka selanjutnya kita akan menghitung suku kedua dalam polarisasi dimana $[q \rightarrow -q]$ pada $F_{\mu\nu}(p, p+q)$. Sekarang, kita akan menghitung $F_{\mu\nu}(p, p-q)$ sebagai

$$
-i\int \frac{d^4p}{(2\pi)^4} Tr[G_D(p)J_\mu G_F(p-q)J_\nu + \frac{1}{2}G_D(p)J_\mu G_D(p-q)J_\nu].
$$

dengan mensubsitusikan $G_D(p)$, $G_F(p-q)$, $G_D(p-q)$, $J_\mu$ maka akan didapatkan persamaan yang dinyatakan sebagai

$$
\begin{aligned}
&= \; -i\int \frac{d^4p}{(2\pi)^4} Tr[g_D(p)(\not{p}+M)(\gamma_\mu C_V - C_A\gamma_\mu\gamma_5)g_F(p-q)(\not{p}-\not{q}+M)(\gamma_\nu C_V - C_A\gamma_\nu\gamma_5) \\
&+ \; \frac{1}{2}g_D(p)(\not{p}+M)(\gamma_\mu C_V - C_A\gamma_\mu\gamma_5)g_D(p-q)(\not{p}-\not{q}+M)(C_V\gamma_\mu - C_A\gamma_\mu\gamma_5)], \\
&= \; -i\int \frac{d^4p}{(2\pi)^4}[g_D(p)g_F(p-q) + \frac{1}{2}g_D(p)g_D(p-q)] \\
&\times \; Tr[(\not{p}+M)(\gamma_\mu C_V - \gamma_\mu\gamma_5 C_A)(\not{p}-\not{q}+M)(\gamma_\nu C_V - C_A\gamma_\nu\gamma_5)]. \qquad \text{(A.17)}
\end{aligned}
$$

Persamaan ini juga dapat ditulis dalam bentuk yang lain, seperti yang dibawah ini

$$
-i\int \frac{d^4p}{(2\pi)^4}[g_D(p)g_F(p-q) + \frac{1}{2}g_D(p)g_D(p-q)]\mathbf{F}_{\mu\nu}(\mathbf{p}, \mathbf{p}-\mathbf{q}).
$$

## A.3 Perhitungan Trace $F_{\mu\nu}(p, p-q)$

$$
\begin{aligned}
F_{\mu\nu}(p, p-q) \;&=\; Tr[(\not{p}+M)(\gamma_\mu C_V - C_A\gamma_\mu\gamma_5)(\not{p}-\not{q}+M)(\gamma_\nu C_V - C_A\gamma_\nu\gamma_5)], \\
&= \; C_V^2 Tr[(\not{p}+M)\gamma_\mu(\not{p}-\not{q}+M)\gamma_\nu], \\
&- \; C_V C_A Tr[(\not{p}+M)\gamma_\mu(\not{p}-\not{q}+M)\gamma_\nu\gamma_5]
\end{aligned}
$$

$$
\begin{aligned}
&- \quad C_V C_A Tr[(\not{p}+M)\gamma_\mu\gamma_5(\not{p}-\not{q}+M)\gamma_\nu] \\
&+ \quad C_A^2 Tr[(\not{p}+M)\gamma_\mu\gamma_5(\not{p}-\not{q}+M)\gamma_\nu\gamma_5], \\
&= \quad C_V^2 \mathbf{F}^{\mathbf{V}}_{\mu\nu}(\mathbf{p},\mathbf{p}-\mathbf{q}) + C_A^2 \mathbf{F}^{\mathbf{A}}_{\mu\nu}(\mathbf{p},\mathbf{p}-\mathbf{q}) \\
&- \quad 2C_V C_A \mathbf{F}^{\mathbf{V}} - \mathbf{A}_{\mu\nu}(\mathbf{p},\mathbf{p}-\mathbf{q}). \qquad \text{(A.18)}
\end{aligned}
$$

dengan cara yang sama seperti yang diawal maka,$F^V_{\mu\nu}$, $F^{V-A}_{\mu\nu}$,$F^A_{\mu\nu}$ dapat dihitung.

### A.3.1 Bagian Vektor

$$
\begin{aligned}
F^V_{\mu\nu}(p,p-q) &= Tr[(\not{p}+M)\gamma_\mu(\not{p}-\not{q}+M)\gamma_\nu], \\
&= Tr[\not{p}\gamma_\mu\not{p}\gamma_\nu] - Tr[\not{p}\gamma_\mu\not{q}\gamma_\nu] \\
&+ Tr[\not{p}\gamma_\mu M\gamma_\nu] + Tr[M\gamma_\mu\not{p}\gamma_\nu] \\
&- Tr[M\gamma_\mu\not{q}\gamma_\nu] + M^2 Tr[\gamma_\mu\gamma_\nu], \\
&= Tr[\gamma_\rho p^\rho\gamma_\mu\gamma_\sigma p^\sigma\gamma_\nu] - Tr[\gamma_\rho p^\rho\gamma_\mu\gamma_\sigma q^\sigma\gamma_\nu] \\
&+ Tr[\gamma_\rho p^\rho\gamma_\mu M\gamma_\nu + Tr[M\gamma_\mu\gamma_\sigma\gamma_\nu] \\
&- Tr[M\gamma_\mu\gamma_\sigma q^\sigma\gamma_\nu] + 4M^2 g_{\mu\nu}, \qquad \text{(A.19)}
\end{aligned}
$$

$$
\begin{aligned}
F^V_{\mu\nu}(p,p-q) &= p^\rho p^\sigma Tr[\gamma_\mu\gamma_\rho\gamma_\sigma\gamma_\nu] - p^\rho q^\sigma Tr[\gamma_\mu\gamma_\rho\gamma_\sigma\gamma_\nu] + 4M^2 g_{\mu\nu}, \\
&= 4p^\rho p^\sigma[g_{\mu\rho}g_{\nu\sigma} - g_{\mu\nu}g_{\rho\sigma} + g_{\mu\sigma}g_{\rho\nu}] \\
&- 4p^\rho q^\sigma[g_{\mu\rho}g_{\nu\sigma} - g_{\mu\nu}g_{\rho\sigma} + g_{\mu\sigma}g_{\rho\nu}] + 4M^2 g_{\mu\nu}, \\
&= 8p_\mu p_\nu - 4p_\mu q_\nu - 4p_\nu q_\mu + 4p.q g_{\mu\nu}, \qquad \text{(A.20)}
\end{aligned}
$$

$$
F^V_{\mu\nu} = 4[2p_\mu p_\nu - p_\mu q_\nu - p_\nu q_\mu + p.q g_{\mu\nu}]. \qquad \text{(A.21)}
$$

### A.3.2 Bagian Aksial

$$
\begin{aligned}
F^A_{\mu\nu}(p,p-q) &= Tr[(\not{p}+M)\gamma_\mu\gamma_5(\not{p}-\not{q}+M)\gamma_\nu\gamma_5], \\
&= Tr[(\not{p}+M)\gamma_\mu\gamma_5(\not{p}-\not{q}-M)\gamma_5\gamma_\nu], \\
&= Tr[(\not{p}+M)\gamma_\mu(\not{p}-\not{q}-M)\gamma_\nu], \\
&= Tr[\not{p}\gamma_\mu\not{p}\gamma_\nu] - Tr[\not{p}\gamma_\mu\not{q}\gamma_\nu] - M^2 Tr[\gamma_\mu\gamma_\nu], \\
&= Tr[\gamma_\rho p^\rho\gamma_\mu\gamma_\sigma p^\sigma\gamma_\nu] - Tr[\gamma_\rho p^\rho\gamma_\mu\gamma_\sigma q^\sigma\gamma_\nu] - 4M^2 g_{\mu\nu}, \\
&= p^\rho p^\sigma Tr[\gamma_\rho\gamma_\mu\gamma_\sigma\gamma_\nu] - p^\rho q^\sigma Tr[\gamma_\rho\gamma_\mu\gamma_\sigma\gamma_\nu] - 4M^2 g_{\mu\nu}, \qquad \text{(A.22)}
\end{aligned}
$$

$$
\begin{aligned}
F^A_{\mu\nu}(p,p-q) &= 4p^\rho p^\sigma[g_{\rho\mu}g_{\sigma\nu} - g_{\rho\sigma}g_{\mu\nu} + g_{\rho\nu}g_{\mu\sigma}] \\
&- 4p^\rho q^\sigma[g_{\rho\mu}g_{\sigma\nu} - g_{\rho\sigma}g_{\mu\nu} + g_{\rho\nu}g_{\mu\sigma}] - 4M^2 g_{\mu\nu}, \\
&= 4p_\mu p_\nu - 4p_\mu p^\mu g_{\mu\nu} + 4p_\nu \mu - 4p_\mu \nu + 4p_\mu q^\mu g_{\mu\nu} - 4p_\nu q_\mu - 4M^2 g_{\mu\nu}, \\
&= 8p_\mu p_\nu - 4p_\mu q_\nu - 4p_\nu q_\mu + 4p.q g_{\mu\nu} - 8M^2 g_{\mu\nu}, \qquad \text{(A.23)}
\end{aligned}
$$

Dengan demikian persamaan diatas, dapat ditulis sebagai berikut

$$
F^A_{\mu\nu}(p,p-q) = 4[2p_\mu p_\nu - p_\mu q_\nu - p_\nu q_\mu + p.q g_{\mu\nu} - 2M^2 g_{\mu\nu}]. \qquad \text{(A.24)}
$$

### A.3.3 Bagian Vektor-Aksial

$$
\begin{aligned}
F^{V-A}_{\mu\nu}(p,p-q) &= \frac{1}{2}Tr[(\not{p}+M)\gamma_\mu(\not{p}-\not{q}+M)\gamma_\nu\gamma_5] + \frac{1}{2}Tr[(\not{p}+M)\gamma_\mu(\not{p}-\not{q}+M)\gamma_5\gamma_\nu], \\
&= \frac{1}{2}Tr[(\not{p}+M)\gamma_\mu(\not{p}-\not{q}-M)\gamma_5\gamma_\nu] + \frac{1}{2}Tr[(\not{p}+M)\gamma_\mu(\not{p}-\not{q}+M)\gamma_5\gamma_\nu], \\
&= Tr[\not{p}\gamma_\mu(\not{p}-\not{q})\gamma_5\gamma_\nu], \\
&= Tr[\not{p}\gamma_\mu\not{p}\gamma_5\gamma_\nu] - Tr[\not{p}\gamma_\mu\not{q}\gamma_5\gamma_\nu], \\
&= Tr[\gamma_\rho p^\rho\gamma_\mu\gamma_\sigma p^\sigma\gamma_5\gamma_\nu] - Tr[\gamma_\rho p^\rho\gamma_\mu\gamma_\sigma p^\sigma\gamma_5\gamma_\nu], \\
&= p^\rho p^\sigma Tr[\gamma_\rho\gamma_\mu\gamma_\sigma\gamma_5\gamma_\nu] - p^\rho q^\sigma Tr[\gamma_\rho\gamma_\mu\gamma_\sigma\gamma_5\gamma_\nu], \\
&= Tr[\gamma_\rho\gamma_\mu\gamma_\sigma\gamma_5\gamma_\nu][p^\rho p^\sigma - p^\rho q^\sigma], \qquad \text{(A.25)}
\end{aligned}
$$

Dengan menggunakan teorema trace

$$
Tr[\gamma_\rho\gamma_\mu\gamma_\sigma\gamma_5\gamma_\nu] = -4i\epsilon_{\rho\mu\sigma\nu}.
$$

Dengan demikian persamaan diatas, dapat ditulis sebagai berikut :

$$
\begin{aligned}
F^{V-A}_{\mu\nu}(p,p-q) &= 4i\epsilon_{\rho\mu\sigma\nu}[p^\rho p^\sigma - p^\rho q^\sigma], \\
F^{V-A}_{\mu\nu}(p,p-q) &= 4i\epsilon_{\rho\mu\sigma\nu}[p^\rho p^\sigma - p^\rho q^\sigma]. \qquad \text{(A.26)}
\end{aligned}
$$

# Lampiran B

# Propagator

## B.1 Propagator

$$
\begin{aligned}
g_D(p)g_F(p\pm q) &= \frac{i\pi}{\overrightarrow{E_p}}\delta(p_0 - \overrightarrow{E_p})\Theta(p_F - |\overrightarrow{p}|)\delta(p_0 \pm q_0 - E_{\overrightarrow{p}\pm\overrightarrow{q}}) \\
&\times [\frac{\wp}{(p\pm q)^2 - M^2} - \frac{i\pi}{2E_{\overrightarrow{p}\pm\overrightarrow{q}}}],
\end{aligned} \tag{B.1}
$$

dimana $\wp$ menyatakan *Principle Value.*

$$
\begin{aligned}
g_D(p)g_D(p\pm q) &= -\frac{(\pi)^2}{E_{\overrightarrow{p}}E_{\overrightarrow{p}\pm\overrightarrow{q}}}\delta(p_0 - E_{\overrightarrow{p}})\Theta(p_F - |\overrightarrow{p}|) \\
&\times \delta(p_0 \pm q_0 - E_{\overrightarrow{p}\pm\overrightarrow{q}})\Theta(p_F|\overrightarrow{p} \pm \overrightarrow{q}|),
\end{aligned} \tag{B.2}
$$

Setelah menghitung propagator $F_{\mu\nu}$, $G(p)$, selanjutnya kita akan menghitung polarisasi tensor $\Pi_{\mu\nu}$

$$
\begin{aligned}
\Pi^D_{\mu\nu}(q) &= -i\int \frac{d^4p}{(2\pi)^4}[g_D(p)g_F(p+q) + \frac{1}{2}g_D(p)g_D(p+q)] \\
&\times F_{\mu\nu}(p, p+q) + [q \to -q].
\end{aligned} \tag{B.3}
$$

Sekarang, kita akan mensubsitusikan propagator ke dalam polarisasi, maka akan diperoleh sebagai

$$
\begin{aligned}
\Pi^D_{\mu\nu}(q) &= -i\int \frac{d^4p}{(2\pi)^4}[\frac{i\pi}{E_{\overrightarrow{p}}}\delta(p_0 - E_{\overrightarrow{p}})\delta(p_F - |\overrightarrow{p}|)(\frac{1}{(p+q)^2 - M^2 + i\epsilon}) \\
&- \frac{\pi^2}{2E_{\overrightarrow{p}}E_{\overrightarrow{p}+\overrightarrow{q}}}\delta(p_0 - E_{\overrightarrow{p}})\Theta(p_F - |\overrightarrow{p}|) \\
&\times \delta(p_0 + q_0 - E_{\overrightarrow{p}+\overrightarrow{q}})\Theta(p_F - |\overrightarrow{p} + \overrightarrow{q}|)]\mathbf{F}_{\mu\nu}(\mathbf{p}, \mathbf{p}+\mathbf{q}) \\
&- i\int \frac{d^4p}{(2\pi)^4}[i\pi\delta(p_0 - E_{\overrightarrow{p}})(\frac{1}{(p-q)^2 - M^2 + i\epsilon})
\end{aligned}
$$

$$
\begin{aligned}
&- \frac{\pi^2}{2E_{\vec{p}}E_{\vec{p}-\vec{q}}}\delta(p_0 - E_{\vec{p}})\Theta(p_F - |\vec{p}|)\delta(p_0 - q_0 - E_{\vec{p}-\vec{q}}) \\
&\times \Theta(p_F - |\vec{p} - \vec{q}|)]\mathbf{F}_{\mu\nu}(\mathbf{p}, \mathbf{p} - \mathbf{q}). \qquad \text{(B.4)}
\end{aligned}
$$

## B.2 Bagian Komponen Rill dan Imajiner

### B.2.1 Bagian Pertama

$$
\begin{aligned}
\Pi^{D1}_{\mu\nu}(q) &= \frac{\pi}{E_{\vec{p}}}\int \frac{d^4p}{(2\pi)^4}\delta(p_0 - E_{\vec{p}})\Theta(p_F - |\vec{p}|)\frac{1}{(p+q)^2 - M^2 + i\epsilon}\mathbf{F}_{\mu\nu}(\mathbf{p}, \mathbf{p} + \mathbf{q}) \\
&+ \frac{\pi}{E_{\vec{p}}}\int \frac{d^4p}{(2\pi)^4}\delta(p_0 - E_{\vec{p}})\Theta(p_F - |\vec{p}|)\frac{1}{(p-q)^2 - M^2 + i\epsilon}\mathbf{F}_{\mu\nu}(\mathbf{p}, \mathbf{p} - \mathbf{q}).
\end{aligned}
$$

### B.2.2 Bagian Imajiner

$$
\begin{aligned}
\Pi^{DI}_{\mu\nu}(q) &= \frac{i\pi^2}{2E_{\vec{p}}E_{\vec{p}+\vec{q}}}\int \frac{d^4p}{(2\pi)^4}F_{\mu\nu}(p, p+q)\delta(p_0 - E_{\vec{p}})\Theta(p_F - |\vec{p}|) \\
&\times \delta(p_0 + q_0 - E_{\vec{p}+\vec{q}})\Theta(p_F - |\vec{p} + \vec{q}|) \\
&+ \frac{i\pi^2}{2E_{\vec{p}}E_{\vec{p}+\vec{q}}}\int \frac{d^4p}{(2\pi)^4}F_{\mu\nu}(p, p-q)\delta(p_0 - E_{\vec{p}})\Theta(p_F - |\vec{p}|) \\
&\times \delta(p_0 - q_0 - E_{\vec{p}-\vec{q}})\Theta(p_F - |\vec{p} - \vec{q}|). \qquad \text{(B.5)}
\end{aligned}
$$

Sebelum, kita menghitung komponen bagian imajiner, ternyata didalam suku bagian pertama masih ada suku yang imajiner. Hal itu dapat dilihat pada bagian pertama yang mengandung suku $\frac{1}{(p\pm q)^2 - M^2 + i\epsilon}$, suku dalam persamaan tersebut dapat ditulis dalam bentuk sebagai berikut

$$
\frac{1}{(p\pm q)^2 - M^2 + i\epsilon} \equiv \frac{\wp}{(p\pm q)^2 - M^2} - \frac{i\pi}{2E_{\vec{p}\pm\vec{q}}}\delta((p_0 \pm q_0) - E_{\vec{p}+\vec{q}}) \qquad \text{(B.6)}
$$

Bila pers.(B.6) disubsitusikan ke $\Pi^{DR}_{\mu\nu}(q))$ maka akan didapatkan bentuk persamaan sebagai

$$
\begin{aligned}
\Pi^{D1}_{\mu\nu}(q) &= \frac{\pi}{(2\pi)^4}\int \frac{d^4p}{E_{\vec{p}}}[(\frac{\wp}{(p+q)^2 - M^2} - \frac{i\pi}{2E_{\vec{p}+\vec{q}}}\delta(p_0 + q_0 - E_{\vec{p}+\vec{q}})) \\
&\times \delta(p_0 - E_{\vec{p}})\Theta(p_F - |\vec{p}|)]\mathbf{F}_{\mu\nu}(\mathbf{p}, \mathbf{p} + \mathbf{q}) \\
&+ \frac{\pi}{(2\pi)^2}\int \frac{d^4p}{E_{\vec{p}}}[(\frac{\wp}{(p-q)^2 - M^2} - \frac{i\pi}{2E_{\vec{p}+\vec{q}}}\delta(p_0 + q_0 - E_{\vec{p}-\vec{q}})) \\
&\times \delta(p_0 - E_{\vec{p}})\Theta(p_F - |\vec{p}|)]\mathbf{F}_{\mu\nu}(\mathbf{p}, \mathbf{p} - \mathbf{q}). \qquad \text{(B.7)}
\end{aligned}
$$

Lihat, suku pertama pada $\Pi^{D1}_{\mu\nu}(q)$) yakni, $[\frac{\wp}{(p+q)^2-M^2}-\frac{i\pi}{2E_{\vec{p}+\vec{q}}}\delta(p_0+q_0-E_{\vec{p}+\vec{q}})]\delta(p_0-E_{\vec{p}})\Theta(p_F-|\vec{p}|)$ masih mengandung suku imajiner sehingga dapat ditulis:

$$\frac{\wp}{(p+q)^2-M^2}\delta(p_0-E_{\vec{p}})\Theta(p_F-|\vec{p}|)-\frac{i\pi}{2E_{\vec{p}+\vec{q}}}\delta(p_0+q_0-E_{\vec{p}+\vec{q}})\delta(p_0-E_{\vec{p}}) \\ \times\Theta(p_F-|\vec{p}|).$$

Bila persamaan ini disubsitusikan ke $\Pi^{D1}_{\mu\nu}$ maka akan diperoleh persamaan sebagai

$$\begin{aligned}
\Pi^{D1}_{\mu\nu}(q) &= \frac{\pi}{E_{\vec{p}}}\int\frac{d^4p}{(2\pi)^4}(\frac{\wp}{(p+q)^2-M^2})\delta(p_0-E_{\vec{p}})\Theta(p_F-|\vec{p}|)F_{\mu\nu}(p,p+q) \\
&- \frac{i\pi^2}{2E_{\vec{p}}E_{\vec{p}+\vec{q}}}\int\frac{d^4p}{(2\pi)^4}\Theta(p_F-|\vec{p}|)\delta(p_0-E_{\vec{p}})\delta(p_0+q_0-E_{\vec{p}+\vec{q}})F_{\mu\nu}(p,p+q) \\
&+ \frac{\pi}{E_{\vec{p}}}\int\frac{d^4P}{(2\pi)^4}(\frac{\wp}{(p-q)^2-M^2})\delta(p_0-E_{\vec{p}})\Theta(p_F-|\vec{p}|)\delta(p_0-q_0-E_{\vec{p}-\vec{q}}) \\
&\times F_{\mu\nu}(p,p-q).
\end{aligned} \tag{B.8}$$

Dari persamaan ini dapat dilihat bahwa bagian pertama masih mengandung bagian imajiner secara eksplisit, sehingga $\Pi^{ID}_{\mu\nu}(q)$ keseluruhan dapat ditulis sebagai

$$\begin{aligned}
\Pi^{ID}_{\mu\nu}(q) &= \frac{i\pi^2}{(2\pi)^4}\int\frac{d^4p}{2E_{\vec{p}}E_{\vec{p}+\vec{q}}}F_{\mu\nu}(p,p+q)\delta(p_0-E_{\vec{p}})\Theta(p_F-|\vec{p}|) \\
&\times \delta(p_0+q_0-E_{\vec{p}+\vec{q}})\Theta(p_F-|\vec{p}+\vec{q}|) \\
&- \frac{i\pi^2}{(2\pi)^4}\int\frac{d^4p}{2E_{\vec{p}}E_{\vec{p}+\vec{q}}}F_{\mu\nu}(p,p+q)\delta(p_0-E_{\vec{p}})\Theta(p_F-|\vec{p}|) \\
&\times \delta(p_0+q_0-E_{\vec{p}-\vec{q}})+[q\rightarrow-q].
\end{aligned} \tag{B.9}$$

$$\begin{aligned}
\Pi^{ID}_{\mu\nu}(q) &= \frac{i\pi^2}{(2\pi)}\int\frac{d^4p}{2E_{\vec{p}}E_{\vec{p}+\vec{q}}}F_{\mu\nu}(p,p+q)\delta(p_0-E_{\vec{p}})\Theta(p_F-|\vec{p}|) \\
&\times \delta(p_0+q_0-E_{\vec{p}+\vec{q}})[\Theta(p_F-|\vec{p}+\vec{q}|)-1], \\
&+ \frac{i\pi^2}{(2\pi)^4}\left[\int\left[\frac{d^4p}{2E_{\vec{p}}E_{\vec{p}+\vec{q}}}F_{\mu\nu}(p,p-q)\delta(p_0-E_{\vec{p}})\Theta(p_F-|\vec{p}|),\right.\right. \\
&\times \delta(p_0-q_0-E_{\vec{p}-\vec{q}})[\Theta(p_F-|\vec{p}-\vec{q}|)-1].
\end{aligned} \tag{B.10}$$

Hubungan $F_{\mu\nu}(p,p+q)$ dan $F_{\mu\nu}(p,p-q)$, dengan menggunakan konservasi arus: $q^\mu F_{\mu\nu}=0$ maka, akan diperoleh

$$F_{\mu\nu}(p,p+q) = F_{\mu\nu}(p,p-q), \tag{B.11}$$

$$
\begin{aligned}
\Pi^{ID}_{\mu\nu}(q) &= \frac{i\pi^2}{(2\pi)^4}\int \frac{d^4p}{2E_{\vec{p}}E_{\vec{p}+\vec{q}}} F_{\mu\nu}(p,p+q)\delta(p_0-E_{\vec{p}})\Theta(p_F-|\vec{p}|)\\
&\times [\delta(p_0+q_0-E_{\vec{p}+\vec{q}})[\Theta(p_F-|\vec{p}+\vec{q}|)-1]\\
&+ \delta(p_0-q_0-E_{\vec{p}+\vec{q}})[\Theta(p_F-|\vec{p}-\vec{q}|)-1]]\\
&= \frac{i\pi^2}{(2\pi)^4}\int \frac{d^4p}{2E_{\vec{p}}E_{\vec{p}+\vec{q}}} F_{\mu\nu}(p,p+q)\delta(p_0-E_{\vec{p}})\Theta(p_F-|\vec{p}|)\\
&\times [\delta(p_0+q_0-E_{\vec{p}+\vec{q}})[1-\Theta(p_F-|\vec{p}+\vec{q}|)]\\
&+ \delta(p_0-q_0-E_{\vec{p}+\vec{q}})[1-\Theta(p_F-|\vec{p}-\vec{q}|)]]. \qquad \text{(B.12)}
\end{aligned}
$$

Perhatikan, suku kedua yang didalam [ ], bila $p \rightarrow p+2q$ maka akan diperoleh persamaan sebagai

$$
\begin{aligned}
\Pi^{ID}_{\mu\nu}(q) &= \frac{-i\pi^2}{(2\pi)^4}\int \frac{d^4p}{2E_{\vec{p}}E_{\vec{p}+\vec{q}}} F_{\mu\nu}(p,p+q)\delta(p_0-E_{\vec{p}})\Theta(p_F-|\vec{p}|)\\
&\times [\delta(p_0+q_0-E_{\vec{p}+\vec{q}})[1-\Theta(p_F-|\vec{p}+\vec{q}|)]\\
&+ \delta(p_0-q_0-E_{\vec{p}+\vec{q}})[1-\Theta(p_F-|\vec{p}+\vec{q}|)]]\\
&= \frac{-i\pi^2}{(2\pi)^4}\int \frac{d^4p}{2E_{\vec{p}}E_{\vec{p}+\vec{q}}} F_{\mu\nu}\delta(p_0-E_{\vec{p}})\Theta(p_F-|\vec{p}|)\\
&\times \Theta(|\vec{p}+\vec{q}|)[\delta(p_0+q_0-E_{\vec{p}+\vec{q}})+\delta(p_0-q_0-E_{\vec{p}+\vec{q}})] \qquad \text{(B.13)}
\end{aligned}
$$

Dengan syarat, $q_0 > 0$ maka $E_{\vec{p}+\vec{q}} - p_0 > 0$, sehingga diperoleh persamaan untuk bagian polarisasi imajiner sebagai

$$
\begin{aligned}
\Pi^{ID}_{\mu\nu}(q) &= \frac{-i\pi^2}{(2\pi)^4}\int \frac{d^4p}{2E_{\vec{p}}E_{\vec{p}+\vec{q}}} F_{\mu\nu}(p,p+q)\delta(p_0-E_{\vec{p}})\Theta(p_F-|\vec{p}|)\\
&\times \Theta(|\vec{p}+\vec{q}|)\delta(p_0+q_0-E_{\vec{p}+\vec{q}}). \qquad \text{(B.14)}
\end{aligned}
$$

## B.3 Perhitungan Polarisasi Imajiner

$$
\begin{aligned}
\Pi^{ID}_{\mu\nu}(q) &= \frac{-i\pi^2}{(2\pi)^4}\int \frac{d^4p}{2E_{\vec{p}}E_{\vec{p}+\vec{q}}} F_{\mu\nu}(p,p+q)\delta(p_0-E_{\vec{p}})\Theta(p_F-|\vec{p}|)\\
&\times \Theta(|\vec{p}+\vec{q}|)\delta(p_0+q_0-E_{\vec{p}+\vec{q}}). \qquad \text{(B.15)}
\end{aligned}
$$

Bila pers.(A.8) disubsitusikan ke pers.(B.15) maka akan diperoleh persamaan sebagai berikut:

$$
\Pi^{ID}_{\mu\nu}(q) = \frac{-i\pi^2}{(2\pi)^4}\int \frac{d^4p}{2E_{\vec{p}}E_{\vec{p}+\vec{q}}}[C_V^2 F^V_{\mu\nu}(p,p+q) + C_A^2 F^A_{\mu\nu}(p,p+q)
$$

$$
\begin{aligned}
&- \quad 2C_V C_A F_{\mu\nu}^{V-A}(p,p+q)]\delta(p_0 - E_{\vec{p}})\Theta(p_F - |\vec{p}|) \\
&\times \quad \Theta(|\vec{p} + \vec{q}|)\delta(p_0 + q_0 - E_{\vec{p}+\vec{q}}). \qquad \text{(B.16)}
\end{aligned}
$$

Pers.(B.16) juga dapat ditulis:

$$
\Pi_{\mu\nu}^{ID}(q) = C_V^2 \mathbf{\Pi_{\mu\nu}^{IDV}} + C_A^2 \mathbf{\Pi_{\mu\nu}^{IDA}} - 2C_V C_A \mathbf{\Pi_{\mu\nu}^{IDV-A}}, \qquad \text{(B.17)}
$$

dimana

$$
\begin{aligned}
\Pi_{\mu\nu}^{IDV}(q) &= \frac{-i\pi^2}{(2\pi)^4} \int \frac{d^4p}{2E_{\vec{p}} E_{\vec{p}+\vec{q}}} F_{\mu\nu}^{V}(p,p+q)\delta(p_0 - E_{\vec{p}})\Theta(p_F - |\vec{p}|) \\
&\times \quad \Theta(|\vec{p} + \vec{q}| - p_F)\delta(p_0 + q_0 - E_{\vec{p}+\vec{q}}), \qquad \text{(B.18)}
\end{aligned}
$$

$$
\begin{aligned}
\Pi_{\mu\nu}^{IDA}(q) &= \frac{-i\pi^2}{(2\pi)^4} \int \frac{d^4p}{2E_{\vec{p}} E_{\vec{p}+\vec{q}}} F_{\mu\nu}^{A}(p,p+q)\delta(p_0 - E_{\vec{p}})\Theta(p_F - |\vec{p}|) \\
&\times \quad \Theta(|\vec{p} + \vec{q}| - p_F)\delta(p_0 + q_0 - E_{\vec{p}+\vec{q}}), \qquad \text{(B.19)}
\end{aligned}
$$

$$
\begin{aligned}
\Pi_{\mu\nu}^{IDV-A}(q) &= \frac{-i\pi^2}{(2\pi)^4} \int \frac{d^4p}{2E_{\vec{p}} E_{\vec{p}+\vec{q}}} F_{\mu\nu}^{V-A}(p,p+q)\delta(p_0 - E_{\vec{p}})\Theta(p_F - |\vec{p}|) \\
&\times \quad \Theta(|\vec{p} + \vec{q}| - p_F)\delta(p_0 + q_0 - E_{\vec{p}+\vec{q}}). \qquad \text{(B.20)}
\end{aligned}
$$

## B.4 Polarisasi $\Pi_{\mu\nu}^{IDV}$

$$
\begin{aligned}
\Pi_{\mu\nu}^{IDV} &= \frac{-i}{8E_{\vec{p}} E_{\vec{p}+\vec{q}}} \int \frac{dp_0 d^3p}{(2\pi)^4} F_{\mu\nu}^{V}(p,p+q)\delta(p_0 - E_{\vec{p}})\Theta(p_F - |\vec{p}|) \\
&\times \quad \delta(p_0 + q_0 - E_{\vec{p}+\vec{q}})\Theta(|\vec{p} + \vec{q}| - p_F), \qquad \text{(B.21)}
\end{aligned}
$$

dimana

$$
F_{\mu\nu}^{V}(p,p+q) = 4[2p_\mu p_\nu + p_\mu q_\nu + p_\nu q_\mu - p.q g_{\mu\nu}]. \qquad \text{(B.22)}
$$

Kerangka yang dipilih adalah $q_\mu = (q_0, |\vec{q}|, 0, 0)$,$p_\mu = (E, p_x, p_y, p_z)$ dimana, $p_x = |\vec{p}|cos\theta$, $p_y = |\vec{p}|sin\theta cos\phi$, $p_z = |\vec{p}|sin\theta sin\phi$. Dengan demikian, kita dapat: $p{\cdot}q = Eq_0 - \vec{p}{\cdot}\vec{q}$ Bila Pers.(B.22) diselesaikan dengan menggunakan kerangka ini maka akan didapat persamaan sebagai

$$
F_{00} = 4[2E^2 + Eq_0 + |\vec{p}||\vec{q}|cos\theta]
$$

$$
\begin{aligned}
F_{11} &= 4[2|\overrightarrow{p}|^2 cos^2\theta + |\overrightarrow{p}||\overrightarrow{q}|cos\theta + Eq_0] \\
F_{22} &= 4[2|\overrightarrow{p}|^2 sin^2\theta cos^2\phi + Eq_0 + |\overrightarrow{p}||\overrightarrow{q}|cos\theta] \\
F_{33} &= 4[2|\overrightarrow{p}|^2 sin^2\theta sin^2\phi + Eq_0 + |\overrightarrow{p}||\overrightarrow{q}|cos\theta] \\
&\vdots
\end{aligned}
\tag{B.23}
$$

Bila masing-masing $F_{00}$, $F_{22}$, $F_{33}$, ... disubsitusikan ke Pers.(B.18) maka hasilnya dapat disusun dalam bentuk matriks sebagai

$$
\Pi_{\mu\nu}^{IDV} = \begin{pmatrix} \Pi_{00} & \Pi_{01} & 0 & 0 \\ \Pi_{10} & \Pi_{11} & 0 & 0 \\ 0 & 0 & \Pi_{22} & 0 \\ 0 & 0 & 0 & \Pi_{33} \end{pmatrix}.
$$

Dengan pertimbangan *konservasi arus*, dimana $q_\mu \Pi_{\mu nu}^{IDV} = 0$, maka

$$
\begin{aligned}
q_0\Pi_{00} + |\overrightarrow{q}|\Pi_{01} = 0 \longrightarrow \Pi_{01} = -\frac{q_0}{|\overrightarrow{q}|}\Pi_{00}, \\
q_0\Pi_{10} + |\overrightarrow{q}|\Pi_{11} = 0 \longrightarrow \Pi_{11} = -\frac{q_0}{|\overrightarrow{q}|}\Pi_{10},
\end{aligned}
\tag{B.24}
$$

karena, $\Pi_{01} = \Pi_{10}$ maka Pers.(B.24) dapat ditulis sebagai

$$
\Pi_{11} = \frac{q_0^2}{|\overrightarrow{q}|^2}\Pi_{00}, \tag{B.25}
$$

Dengan menggunakan hubungan, $\Pi_L = \Pi_{00} - \Pi_{11}$, $q_\mu^2 = |\overrightarrow{q_0}|^2 - |\overrightarrow{q}|^2$ maka:

$$
\begin{aligned}
\Pi_L &= \Pi_{00} - \Pi_{11}, \\
&= \Pi_{00} - \frac{q_0^2}{|\overrightarrow{q}|^2}\Pi_{00}, \\
&= -\frac{q_\mu^2}{|\overrightarrow{q}|^2}\Pi_{00}.
\end{aligned}
\tag{B.26}
$$

Sedangkan,$\Pi_T = \Pi_{22} = \Pi_{33}$. Polarisasi vektor dapat dibagi menjadi dua yaitu longitudinal dan transversal.

### B.4.1 Polarisasi Longitudinal

Dengan menggunakan konservasi arus diperoleh persamaan $\Pi_L = -\frac{q_\mu^2}{|\overrightarrow{q}|^2}\Pi_{00}$, sehingga $\Pi_{00}$ yang merupakan polarisasi longitudinal dapat dihitung dengan mudah,

yakni sebagai

$$\begin{aligned}\Pi_{00} &= \frac{-i}{(2\pi)^4}\int \frac{dp_0 d^3p}{4E_{\overrightarrow{p}}E_{\overrightarrow{p}+\overrightarrow{q}}} F_{00}(p,p+q)\delta(p_0 - E_{\overrightarrow{p}})\Theta(p_F - |\overrightarrow{p}|)\\ &\times\ \delta(p_0 + q_0 - E_{\overrightarrow{p}+\overrightarrow{q}})\Theta(|p + \overrightarrow{q}| - p_F),\\ &= \frac{-i}{(2\pi)^2}\int \frac{d^3p}{8E_{\overrightarrow{p}}E_{\overrightarrow{p}+\overrightarrow{q}}} F_{00}(p,p+q)\Theta(p_F - |\overrightarrow{p}|)\delta(p_0 + q_0 - E_{\overrightarrow{p}+\overrightarrow{q}})\\ &\times\ \Theta(|\overrightarrow{p} + \overrightarrow{q}| - p_F). \end{aligned} \tag{B.27}$$

Dengan menggunakan hubungan $\overrightarrow{p}\cdot\overrightarrow{q} = |\overrightarrow{p}||\overrightarrow{q}|cos\theta = |\overrightarrow{p}||\overrightarrow{q}|x$ dan fungsi $\delta$ serta dengan memisalkan, $g(x) = p_0 + q_0 - E_{\overrightarrow{p}+\overrightarrow{q}} = 0$ maka diperoleh $E_{\overrightarrow{p}+\overrightarrow{q}} = E + q_0$ melalui konservasi arus. Dengan demikian maka hasil dari $\overrightarrow{p}\cdot\overrightarrow{q} = Eq_0 + \frac{q_\mu^2}{2}$. Dengan menggunakan hubungan tersebut maka didapatkan polarisasi longitudinal sebagai berikut

$$\Pi_L = \frac{q_\mu^2}{2\pi|\overrightarrow{q}|^3}[\frac{1}{4}(E_F - E^\star) + \frac{q_0}{2}(E_F^2 - E^{\star 2}) + \frac{1}{3}(E_F^3 - E^{\star 3})]. \tag{B.28}$$

### B.4.2 Polarisasi Transversal

Dengan menggunakan hubungan $\Pi_T = \Pi_{22} = \Pi_{33}$, maka persamaan $\Pi_{22}$ dapat diselesaikan sebagai berikut

$$\begin{aligned}\Pi_{22} &= \frac{-i}{(2\pi)^2}\int \frac{d^3p}{8E_{\overrightarrow{p}}E_{\overrightarrow{p}+\overrightarrow{q}}} F_{22}(p,p+q)\Theta(p_F - |\overrightarrow{p}|)\delta(p_0 + q_0 - E_{\overrightarrow{p}+\overrightarrow{q}})\\ &\times\ \Theta(|\overrightarrow{p} + \overrightarrow{q}| - p_F),\\ &= \frac{-i}{(2\pi)}\int \frac{|p|EdE}{2E_{\overrightarrow{p}}E_{\overrightarrow{p}+\overrightarrow{q}}}[2|\overrightarrow{p}|^2 sin^2\theta cos^2\phi + Eq_0 + \overrightarrow{p}\cdot\overrightarrow{q}]\\ &\times\ \delta(p_0 + q_0 - E_{\overrightarrow{p}+\overrightarrow{q}})\theta(p_F - |\overrightarrow{p}|)\theta(|\overrightarrow{p} + \overrightarrow{q}|),\\ &= \frac{-i}{(\pi)^2}\int \frac{|p|EdE}{2E_{\overrightarrow{p}}E_{\overrightarrow{p}+\overrightarrow{q}}}[|\overrightarrow{p}|^2 - |\overrightarrow{p}|^2 cos^2\theta + p\cdot q]\\ &\times\ \delta(p_0 + q_0 - E_{\overrightarrow{p}+\overrightarrow{q}})\Theta(p_F - |\overrightarrow{p}|)\Theta(|\overrightarrow{p} + \overrightarrow{q}| - p_F). \end{aligned} \tag{B.29}$$

dimana $|\overrightarrow{p}|cos\theta = \frac{2Eq_0 + q_\mu^2}{2|\overrightarrow{p}|}$ dan $p\cdot q = \frac{-q_\mu^2}{2}$, dengan demikian Pers.(B.29) dapat ditulis sebagai

$$\begin{aligned}\Pi_{22} &= \frac{-i}{2\pi}\int \frac{|p|EdE}{2E_{\overrightarrow{p}}E_{\overrightarrow{p}+\overrightarrow{q}}}[|\overrightarrow{p}|^2 - |\overrightarrow{p}|^2 cos^2\theta + p\cdot q]\\ &\times\ \delta(p_0 + q_0 - E_{\overrightarrow{p}+\overrightarrow{q}})\Theta(p_F - |\overrightarrow{p}|)\Theta(|\overrightarrow{p} + \overrightarrow{q} - p_F). \end{aligned} \tag{B.30}$$

Perhatikan suku pada [ ] dalam Pers.(B.30). Dengan menggunakan relasi konservasi arus dan sedikit manipulasi matematis dengan menggunakan $|\overrightarrow{p}|cos\theta = \frac{2Eq_0+q_\mu^2}{2|\overrightarrow{p}|}$, maka akan diperoleh persamaan polarisasi transversal sebagai

$$\begin{aligned} \Pi_T &= \Pi_{22} \\ &= \frac{1}{4\pi|\overrightarrow{q}|}[(M^2 + \frac{q_\mu^4}{4|\overrightarrow{q}^2} + \frac{q_\mu^2}{2})(E_F - E^\star) + \frac{q_0 q_\mu}{2|\overrightarrow{q}|^2}(E_F^2 - E^{\star 2}) \\ &+ \frac{q_\mu^2}{3|\overrightarrow{q}|^2}(E_F^3 - E^{\star 3})]. \end{aligned} \tag{B.31}$$

## B.5 Perhitungan $\Pi_{\mu\nu}^{IDA}(q)$

$$\begin{aligned} F_{\mu\nu}^A(q) &= 4[2p_\mu p_\nu + p_\mu q_\nu + p_\nu q_\mu - p{\cdot}q g_{\mu\nu} - 2M^2 g_{\mu\nu}] \\ &= F_{\mu\nu}^V(p, p+q) + g_{\mu\nu}F_A, \end{aligned} \tag{B.32}$$

dimana $F_A = -8M^2$, selanjutnya subsitusikan Pers.(B.32) ke Pers.(B.19), maka akan diperoleh persamaan sebagai

$$\begin{aligned} \Pi_{\mu\nu}^{IDA}(q) &= \frac{-i}{(2\pi)^2}\int \frac{d^3p}{8E_{\overrightarrow{p}}E_{\overrightarrow{p}+\overrightarrow{q}}} F_{\mu\nu}^A(p, p+q)\Theta(p_F - |\overrightarrow{p}|) \\ &\times \delta(p_0 + q_0 - E_{\overrightarrow{p}+\overrightarrow{q}})\Theta(|\overrightarrow{p} + \overrightarrow{q} - p_F), \\ &= \frac{i}{(2\pi)^2}\int \frac{d^3p}{8E_{\overrightarrow{p}}E_{\overrightarrow{p}+\overrightarrow{q}}}(F_{\mu\nu}^V(p, p+q) + g_{\mu\nu}F_A)\Theta(p_F - |\overrightarrow{p}|) \\ &\times \delta(p_0 + q_0 - E_{\overrightarrow{p}+\overrightarrow{q}})\Theta(|\overrightarrow{p} + \overrightarrow{q}| - p_F). \end{aligned} \tag{B.33}$$

Hubungan antara polarisasi aksial dan vektor dapat ditulis sebagai

$$\Pi_{\mu\nu}^{IDA}(q) = \Pi_{\mu\nu}^{IDV}(q) + g_{\mu\nu}\Pi_A. \tag{B.34}$$

Dengan demikian $\Pi_A$ dapat diselesaikan dengan singkat sebagai berikut

$$\begin{aligned} \Pi_A(q) &= \frac{-i}{(2\pi)^2}\int \frac{d^3p}{8E_{\overrightarrow{p}}E_{\overrightarrow{p}+\overrightarrow{q}}} F_A\Theta(p_F - |\overrightarrow{p}|)\delta(p_0 + q_0 - E_{\overrightarrow{p}+\overrightarrow{q}}) \\ &\times \Theta(|\overrightarrow{p} + \overrightarrow{q}| - p_F), \\ &= \frac{i}{2\pi}\int \frac{d^3p}{E_{\overrightarrow{p}}E_{\overrightarrow{p}+\overrightarrow{q}}} M^2\delta(p_0 + q_0 - E_{\overrightarrow{p}+\overrightarrow{q}})\Theta(p_F - |\overrightarrow{p}|) \\ &\times \Theta(|\overrightarrow{p} + \overrightarrow{q}|), \\ &= \frac{i}{2\pi|\overrightarrow{q}|}M^2[E_F - E^\star]. \end{aligned} \tag{B.35}$$

Dengan demikian, maka polarisasi longitudinal dan transversal secara singkat dapat ditulis dalam bentuk persamaan dibawah ini

$$\begin{aligned} \Pi_L^{IDA}(q) &= \Pi_L^{IDV}(q) + \Pi_A(q), \\ \Pi_T^{IDA}(q) &= \Pi_T^{IDV}(q) - \Pi_A(q). \end{aligned} \quad \text{(B.36)}$$

## B.6 Perhitungan $\Pi_{\mu\nu}^{IDV-A}(q)$

$$\begin{aligned} F_{\mu\nu}^{V-A}(p, p+q) &= 4i\epsilon_{\mu\nu\rho 0} q_\rho E + 4i\epsilon_{\mu\nu\rho 1} q_\rho (|\overrightarrow{p}| cos\theta)_1, \\ &= 4i\epsilon_{\mu\nu\rho 0} q_\rho E + 4i\epsilon_{\mu\nu\rho 1} (|\overrightarrow{p}| cos\theta)_1, \\ &= 4i\epsilon_{\mu\nu 10} |\overrightarrow{q}| E + 4i\epsilon_{\mu\nu 01} q_0 (|\overrightarrow{p}| cos\theta)_1, \\ &= 4i\epsilon_{\mu\nu 10} (|\overrightarrow{q}| E - q_0 |\overrightarrow{p}| cos\theta). \end{aligned} \quad \text{(B.37)}$$

Bila Pers.(B.37) disubsitusikan ke dalam Pers.(B.20), maka akan diperoleh persamaan sebagai

$$\begin{aligned} \Pi_{\mu\nu}^{IDV-A}(q) &= \frac{-i}{(2\pi)^2} \int \frac{d^3p}{8E_{\overrightarrow{p}} E_{\overrightarrow{p}+\overrightarrow{q}}} [4i\epsilon_{\mu\nu 10} (|\overrightarrow{q}| E - q_0 |\overrightarrow{p}| cos\theta)] \\ &\times \Theta(p_F - |\overrightarrow{p}|) \Theta(|\overrightarrow{p} + \overrightarrow{q}| - p_F) \delta(p_0 + q_0 - E_{\overrightarrow{p}+\overrightarrow{q}}). \end{aligned} \quad \text{(B.38)}$$

Dengan singkat, maka polarisasi vektor-aksial dinyatakan sebagai

$$\Pi_{\mu\nu}^{IDV-A}(q) = \frac{q_\mu^2}{8\pi |\overrightarrow{q}|^3} [(E_F^2 - E^{\star 2}) + q_0 (E_F - E^\star)]. \quad \text{(B.39)}$$

Sehingga, polarisasi total imajiner yang merupakan kontribusi dari vektor, aksial dan vektor-aksial dapat dinyatakan sebagai berikut:

$$\Pi_{\mu\nu}^{ID}(q) = C_V^2 \Pi_{\mu\nu}^{IDV}(q) + C_A^2 \Pi_{\mu\nu}^{IDA}(q) - 2C_V C_A \Pi_{\mu\nu}^{IDV-A}(q). \quad \text{(B.40)}$$

# Lampiran C

# Kontraksi

## C.1 Kontraksi antara $\Pi^{IDA}_{\mu\nu}(q)$ dan $L_{\mu\nu}$

Tensor neutrino dapat ditulis sebagai

$$L_{\mu\nu} = 8[2k_\mu k_\nu + (k\cdot q)g_{\mu\nu} - (k_\mu q_\nu + q_\mu k_\nu) \mp i\epsilon_{\mu\nu\alpha\beta}k^\alpha q^\beta]. \qquad (C.1)$$

Bila Pers.(B.40) dikontraksikan dengan Pers.(C.1), maka akan diperoleh persamaan sebagai berikut:

$$L_{\mu\nu}\Pi^{\mu\nu}_{ID} = C_A^2 L_{\mu\nu}\Pi^{\mu\nu}_{IDA} + C_V^2 L_{\mu\nu}\Pi^{\mu\nu}_{IDV} - 2C_V C_A L_{\mu\nu}\Pi^{\mu\nu}_{IDV-A}. \qquad (C.2)$$

### C.1.1 Kontraksi $L_{\mu\nu}\Pi^{\mu\nu}_{IDV}$

Jika kita lihat dalam matriks B.4 dimana hanya $\Pi_{00}$, $\Pi_{11}$, $\Pi_{22}$, $\Pi_{33}$ dan $\Pi_{01}$ serta $\Pi_{10}$ yang memberikan kontribusi terhadap polarisasi $\Pi^{iD}_{\mu\nu}$ sedangkan yang lain bernilai nol, maka dengan demikian kontraksi ini dapat dinyatakan sebagai

$$\begin{aligned} L_{\mu\nu}\Pi^{\mu\nu}_{IDV} &= L_{00}\Pi^{00} + L_{10}\Pi^{10} + L_{01}\Pi^{01} + L_{22}\Pi^{11} + L_{22}\Pi^{22} \\ &+ L_{33}\Pi^{33}, \end{aligned} \qquad (C.3)$$

Dengan menggunakan konservasi arus maka kita akan mendapatkan hasil akhir kontraksi tersebut sebagai

$$\begin{aligned} L_{\mu\nu}\Pi^{\mu\nu}_{IDV} &= \frac{-q_\mu^2}{|\overrightarrow{q}|^2}[8(2k_0(k_0 - q_0) + \frac{q_\mu^2}{2})]\Pi_L \\ &- \frac{8q_\mu^2}{|\overrightarrow{q}|^2}[2k_0(k_0 - q_0) + \frac{q_\mu^2}{2} + |\overrightarrow{q}|^2]\Pi_T. \end{aligned} \qquad (C.4)$$

### C.1.2 Kontraksi $L_{\mu\nu}\Pi^{\mu\nu}_{IDA}(q)$

$$\begin{aligned}
L_{\mu\nu}\Pi^{\mu\nu}(q) &= L_{00}\Pi^{00}_A + L_{11}\Pi^{11}_A + L_{01}\Pi^{01}_A + L_{10}\Pi^{10}_A + L_{22}\Pi^{22}_A \\
&+ L_{33}\Pi^{33}_A, \\
&= L_{\mu\nu}\Pi^{\mu\nu}_{IDV} + [L_{00} - L_{11}[L_{22} + L_{33}]]\Pi_A, \\
&= L_L\Pi_L + 2L_T\Pi_T + 8q^2_\mu\Pi_A, \\
&= L_{\mu\nu}\Pi^{\mu\nu}_{IDV} + 8q^2_\mu\Pi_A. \qquad \text{(C.5)}
\end{aligned}$$

dimana, $L_{\mu\nu}\Pi^{\mu\nu}_{IDV}$ dapat dilihat pada pers.(C.4).

### C.1.3 Kontraksi $L_{\mu\nu}\Pi^{\mu\nu}_{IDV-A}(q)$

$$\Pi^{\mu\nu}_{IDV-A}(q) = i\epsilon_{\mu\nu\alpha 0}q^\alpha\Pi_{VA}. \qquad \text{(C.6)}$$

Bila dikontraksi dengan tensor neutrino maka akan didapatkan persamaan sebagai berikut:

$$L_{\mu\nu}\Pi^{\mu\nu}_{IDV-A}(q) = L_{23}\Pi^{23}_{V-A} + L_{32}\Pi^{32}_{V-A}. \qquad \text{(C.7)}$$

Dari Pers.(C.7) dapat dilihat bahwa yang memberikan kontribusi pada kontraksi Pers.(C.7) adalah hanya $L_{23}\Pi^{23}$, $L_{32}\Pi^{32}$ sedangkan yang lain bernilai nol. Hasil akhir perhitungan kontraksi memberikan persamaan sebagai

$$L_{\mu\nu}\Pi^{\mu\nu}_{IDV-A}(q) = \mp[8q^2_\mu B\Pi_A]. \qquad \text{(C.8)}$$

dimana $B = 2k_0 - q_0$.

# Lampiran D

# Lintasan Bebas Rata-rata

## D.1 Perhitungan Tampang Lintang Diferensial

Persamaan tampang lintang seperti yang diperlihatkan pada Pers.(3.2) terdapat bagian imajiner yang berada dalam tanda kurung kurawal. Bagian ini dapat diselesaikan secara detail sebagai berikut

$$\begin{aligned} L_{\mu\nu}\Pi^{\mu\nu} &= Im(L_{\mu\nu}\Pi^{\mu\nu}), \\ &= -8q_\mu^2[AR_1 + R_2 + BR_3]. \end{aligned} \tag{D.1}$$

dimana,

$$\begin{aligned} R_1 &= (C_V^2 + C_A^2)(\Pi_L + \Pi_T), \\ R_2 &= C_V^2\Pi_T + C_A^2(\Pi_T - \Pi_A), \\ R_3 &= \pm 2C_V C_A \Pi_{VA}. \end{aligned} \tag{D.2}$$

Dengan demikian apabila Pers.(D.2) disubsitusikan ke dalam Pers.(D.1) dan ke Pers.(3.2), maka akan diperoleh persamaan tampang lintang sebagai

$$\begin{aligned} \frac{1}{V}\frac{d^3\sigma}{d^2\Omega' dE'_\nu} &= \frac{-G^2}{32\pi^3}\frac{E'_\nu}{E_\nu} Im(L_{\mu\nu}\Pi^{\mu\nu}), \\ &= \frac{G^2}{4\pi^3}\frac{E'_\nu}{E_\nu} q_\mu^2(AR_1 + R_2 + BR_3). \end{aligned} \tag{D.3}$$

dimana $B = 2k_0 - q_0$ dan $A = \frac{2k_0(k_0-q_0)+q_\mu^2/2}{|\overrightarrow{q}|^2}$. Dengan menggunakan relasi $\frac{1}{\lambda} \approx \sigma$ maka dari persamaan diatas dapat dihitung lintas bebas rata-rata.

# Lampiran E

# Konservasi Arus

## E.1 Konservasi Arus

### E.1.1 $q_\mu F^V_{\mu\nu}(p, p+q) = 0$

karena,$q_\mu F^V_{\mu\nu}(p, p+q) = 0$, maka

$$\begin{aligned} q_\mu[2p_\mu p_\nu + p_\mu q_\nu + p_\nu q - \mu - p{\cdot}q g_{\mu\nu}] &= 0 \\ 2p{\cdot}qp_\nu + q{\cdot}pq_\nu + q^2 p_\nu - p{\cdot}qq_\nu &= 0 \\ 2p{\cdot}qp_\nu &= -q^2 p_\nu \\ p{\cdot}q &= \frac{-q^2}{2}, \qquad \text{(E.1)} \end{aligned}$$

atau dengan cara lain Pers.(E.1) dapat ditulis menjadi $2p{\cdot}q = -q^2$ atau $\frac{-2p\cdot q}{q^2} = 1$. Bila disubsitusikan Pers.(E.1) ke dalam persamaan $F_{\mu\nu}(p, p+q)$ maka akan diperoleh

$$\begin{aligned} F^V_{\mu\nu} &= 4[2p_\mu p_\nu + p_\mu q_\nu + p_\nu q_\mu + \frac{q^2}{2} g_{\mu\nu}] \\ &= 4[2p_\mu p - \nu - 2\frac{2p{\cdot}q}{q^2}(p_\mu q_\nu + p_\nu q_\mu) + \frac{q^2}{2} g_{\mu\nu}] \qquad \text{(E.2)} \end{aligned}$$

### E.1.2 $q_\mu F^V_{\mu\nu}(p, p-q) = 0$

$$\begin{aligned} q_\mu[2p_\mu p_\nu - p_\mu q_\nu - p_\nu q_\mu + p{\cdot}q g_{\mu\nu}] &= 0 \\ 2p{\cdot}qp_\nu - p{\cdot}qq_\nu - q^2 p_\nu + p{\cdot}qq_\nu &= 0 \\ p{\cdot}q &= \frac{q^2}{2}, \qquad \text{(E.3)} \end{aligned}$$

Bila Pers.(E.3) disubsitusikan ke dalam $F_{\mu\nu}(p, p-q)$ maka dengan cara yang sama akan diperoleh hasil yang sama dengan Pers.(E.2). Dengan demikian, dengan menggunakan konservasi arus maka dapat dinyatakan relasi sebagai

$$F^V_{\mu\nu}(p, p+q) \quad = \quad F^V_{\mu\nu}(p, p-q). \tag{E.4}$$

# Referensi

[1] C. J. Horowitz and K. Wehberger, Phys. Rev. Lett. C **66** (1991) 272.

[2] R. Niembro, P. Bernardos, M. Lopez-Quelle and S. Marcos, Phys. Rev. C **64** (2001) 055802.

[3] V. N. Fomenko, L. N. Savushkin, H. Toki, to be published, (1998).

[4] S. Reddy and M. Prakash, arxiv:astro-ph/9610115 v2 (1996).

[5] W. Greiner and J. A. Mahrun, Springer-Verlag Berlin Heidelberg, New York, 1996.

[6] A. R. Bodmer, C. E. Price, Nucl. Phys. A **505** (1989) 123.

[7] J. D. Walecka, Oxford University Press, 1995.

[8] A. Bouyssy, S. Marcos and P. V. Thieu, Phys. Rev. Lett. C **50** (1984) 541.

[9] M. Rufa, P.-G. Reinhard, J. A. Mahrun, W. Greiner dan M. R. Strayer, Phys. Rev. C **38** (1988) 390.

[10] P. Finelli, N. Kaiser, D. Vretenar and W. Weise, arXiv:nucl-th/0307069 v2 (2003).

[11] J. L. Friar, D. G. Madland and B. W. Lynn, Phys. Rev. C **53** (1989) 3085.

[12] J. J. Rusnak and R. J. Furnstahl, arxiv:nucl-th/9708040 (1999).

[13] B. A. Nikolaus, *et. al.*, (1992).

[14] Y. Sugahara, H. Toki, Nucl. Phys. A. **579** (1994) 557.

[15] S. Gmuca, Z. Phys-Hadron and Nuclei. A **342** (1992) 387.

[16] B. D. Serot and J. D. Walecka, arxiv:nucl-th/9701058 v1 (1986).

[17] C. Akcay, Thesis, College of William and Mary, Virginia, 2002.

[18] K. Chung, C. S. Wang, A. J. Santiago and J. W. Zhang, Eur. Phys. J. A **9** (2000) 453.

[19] S. Akcay, Thesis, College of William and Mary, Virginia, 2002.

[20] B. Todd, to be published, (2002).

[21] M. Rufa, Dissertasion, Frankfurt Universiteit, 1989.

[22] R. J. Furnstahl and B. D. Serot, Phys. Lett. B**104** (1981) 339.

[23] T. Hoch, *et. al.*, Phys. Reports. **242** (1994) 253.

[24] J. I. Kapusta, Cambridge University Press, 1989.

[25] S. Gasiorowicz, Jhon Wiley and Sons. Inc, 1966.

[26] L. Mornas, arxiv:nucl-th/0210035 v2 (2002).

[27] L. Mornas and A. Perez, arxiv:nucl-th/0106058 v1 (2001).

[28] H. Kim, J. Piekarewicz and C. J. Horowitz, Phys. Rev. C **55** 1.

[29] M. Prakash and J. M. Lattimer, arxiv:astro-ph/0103095 (2001).

[30] U. Lombardo, C. Shen, N. V. Giai, W. Zuo, arxiv:nucl-th/0212037 v3 (2002).

[31] A. K. Dutt-Mazumder, A. Kundu, B. Dutta-Roy, T. De, Phys. Rev. C **53** (1996) 790.

[32] H. Dobereiner and P. -G. Reinhard, Phys. Lett. B **227** (1989) 305.

[33] H. Kurasawa and T. Suzuki, Phys. Lett. **154** B (1985) 16.

[34] M. Nakano, N. Noda, T. Mitsumori, K. Koide, H. Kouno, A. Hasegawa, L. Liu, Phys. Rev. C **56** (1997) 3287.

[35] C. J. Zhai, *et.al.*, (2001).

[36] K. Lim and C. J. Horowitz, Nucl. Phys. A **501** (1989) 729.

[37] K. Chung, C. S. Wang, A. J. Santiago and J. W. Zhang, arxiv:nucl-Th/0011025 (2002).

[38] A. Bhattavharya and S. Raha, Phys. Rev. C **53** (1996) 522.

[39] S. A. Chin, Ann. Phys. **108** (1977) 301.

[40] E. V. Dalen, A. E. L. Dieperink, J. A. Tjon, arxiv:nucl-th/0303037 v1 (2001).

[41] L. S. Calenza, A. Pantziris and C. M. Shakin, Phys. Rev. C **45** (1992) 205.

[42] P. Arumugam, B. K. Sharma, P. K. Sahu and S. K. Patra, arxiv:nucl-th/030946 v1 (2003).

[43] J. K. Bunta and S. Gmuca, arxiv:nucl-th/0309046 v1 (2003).

[44] S. Alessandro and F. Vissani, arxiv:astro-ph/0302055 v2 (2003).

[45] M. Prakash, *et.al.*, arxiv:astro-ph/0012136 (2000).

[46] C. J. Horowitz, K. Wehberger, Nucl. Phys. A **531** (1991) 665.

[47] C. J. Horowitz, J. Piekarewicz, Phys. Rev. Lett. C **86** (2001) 5647.

[48] J. C. Caillon, P. Gabinski and J. Labasouque, J. Phys. G. Part. Phys. **28** (2002) 189.

[49] T. Burvenich, D. G. Madland, J. A. Mahrun and P. -G. Reinhard, Journal of Nuclear and Radiochemical Science **1** (2002) 191.

[50] M. Prakash, *et.al.*, arxiv:nucl-th/9307006 v1 (1993).

[51] A. L. Fetter and J. D. Walecka, Mc.Graw-Hill Book Company, 1971.

[52] F. Halzen and A. D. Martin, Jhon Wiley, New York, 1984.

[53] G. Fabbri and F. Matera, Phys. Rev. C **54** (1986) 2031.

[54] J. C. Caillon and J. Labarsouque, Phys. Rev. C **61** (1999) 015203.